\documentclass[useAMS,usenatbib]{mnras}

\usepackage{amsmath,amssymb,graphicx,xcolor,catchfilebetweentags,bm,array,pstricks-add,arydshln}
\usepackage{subfigure,rotating,makeidx,changepage,upgreek,longtable,lscape,multirow,multicol,captcont}
\usepackage{calc,pgf,fp,verbatim,lastpage,booktabs,ifthen}
\usepackage{natbib}
\hypersetup{
        colorlinks   = true, 
        urlcolor     = magenta, 
        linkcolor    = blue, 
        citecolor   = blue 
}
\usepackage[flushleft]{threeparttable}
\usepackage[utf8x]{inputenc}
\usepackage[final]{pdfpages}
\usepackage{tabu}%
\newcounter{tkq}[section]
\newenvironment{tkq}[1][]{\refstepcounter{tkq}\par\textbf{Q\thetkq.#1 }\rmfamily}{}
\counterwithin*{tkq}{subsubsection}
\counterwithin*{tkq}{subsection}

\newcommand\xmm{{\it XMM-Newton}}
\newcommand\swift{{\it Swift}}

\newcommand\rosat{{\it ROSAT}}

\newcommand{\szp}{SZ~Psc}

\newcommand{\E}[1]{$\times~10^{#1}$}
\newcommand{\Pten}[1]{$10^{#1}$}
\newcommand\td{$\tau_{d}$}
\newcommand\tr{$\tau_{r}$}
\newcommand\ergs{$\rm{erg}~\rm{s}^{-1}$}       
\newcommand\arcs{\hbox{$^{\prime\prime}$}}    
\newcommand\rsun{$\rm{R}_{\odot}$}           
\newcommand\msun{$\rm{M}_{\odot}$}           
\newcommand\zsun{$\rm{Z}_{\odot}$}           
\newcommand\cts{count~s$^{-1}$}
\newcommand\ftools{{\tt FTOOLS}}
\newcommand\caldb{{\tt CALDB}}

\newcommand\apec{{\tt apec}}
\newcommand\vapec{{\tt vapec}}
\newcommand\tbabs{{\tt tbabs}}
\newcommand\cflux{{\tt cflux}}
\newcommand\xspec{{\tt XSPEC}}
\newcommand\U{\textit{U}}
\newcommand\B{\textit{B}}
\newcommand\V{\textit{V}}
\usepackage{numdef}
\num\def{\W1}{\textit{UVW1}}
\num\def{\W2}{\textit{UVW2}}
\num\def{\M2}{\textit{UVM2}}
\newcommand{\pcite}[1]{\protect\cite{#1}} 

\newcommand\chisq{$\chi^2$}

\newcommand\nh{$\rm{N_{H}}$}

\newcommand{\gsimeq}{\hbox{\raise0.5ex\hbox{$>\lower1.06ex\hbox{$\kern-0.92em{\sim}$}$}}}
\newcommand{\lsimeq}{\hbox{\raise0.4ex\hbox{$<\lower1.06ex\hbox{$\kern-0.92em{\sim}$}$}}}

\makeatletter
\newlength{\abovecaptionskip}%
\makeatother

\newcommand{\tabstart}{ \begin{table*} \begin{threeparttable} }
\newcommand{\tabstartNOSTAR}{\begin{table} \begin{threeparttable}}
\newcommand{\tabnote}{\begin{tablenotes}}
\newcommand{\tabend}{\end{tablenotes} \end{threeparttable} \end{table*}}
\newcommand{\tabendNOSTAR}{\end{tablenotes}\end{threeparttable}\end{table}}

\newcommand{\tabularstart}[2]{ \begin{tabular}{#1} \begin{tabular}{@{}#2} \hline }
\newcommand{\tabularend}{ \hline \end{tabular} \end{tabular} }

\title[Superflare on SZ Psc]{\swift\ and \xmm\ observations of an RS~CVn type eclipsing binary SZ~Psc: Superflare and coronal properties}
\author[Karmakar et al.]{Subhajeet Karmakar\textsuperscript{1,2,3\thanks{E-mail: \href{mailto:subhajeet09@gmail.com}{subhajeet09@gmail.com, sk@mira.org}}}, Sachindra Naik\textsuperscript{2}, Jeewan C. Pandey\textsuperscript{3}, and Igor S. Savanov\textsuperscript{4}\\
\textsuperscript{1} Monterey Institute for Research in Astronomy (MIRA), 200 Eighth Street, Marina, California 93933, USA\\
\textsuperscript{2} Astronomy \& Astrophysics Division, Physical Research Laboratory, Navrangapura, Ahmedabad 380009, India\\
\textsuperscript{3} Aryabhatta Research Institute of Observational Sciences (ARIES), Manora Peak, Nainital 263002, India\\
\textsuperscript{4} Institute of Astronomy, Russian Academy of Sciences, ul. Pyatniskaya 48, Moscow 119017, Russia\\
 } 

\begin{document}
\date{Accepted 2022 October 12. Received 2022 October 12; in original form 2022 January 6}
\pagerange{\pageref{firstpage}--\pageref{draft_end}} \pubyear{2022}
 \maketitle
 \label{firstpage}

\begin{abstract}
  We present an in-depth study of a large and long duration ($>$1.3~days) X-ray flare observed on an RS~CVn type eclipsing binary system \szp\ using observations from  \swift\ observatory. In the 0.35--10~keV energy band, the peak luminosity is estimated to be 4.2$\times$10$^{33}$~\ergs. The quiescent corona of \szp\ was observed $\sim$5.67~d after the flare using \swift\ observatory, and also $\sim$1.4~yr after the flare using the \xmm\ satellite. The quiescent corona is found to consist of three temperature plasma: 4, 13, and 48 MK. High-resolution X-ray spectral analysis of the quiescent corona of \szp\ suggests that the high first ionization potential (FIP) elements are more abundant than the low-FIP elements. The time-resolved X-ray spectroscopy of the flare shows a significant variation in the flare temperature, emission measure, and abundance. The peak values of temperature, emission measure, and abundances during the flare are estimated to be 199$\pm$11~MK, 2.13$\pm$0.05~\E{56}~cm$^{-3}$, 0.66$\pm$0.09~\zsun, respectively. Using the hydrodynamic loop modeling, we derive the loop length of the flare as 6.3$\pm$0.5~\E{11}~cm, whereas the loop pressure and density at the flare peak are derived to be 3.5$\pm$0.7~\E{3}~dyne~cm$^{-2}$ and 8$\pm$2~\E{10}~cm$^{-3}$, respectively. The total magnetic field to produce the flare is estimated to be 490$\pm$60~G. The large magnetic field at the coronal height is supposed to be due to the presence of an extended convection zone of the sub-giant and the high orbital velocity. 
\end{abstract}
 \section{Introduction}
 \label{sec:intro}
  Our current understanding of the mechanisms of stellar flares is developed based on the Sun. Stellar flares generally occur close to the active regions. Coronal plasma near the active regions confines in closed magnetic structure called ``loop''. The loops have a localized magnetic field of the order of a few kilo-Gauss and extend from the lower atmosphere of these active regions to coronal heights. The chromospheric footpoints of these loops are jostled by convective motions, whereas the loops get twisted and distorted depending on the local conditions until they undergo a magnetic reconnection process near the loop tops \citep[][]{Parker-88-ApJ-21}. The reconnection process results in a rapid and transient release of magnetic energy in the stellar corona, and the event is termed a flare. The flaring event is also associated with the acceleration of charged particles, which gyrate downward along the magnetic field lines producing synchrotron radio emissions. When these ionized beams collide with denser material at the chromospheric footpoints, hard X-ray photons are emitted. This also heats the chromospheric footpoints up to tens of MK, causing evaporation of material from the lower atmosphere. As a result, the density in the newly formed coronal loop is increased and emitted in soft X-ray and extreme UV wavelengths. 

Flares produced by the RS Canum Venaticorum (RS CVn) type binaries show many analogies with the solar flares \citep[][]{Hall-76-ASSL-3}. These binaries are close but detached systems, typically consisting of a G--K giant/subgiant and a late-type main-sequence/subgiant companion. Based on the rotation period (P), the RS CVn binaries are subdivided into three following categories: short-period (P~$\leq$~1~d), classical (1~d~$\leq$~P~$<$~14~d), and long-period (P~$\geq$~14~d) binaries. The tidal forces between the components of the RS CVn binaries can cause the rotational period to be synchronized with the orbital period. Moreover, a thicker convection zone of the evolved giant/subgiant component leads to a much higher level of magnetic activities in RS~CVn binaries than in the Sun and other late-type stars. Analysis of flares in cool giants and subgiants is, therefore, very important as it gives us an opportunity to probe the structure and dynamic behavior of the corona of RS~CVn binaries, which in principle can be significantly different from the corona of dwarf stars. Due to the lower surface gravity in the cool giants and subgiants than in the cool main-sequence stars, a larger scale height is yielded, which possibly allows a very extended corona to develop \citep[][]{Ayres-03-ApJ-3}.

  In this paper, we investigate a long-duration flaring event observed on an active, bright RS~CVn-type partial eclipsing double-line spectroscopic binary system SZ~Psc. The system consists of a spotted chromospherically active K1~IV subgiant with a radius 5.1~\rsun\ and mass 1.62 \msun\ and a less active F8~V companion with a radius 1.50~\rsun\ and mass 1.28~\msun\ \citep{Jakate-76-AJ}. Due to the higher brightness level, the K1~IV subgiant is considered as the primary, whereas the less luminous but comparatively hotter F8~V companion is considered as the secondary. The primary subgiant is filling 80--90\% of its Roche lobe. The system is located at a distance of 89.9$_{-0.6}^{+0.7}$~pc \citep[][]{Bailer-Jones-18-AJ-6}. The orbital period of the system is 3.9657 d \citep{Eaton-07-PASP-2}, which is supposed to be almost synchronous with the rotational period of the components of the tidally locked system. In recent years, a tertiary component of \szp\ system has been detected spectroscopically with the estimated mass and orbital period of  0.9~\msun\ and 1283$\pm$10~d, respectively \citep[][]{Xiang-16-MNRAS-4}. The luminosity contribution of the tertiary star to the system is estimated to be 5\% by  \cite{Xiang-16-MNRAS-4} which is in good agreement with the derived values of 3--4\% by \cite{Eaton-07-PASP-2}.

Since the first light variation was detected by \cite{Jensch-34-AN-2}, \szp\ remained an exciting source for photometric and spectroscopic observations throughout the last century \citep{Jakate-76-AJ, Catalano-78-IBVS, Tumer-79-IBVS, Eaton-82-Ap+SS-3, Tunca-84-Ap+SS, Antonopoulou-95-IBVS, Lanza-01-A+A-5}. Starspot modeling performed on \szp\ suggested the presence of several active regions on the surface of the cooler subgiant component \citep{Eaton-79-ApJ-1, Lanza-01-A+A-5, Kang-03-MNRAS-23}. The first Doppler images of SZ~Psc were generated by \cite{Xiang-16-MNRAS-4} using high-resolution optical spectra. These images revealed that the K1~IV star exhibits pronounced high-latitude spots as well as numerous intermediate- and low-latitude spot groups during the entire observing season. High level of chromospheric activity associated with the K1~IV primary component was also demonstrated by strong chromospheric emission in Mg~{\sc ii} h \& k, Ca {\sc ii} H \& K, H$\alpha$, and Ca {\sc ii} infrared-triplet lines \citep[e.g.,][]{Jakate-76-AJ, Bopp-81-PASP-8, Zhang-08-A+A-292, Cao-12-A+A-12}. H$\alpha$ outbursts and flare like events were detected by \cite{Bopp-81-PASP-8}, \cite{Ramsey-81-PASP-4}, and \cite{Huenemoerder-84-AJ-3}. The flaring events on \szp\ in the ultraviolet waveband were observed by \cite{Doyle-94-A+A-6}. The authors also found variation in Mg~{\sc ii} strength, possibly phase-dependent, and an apparent eclipse of a plage in Mg~{\sc ii}. However, excess absorption features in the subtracted H$\alpha$ profiles caused by prominence-like material have been discussed by \cite{Zhang-08-A+A-292} and \cite{Cao-12-A+A-12}. Several optical chromospheric activity indicators were analyzed by \cite{Cao-19-MNRAS-21} using the spectral subtraction technique, and during their observation, a series of possibly associated magnetic activity phenomena, including flare-related prominence activation, optical flare, and post-flare loops, were detected. The corona of SZ~Psc was studied for the first time by \cite{Walter-81-ApJ-8} in soft X-ray band using the imaging proportional counter (IPC) of the \textit{Einstein Observatory}. The authors also demonstrated that the RS~CVn systems, as a class, are the producers of copious soft X-ray emissions. In recent years, X-ray flaring activity has been observed on \szp\ using the Gas Slit Camera (GSC) of the Monitor of All-sky X-ray Image (MAXI). From the observations on 2009 September 28, \cite{Tsuboi-16-PASJ-4} analyzed a flare observed on \szp\ in 2--20~keV energy range with the flare energy of 5~\E{32}~\ergs. On 2011 November 5, another flare was detected with a larger X-ray luminosity of 3~\E{33}~\ergs\ \citep[][]{Negoro-11-ATel-7}. These detections of large flaring events show that the corona of \szp\ is active. Therefore, further investigation of the extreme events in \szp\ would help us understand the dynamic behavior of the corona of an evolved RS~CVn type binary system. 

\begin{table*}
\centering
\caption{The log of observations of \szp\ from \swift\ and  \xmm\ observatories.}
\setlength{\tabcolsep}{11.5pt}
\begin{tabular}{ccccccccr}
\toprule[1.5pt]
Observatory &  Obs. ID & Obs. Start time &  Phase$^\dagger$  &\multicolumn{4}{c}{Exposure Time}  & Offset                     \\ 
          &          &    (UTC)        &  ($\phi$)&\multicolumn{4}{c}{(ks)}        & \multicolumn{1}{c}{($'$)}  \\ 
\hline\hline 		

\multicolumn{1}{c}{\textit{Swift}} &&&& BAT & XRT & UVOT \\
\hline
&   00030738051 &   2015-01-15 04:27:58  &   0.954  & 6.98  &  --    &  --     &&   3.41  \\
&	00625898000	&	2015-01-15 08:52:56  &   0.143  &12.52  &  5.31  &	4.91   &&   3.41  \\
&	00625898001	&	2015-01-15 17:04:57  &   0.226  & 0.18  &  2.00  &	1.99   &&   2.05  \\
&	00625898002	&	2015-01-16 01:06:59  &   0.310  & 0.15  &  2.00  &	1.99   &&   2.23  \\
&   00084556002 &   2015-01-16 03:00:58  &   0.335  & 0.12  &  --    &  --     &&   2.23  \\
&	00625898003	&	2015-01-16 04:10:29  &   0.343  & --    &  1.99  &	1.98   &&   1.74  \\
&	00625898004	&	2015-01-16 07:37:31  &   0.378  & --    &  1.99  &	1.97   &&   0.62  \\
&	00625898005	&	2015-01-16 14:10:20  &   0.447  & --    &  2.00  &	1.99   &&   0.23  \\
&	00033606001	&	2015-01-21 00:54:59  &   0.569  & --    &  0.22  &	0.22   &&   3.94  \\
&	00033606002	&	2015-01-21 00:59:59  &   0.570  & --    &  2.00  &	1.98   &&   5.88  \\
\hline
\multicolumn{1}{c}{\xmm} &&&& PN & MOS & RGS &OM \\
\hline
&	0785140201  &	2016-05-26 20:49:07  &   0.594  & 13.47  &  13.65  & 13.78 & 3.89 &   0.05   \\
\bottomrule[1.5pt]
\label{tab:log}	
\end{tabular}
\vspace{-3.5mm}
\begin{tablenotes}
\item $^\dagger$ -- The phase is computed corresponding to the start time of each observation ID. The ephemeris has been adopted from \cite{Eaton-07-PASP-2}. 
\end{tablenotes}
\end{table*}


In this paper, we present a detailed time-resolved analysis of an X-ray superflare observed on \szp\ using the \textit{Neil Gehrels Swift Observatory} (hereafter \swift). We have also made use of the observations from the \xmm\ observatory obtained almost 1.4 yrs after the flaring event. This flare is identified as one of the largest and longest-duration flares ever observed on \szp. We have organized the paper as follows: The observations and data reduction procedure are discussed in \S~\ref{sec:obs}. The light curves and the phase-folded light curves have been discussed in \S~\ref{sec:lc} and \S~\ref{sec:phase}, respectively. We discussed the spectral analysis of the quiescent corona and the time-resolved analysis of the flaring corona in \S~\ref{sec:xray-spectra}. The loop modeling has been presented in \S~\ref{sec:loop-modeling}, whereas the coronal loop-properties are described in \S~\ref{sec:loop-prop}. Finally, in \S~\ref{sec:discussion}, we have discussed our results and presented the conclusions.

\section{Observations and Data Reduction} \label{sec:obs}
We have used X-ray, UV, and optical observations of \szp\ using \swift\ and \xmm\ observatories. The observations and the data reduction procedures adopted for each X-ray observatory are described below.

\subsection{\swift}
A flaring event on \szp\ triggered \swift's Burst Alert Telescope \citep[BAT;][]{Barthelmy-05-SSRv-33} as an Automatic Target. All the timing analysis in this paper is referenced to the BAT trigger time 2015 January 15 UT 09:08:42 \citep[reported by][]{DElia-15-GCN-1, Drake-15-ATel-4}, we refer to this as T0. In this work, we have utilised ten observation IDs from the \swift\ observatory, which include five observation IDs for BAT, eight observation IDs for X-Ray Telescope \citep[XRT;][]{Burrows-05-SSRv-26}, and eight observation IDs for ultraviolet/optical telescope \citep[UVOT;][]{Roming-05-SSRv-9}. A detailed log of these observations is given in Table~\ref{tab:log}. 

\subsubsection{BAT data}
The BAT is a large field of view, coded aperture imaging instrument which is highly sensitive in the 14--150~keV range. 
Although \szp\ was within the field of view of BAT between T0--16.861 and T0+110.024~ks, the source was bright enough to be detectable in hard X-ray between  T0--11.101~ks and  T0+57.659~ks. In order to perform spectral analysis, the source needs to be bright enough for which spectrum can be extracted with significant counts per bin. Therefore, for spectral analysis of \szp, we could only use the BAT observations from T0--0.127 to T0+0.957~ks.
We used the standard BAT pipeline software within \ftools\footnote{The \ftools\ software package provides mission-specific data analysis procedures; a full description of the procedures mentioned here can be found at \href{https://heasarc.gsfc.nasa.gov/docs/software/ftools/ftools_menu.html}{https://heasarc.gsfc.nasa.gov/docs/software/ftools/ftools\_menu.html}} version 6.29 with the latest \caldb\ version `BAT (20171016)'\footnote{For latest \caldb\ versions, please see \href{https://heasarc.gsfc.nasa.gov/docs/heasarc/caldb/swift/}{https://heasarc.gsfc.nasa.gov/docs/heasarc/caldb/swift/}} to correct the energy from the efficient but slightly non-linear energy assignment made on board. We used the {\tt batbinevt} task to extract the light curves in the 14--150~keV range. We mask-weighted the spectra in the 14--50~keV band, generated using the {\tt batmaskwtevt} and  {\tt batbinevt} tasks, with an energy bin of 80 channels. The BAT ray-tracing columns in spectral files were updated using the {\tt batupdatephakw} task. We applied the systematic error vector to the spectra from the calibration database using the {\tt batphasyserr} task. The BAT detector response matrix was computed using the {\tt batdrmgen} task. We created sky images in two broad energy bins using the tasks {\tt batbinevt} and {\tt batfftimage}. The {\tt batcelldetect} task was used to find the flux at the source position after removing a fit to the diffuse background and the contribution of bright sources in the field of view. The spectral analysis of all the BAT spectra was carried out in 14--50~keV range using the \xspec\ package \citep[version\footnote{\href{https://heasarc.gsfc.nasa.gov/xanadu/xspec/}{https://heasarc.gsfc.nasa.gov/xanadu/xspec/}} 12.12.0;][]{Arnaud-96-ASPC-2}.

\subsubsection{XRT data}
After the trigger, a slew placed the source in the apertures of the narrow-field instrument XRT, which started observing \szp\ from T0+0.381~ks. From the beginning of the XRT observation of \szp\ till $\sim$T0+110~ks, there are several data gaps mostly within the range of 15~min to 1.5~hr, except for three occasions where the data-gaps are for 3.5, 4.5, and 7.4~hr. After T0+110~ks, \swift\ returned again to the field of \szp\ after 5.67 days (i.e~T0+488.981~ks) and observed until T0+495.881~ks. During the $\sim$500~ks of total {\it Swift}/XRT observation, excluding the data gaps, \szp\ has a net effective exposure time of $\sim$17.5 ks. 

The XRT observation of \szp\ was in the energy range of 0.3--10 keV, using CCD detectors with an energy resolution of $\approx$140~eV at the Fe~K (6 keV) region. However, in this study, we have ignored data in the 0.3--0.35 keV range due to the known charge trapping effects of the XRT \citep[][]{Pagani-11-A+A-14}. We used the \swift ~{\tt xrtpipeline} task (version 0.13.6) to produce the cleaned and calibrated event files. All the data were reduced using the latest calibration files from the \caldb\ version `XRT (20210915)' release\footnote{For latest \caldb\ versions, please see \href{https://heasarc.gsfc.nasa.gov/docs/heasarc/caldb/swift/}{https://heasarc.gsfc.nasa.gov/docs/heasarc/caldb/swift/}}. The cleaned event lists generated with this pipeline are free from the effects of hot pixels and the bright Earth.

As \szp\ was very bright in X-rays during the flare, the data were recorded in Windowed Timing (WT) mode throughout the XRT observations. From the cleaned event list, images, light curves, and spectra for each observation were obtained using {\tt XSELECT} (version V2.4m) package\footnote{See \href{https://swift.gsfc.nasa.gov/analysis/}{https://swift.gsfc.nasa.gov/analysis/}}. We used grade 0--2 events in WT mode to optimize the effective area and hence the number of collected counts. As the peak count rate of the WT mode data of \szp\ is less than 100 counts per second, it was not necessary for further analysis to correct the data for pile-up \citep[][]{Romano-06-A+A-13}. The WT mode data give a one-dimensional image strip. In order to extract the source products, we considered a rectangular 40$\times$20~pixel region, i.e., 40 pixels long along the image strip and 20 pixels wide (where 1 pixel corresponds to 2\arcsec\!.36). In order to extract the background products, we selected a 40$\times$20~pixel rectangular region in the fainter end of the image strip. Taking into account the mirror effective area, filter transmission, vignetting correction, and point-spread function correction \citep[PSF;][]{Moretti-05-SPIE-11} as well as the exposure map correction, the \textit{ancillary response files} for the WT mode were generated using the task {\tt xrtmkarf}. In order to perform the spectral analysis, we used the latest \textit{response matrix files} \citep{Godet-09-A+A-3}, i.e., {\tt swxwt0to2s6\_20131212v015.rmf} for WT mode. All the XRT spectra were binned to contain more than 20 counts per bin using the {\tt grppha} task. The spectral analysis of all the XRT spectra was carried out in an energy range of 0.35--10~keV using \xspec. 

\subsubsection{UVOT data}
The \swift/UVOT began observing \szp\ from T0+0.363~ks in all six filters, which include three optical filters (i.e., \U, \B, and \V) and three UV filters (i.e., \W1, \W2, \M2). Among the UV filters, the image observed with the \W1\ filter is completely saturated, whereas the images observed with the \M2\ and \W2\ filters are partially saturated. The useful non-saturated observations with the \M2\ filter were found between T0+28.795~ks and T0+494.879~ks. Similarly, for the \W2\ filter, the observations between T0+490.065~ks and T0+495.824~ks were found to be unsaturated. The images observed with the  \U, \B, and \V\ optical filters were not saturated during the flare peak. However, \szp\ was not observed in any of the optical filters after T0+12.141~ks. The data reduction was performed using the latest calibration files from the \caldb\ version `UVOT (20201215)' release\footnote{For latest \caldb\ versions, please see \href{https://heasarc.gsfc.nasa.gov/docs/heasarc/caldb/swift/}{https://heasarc.gsfc.nasa.gov/docs/heasarc/caldb/swift/}}. In order to perform photometry for the non-saturated images, we used a circular region of 30\arcs\ radius co-axially centered on the intensity peaks. We also considered another circular region of the same radius from a blank area of the sky near the source position. The light curves were extracted using the {\tt uvotmaghist} task.

\begin{figure*} 
  \centering
\includegraphics[width=0.7\textwidth, angle=-90]{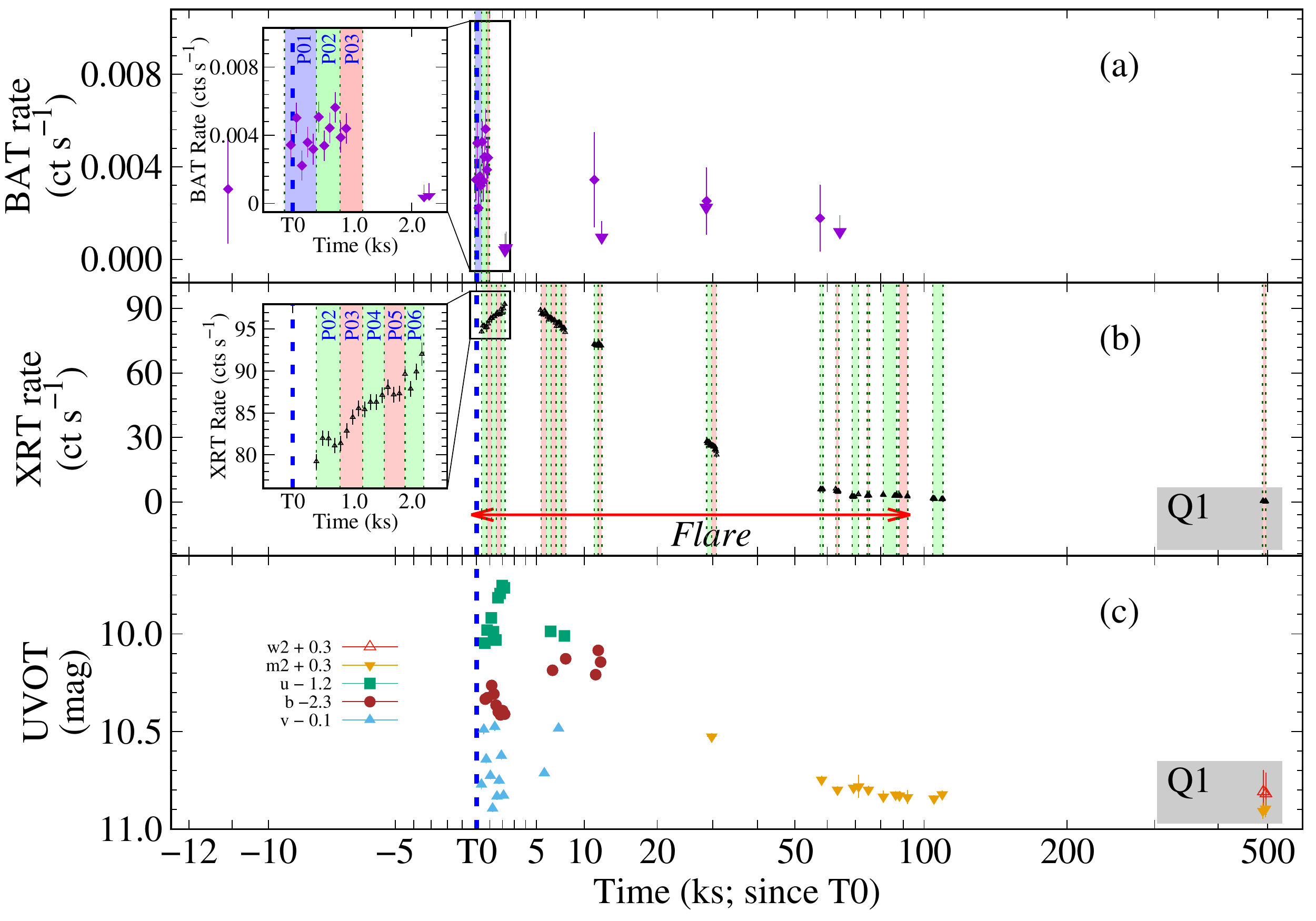}
\caption{
  \swift\ light curve of \szp\ have been shown. From top to bottom panels: (a) BAT, (b) XRT, and (c) UVOT light curves have been shown. Temporal binning for BAT light curves is 50~s, whereas the binning for XRT light curves is 200~s. The BAT light curve was extracted in the 14--150~keV energy band, whereas the XRT light curve was extracted in the 0.35--10~keV band. The non-saturated UVOT light curves for two UV filters (\W2, \M2) and three optical filters (\U, \B, \V) are shown in the bottom panel. The blue dashed vertical line in each panel indicates the trigger time, whereas black dotted vertical lines show the time intervals for which time-resolved spectroscopy was performed. The blue shaded region shows the segment for which only BAT spectral analysis is performed. Alternate orange and green shaded regions show the segments for which XRT (and XRT+BAT, wherever available) spectral analysis has been performed. The insets in panels (a) and (b) are the close-up views of the segments. The details of the close-up view are given in the text. The red horizontal arrow in panel (b) indicates the tentative duration of the flare in the soft X-ray energy band. The grey shaded regions in panels (b) and (c) show the quiescent state segment Q1, which has been analyzed in Section~\ref{sec:xray-spectra}.
}
\label{fig:lc_swift}
\end{figure*}

\subsection{\xmm}
Approximately 1.4 years after T0, \xmm\ observation of \szp\ was carried out on 2016 May 26 UT 20:49:07 (PI. Schmitt, J; ID: 0785140201). \xmm\ satellite has onboard  three  co-aligned X-ray telescopes \citep[][]{Jansen-01-A+A-3} and one co-aligned 30-cm optical/UV telescope, also known as optical monitor \citep[OM;][]{MasonK-01-A+A}. The X-ray telescopes consist of three European Photon Imaging Cameras \citep[EPIC;][]{Turner-01-A+A-7, Struder-01-A+A-2}, and two Reflection Grating Spectrometers \citep[RGS;][]{den-HerderJ-01-A+A}. For the first $\sim$10~ks, \szp\ was simultaneously observed with EPIC and RGS instruments, whereas the source was observed with all the instruments (EPIC, RGS, and OM) only for the last $\sim$4~ks. A log of the \xmm\ observation is given in Table~\ref{tab:log}.

\subsubsection{EPIC data}
The EPIC instrument onboard \xmm\ satellite contains two nearly-identical MOS~1 and MOS~2 \citep{Turner-01-A+A-7} detectors and one PN \citep{Struder-01-A+A-2} detector. It provides imaging and spectroscopy in the energy range of 0.15--15 keV with an angular resolution of 4\arcsec\!.5--6\arcsec\!.6 and a spectral resolution ($E/\Delta E$) of 20--50. \szp\ was observed for $\sim$14~ks using EPIC PN, MOS~1, and MOS~2 detectors. The EPIC data were reduced using standard \xmm\ Science Analysis System ({\tt SAS}) software\footnote{See SAS threads at \href{https://www.cosmos.esa.int/web/xmm-newton/sas-threads}{https://www.cosmos.esa.int/web/xmm-newton/sas-threads}} version 20.0.0, utilizing the latest version of calibration files. The pipeline processing of raw EPIC \textit{observation data files} (ODFs) was done using the {\tt epchain} and {\tt emchain} tasks, which allow calibrations both in energy and astrometry of the events registered in each CCD chip. We restricted our analysis to the 0.35--10 keV range as the background contribution is particularly relevant at high energies, where stellar sources have very little flux and are often undetectable. Event list files were extracted using the {\tt SAS} task {\tt evselect}. We used the {\tt epatplot} task to check for pile-up effects and found that the data are not affected due to photon pile-up and high background proton flare. The background was chosen from several source-free regions on the detectors around the source. We used {\tt epiclccorr} task to correct for the good time intervals, dead time, exposure, PSF, quantum efficiency, and background subtraction. We used the SAS task {\tt especget} to generate the spectra, which also computes the photon redistribution as well as the ancillary matrix. Finally, the spectra were rebinned to have a minimum of 20 counts per spectral bin using \ftools\ task {\tt grppha}. The spectral analysis of EPIC data has been carried out using \xspec\ in an energy range of 0.35--10 keV.

\subsubsection{RGS data}
The RGS consists of two identical grating spectrometers, RGS~1 and RGS~2. These spectrometers share about half of the photons in the converging beams that feed the X-ray telescopes with MOS~1 and MOS~2 detectors. In each spectrometer, the dispersed photons are recorded by a strip of eight CCD MOS chips. One of these chips has failed in each of the spectrometers, leading to gaps in the spectra. However, these gaps affect different spectral regions. \szp\ was observed simultaneously with EPIC and RGS instruments for $\sim$14~ks. The RGS instrument provides spectral resolution of $\approx$70--500 from 5--35 \AA\ (0.35--2.5 keV). In order to reduce the RGS data, we used SAS task {\tt rgsproc}. Standard processing was performed for spectral extraction and response matrix generation. In order to extract light curves, we used the SAS task {\tt rgslccor}. Using the \ftools\ task {\tt grppha}, all RGS spectra were binned to contain more than 10 counts per bin. The spectral analysis of RGS data has been carried out using \xspec\ in an energy range of 0.35--2.5 keV.  

\subsubsection{OM data}
The OM observation of \szp\ was started on 2016 May 26 UT 23:52:30, which is $\sim$10~ks after the beginning of the EPIC and RGS observations. The OM telescope module consists of a modified 30 cm Ritchey-Chretien telescope with a focal ratio of f/12.7. The wavelength coverage is between 170~nm and 650~nm in a 17\arcmin\ square field of view. The OM has six broadband filters (\U, \B, \V, \W1, \M2, and \W2), one white, one magnifier, and two grisms (UV and optical). \szp\ was observed only in the \W2\ filter for $\sim$4~ks until 2016 May 27 UT 00:57:31. The OM data for \szp\ only contained image mode and fast mode data. The OM data has been processed by the SAS tasks {\tt omichain}\footnote{\href{https://www.cosmos.esa.int/web/xmm-newton/sas-thread-omi}{https://www.cosmos.esa.int/web/xmm-newton/sas-thread-omi}} and {\tt omfchain}\footnote{\href{https://www.cosmos.esa.int/web/xmm-newton/sas-thread-omf}{https://www.cosmos.esa.int/web/xmm-newton/sas-thread-omf}} for reducing the image mode and fast mode data, respectively. The light curve is extracted from the fast-mode data. The mean background-subtracted source count rate is estimated to be 40.7~\cts, whereas the background count rate is estimated to be 0.3~\cts.


\section{Light curves} \label{sec:lc}
\subsection{Hard X-ray light curve}
The hard X-ray light curve in the 14--150~keV energy range obtained from \swift\ BAT is shown in the top panel of Figure~\ref{fig:lc_swift}. The inset in the panel shows a close-up view of the BAT light curve for the time interval marked by the rectangular region. The vertical blue dashed line shows the trigger time. The blue shaded region shows the time interval for which only BAT observation is available. The green and orange shaded regions show the time intervals for which both BAT+XRT observations are available. \szp\ was significantly bright in hard X-ray from $\sim$T0--0.127~ks. While the flare triggered BAT at T0, the BAT count rate reached up to 0.005$\pm$0.001~\cts. After the trigger, the BAT count rate gradually increased for $\sim$1~ks and reached a maximum count rate of 0.006$\pm$0.001~\cts. However, after a gap of $\sim$1~ks, the count rate was found to have dropped and reached only $\lsimeq$0.001~\cts. During the rest of the observation, the BAT light curve either shows detections with large uncertainties ($\sim$0.002--0.003~\cts) with a poor signal-to-noise ratio (S/N ratio $\sim$1--1.3) or shows non-detection with the upper limit ranging from $\sim$0.001--0.003~\cts. Therefore, the data quality does not allow us to identify any variability or non-variability of the BAT light curve from T0+0.957~ks to the end of the observation. 

\begin{figure}
  \centering
  \includegraphics[height=8.6cm, angle=-90, trim={0 1.1cm 0 0}, clip]{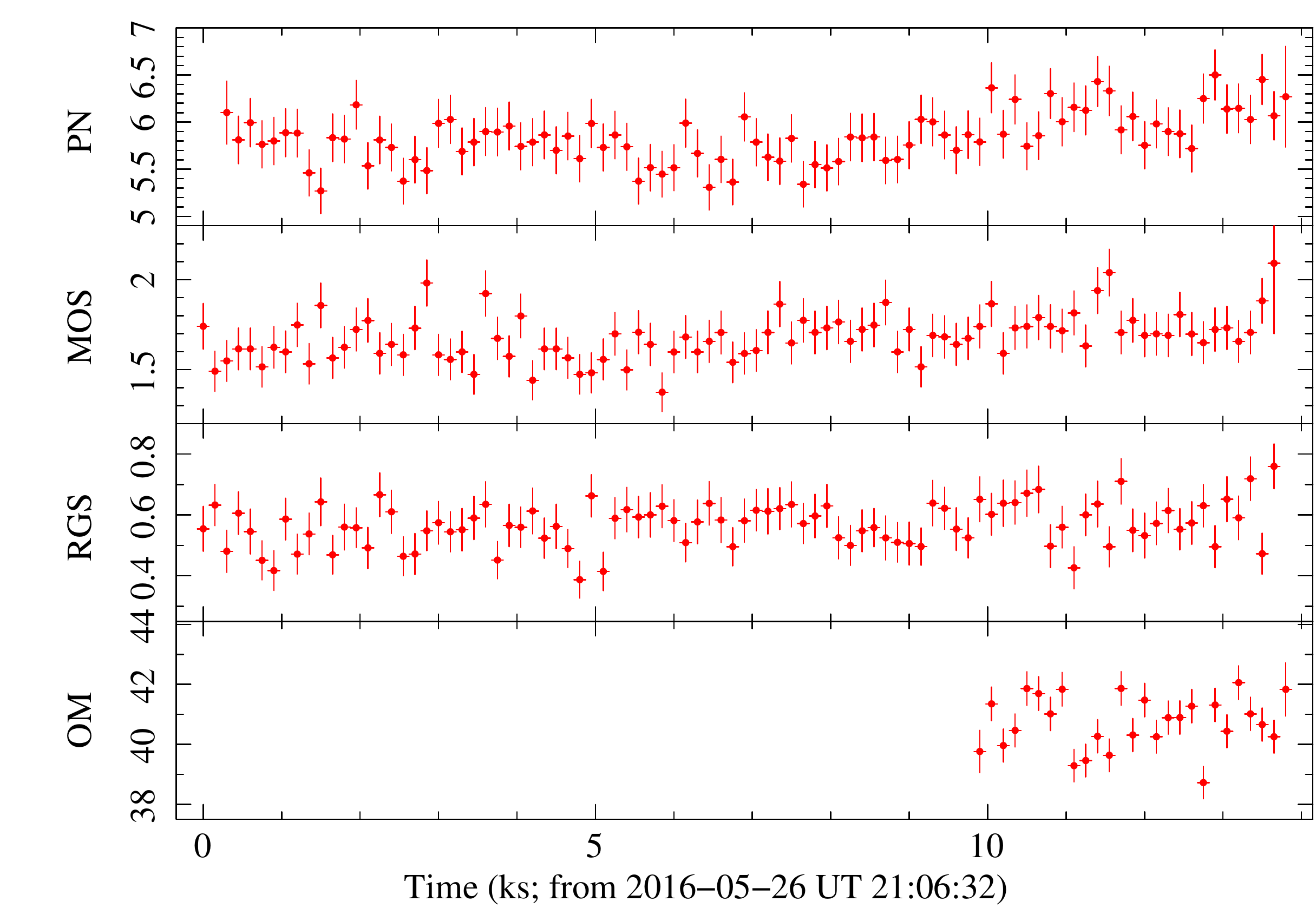}
  \caption{The \xmm\ light curves of the Q2 segment have been shown. The light curves obtained from EPIC~PN, EPIC~MOS, RGS, and OM instruments are shown from top to bottom panels, respectively. All the light curves are binned at 150~s. In the bottom panel, the UV light curve obtained from the OM observation is in the \W2\ filter.}
\label{fig:lc_xmm}
\end{figure}

\subsection{Soft X-ray light curve}
The 0.35--10~keV soft X-ray light curve obtained from \swift\-XRT observations of \szp\ is shown in  Figure~\ref{fig:lc_swift}(b). The inset in Figure~\ref{fig:lc_swift}(b) shows the close-up view of the XRT light curve for the same time interval as shown in the close-up view of the BAT light curve in Figure~\ref{fig:lc_swift}(a). The \swift\ XRT observed the flare on \szp\ from T0+0.381~ks in the WT mode, with an initial XRT count rate of $\sim$79 \cts. The XRT count rate increased up to $\sim$92~\cts until the observation was interrupted at around T0+2.21~ks due to the orbit of the \swift\ satellite. Although the XRT count rate showed an upward trend, the simultaneous BAT observations indicated that \szp\ was already faint in hard X-ray. After 3.24~ks of data-gap, while \swift\ XRT started observing \szp, the source was showing a decay from an initial count rate of $\sim$90~\cts. Until T0+7.971~ks, the XRT count rate decreased to $\sim$79~\cts. The \swift\ XRT count rate was found to decline rapidly and dropped to 28~\cts\ by T0+30~ks, whereas the count rate was $\sim$6~\cts\ by T0+57~ks. After a gap of $\sim$16 ks, the XRT observations for the time interval T0+58 ks to T0+92 ks showed a much slower decay of the flare during which the count rate decreased from  $\sim$7 \cts ~to 3 \cts. After $\sim$109~ks from T0, the soft X-ray count rate decreased to 1.3 \cts. The \swift\ returned to the field of SZ~Psc after 5.67 days from the trigger, where the  XRT count rate had dropped to $\sim$0.5 \cts. This epoch is shown with a grey shaded rectangle and marked as `Q1' in Figure~\ref{fig:lc_swift}, and appears to represent the quiescent corona of \szp. 

Due to the presence of data gaps in the light curves, it is difficult to confirm whether this is a single flaring event that occurred on \szp. For our analysis, however, we assume that the data from the {\it Swift} observations correspond to a single flare. The flare also shows an exponential rise and exponential decay. The e-folding rise time (\tr) and decay time (\td) of the flare in the XRT band are derived to be 14.4$\pm$0.5 and 21.4$\pm$0.3~ks, respectively. This shows a sharp rise and slow decay of the flare. Both of these values are comparable to or more than those of the observed flares in other G-K dwarfs, RS~CVn binaries, and dMe stars \citep[e.g.][]{Schmitt-94-LNP-6, Osten-99-ApJ, Pandey-12-MNRAS-8}. The flare duration is estimated to be $>$1.3~days, which is among the longest duration X-ray flares ever observed on \szp\ thus far.  

The background-subtracted 0.35--10 keV EPIC PN light curve of \szp\ is shown in the top panel of Figure~\ref{fig:lc_xmm}. In the second panel from the top, we plotted the MOS~1~+~MOS~2~(=~MOS,~hereafter) light curve in the same energy band. In the third panel of Figure~\ref{fig:lc_xmm}, 0.35--2.5 keV X-ray light curve obtained from combined RGS instruments (=~RGS~1~+~RGS~2) is shown. All of these light curves indicate that \szp\ remained at a  constant flux level during the $\sim$14~ks of observation in the soft X-ray energy band. In order to compare the \xmm\  observation with the \swift\ observations, we converted the \xmm\ PN count rate to an equivalent \swift\ WT count rate using the Mission Count Rate Simulator {\tt webpimms}\footnote{see~\href{https://heasarc.gsfc.nasa.gov/cgi-bin/Tools/w3pimms/w3pimms_pro.pl}{https://heasarc.gsfc.nasa.gov/cgi-bin/Tools/w3pimms/w3pimms\_pro.pl} for latest version of \texttt{webpimms}} (version PIMMS v4.11b). The corona of \szp\ is assumed to consist of astrophysical plasma. For the conversion, we used the coronal temperature components of \szp\ as estimated in this study and described in Section~\ref{subsec:qs}. We found that the converted count rate of \xmm\ PN is comparable within the 1$\sigma$ uncertainty level of the XRT count rate during the Q1 segment. Although the \xmm\ observation was carried out $\sim$1.4~yr after T0, due to the comparable count rate with the Q1 epoch, we consider this segment as the quiescent corona of \szp\ and term it as the `Q2' segment.

\begin{figure*}
\includegraphics[height=0.85\textwidth,angle=-90]{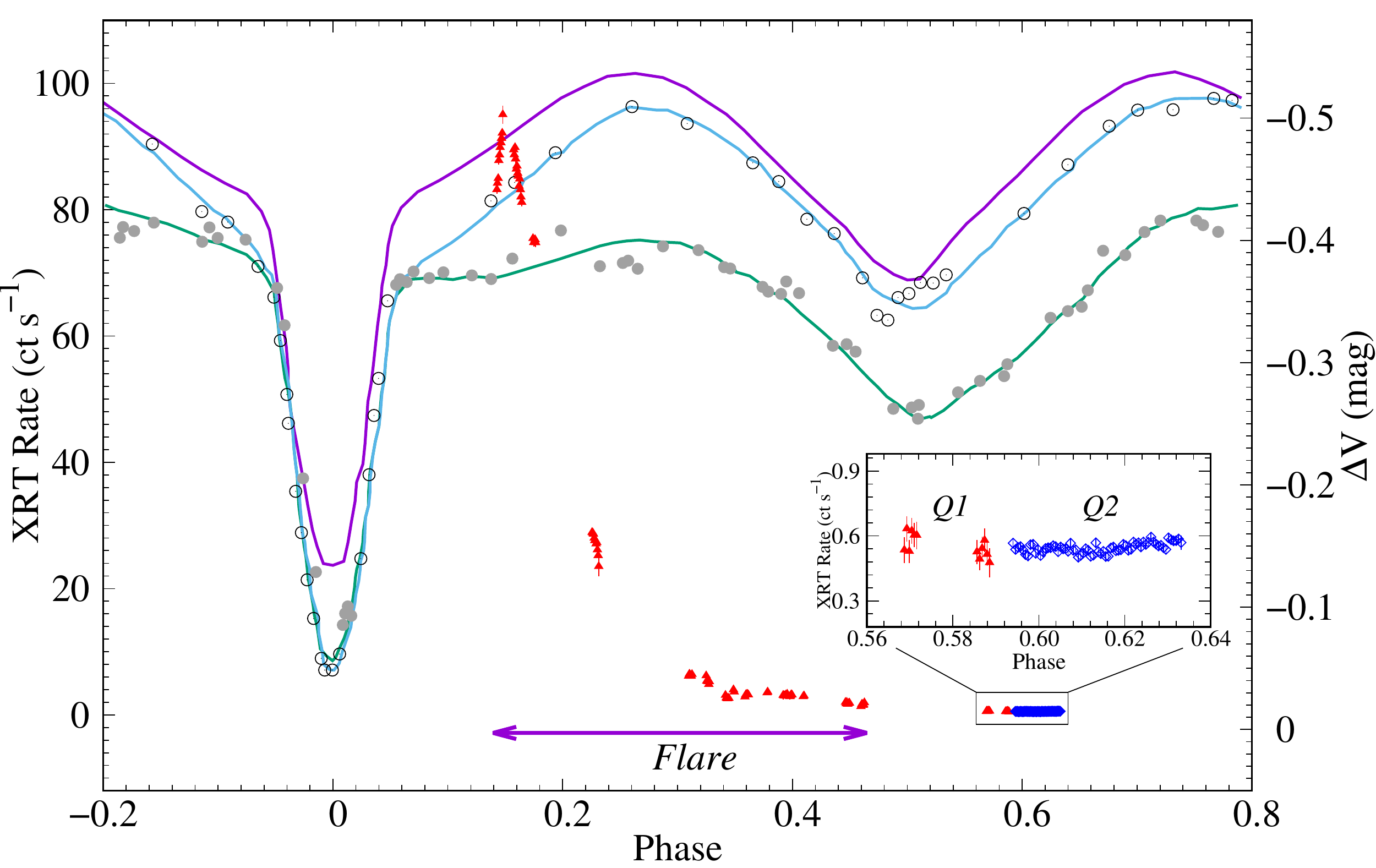}
\caption{Change in soft X-ray count rate and V-band magnitude of \szp\ with orbital phase is shown. 
The left and right sides along Y-axis in the figure indicate the \swift\ XRT count rate and differential V-band magnitude, whereas the X-axis shows the phase of \szp. Data from \swift\ XRT observations in WT mode are shown in solid red triangles. Earlier optical observations of \szp\ are plotted for the same phase in the right-hand Y-axis. The solid grey circles and solid green line represent the observations and modeled light curve of \szp\ from \pcite{Eaton-07-PASP-2}, whereas the black open circles and solid light-blue line represent the same for the observations from \pcite{Eaton-82-Ap+SS-3}. The solid purple line is the modeled light curve for a completely unspotted K star as adopted from Figure~2 of \pcite{Eaton-07-PASP-2}. The purple horizontal line represents the phase range during which the flare on \szp\ was observed with the {\it Swift} observatory. This shows that the flare is out of eclipse. The quiescent observations (Q1 and Q2) from the \swift\ and \xmm\ observations are found to be in the close phase range. The inset in the figure shows the zoomed version of the quiescent phase-folded light curve.}
\label{fig:phase_folded}
\end{figure*}

%
\subsection{UV and optical light curve}
The bottom panel of Figure~\ref{fig:lc_swift} shows the UV and optical light curves obtained from the \swift\ UVOT instruments. For most of the UV and optical filters, the error bars are smaller than the size of the symbols. Since all the images obtained using the \W1\ filter were heavily saturated, we could not extract light curves of \szp\ for this filter. The images obtained with \M2\ filter were not saturated from T0+28.790~ks till the end of the UVOT observation. Observations with \M2\ filter show magnitude variation from 10.2 to 10.5 mag, indicating the late decay phase of the flare. Although the \M2\ light curve seems to show an exponential decay in the later phase of the flare, we could not estimate the decay time since we don't have the information of the flare peak in this filter. The \W2\ observations were not saturated only in the quiescent segment `Q1'. During the observation, the \W2\ magnitude was found to be constant at 10.5$\pm$0.1 mag. 

  Among all the optical observations with \swift\ observatory, \U\ and \V-band light curves covered until T0+7.7 and T0+7.3~ks, whereas \B-band light curve had a longer coverage until T0+11.9~ks. The magnitudes of \U, \B, and \V-band light curves varied in the range of  11.2--10.9,  12.7--12.4, and 11.0--10.6 mag. The \U-band light curve showed a clear rise and decay phase, and the light curve seemed to follow the soft X-ray light curve. Whereas the \B\ and \V-band light curves showed a lot of scatter, and it is difficult to understand the time behavior in \B\ and \V\ optical bands. Due to the data gap and absence of observations in the later part of the flare, a reliable estimation of e-folding decay time was not possible in any of the optical bands. 
  During the Q2 segment, \szp\ was also observed in UV with the OM instrument of the \xmm\ satellite. The observation was carried out only in \W2\ waveband and is shown in the bottom panel of Figure~\ref{fig:lc_xmm}. This shows that the \W2-band light curve remains constant during the Q2 segment.

\section{Phase-folded light curve} \label{sec:phase} 
As SZ~Psc is an eclipsing binary, there is a certain chance that the {\it Swift} observations of the flare could have been affected due to the eclipse. In order to identify whether the flare is eclipsed or not, we phase folded the light curve using the ephemeris HJD~\!=~\!2,449,284.4483~\!+~\!3.96566356~\!$\times$~\!phase, adopted from \cite{Eaton-07-PASP-2}. Figure~\ref{fig:phase_folded} shows phase-folded light curves obtained from the XRT observations of \szp\ in WT mode (solid red triangles). For comparison, we also plotted the phase-folded $\Delta$V magnitude from earlier observations and shown on the right-hand y-axis. The solid grey circles and solid green line represent the observations and modeled light curve of \szp\ from \cite{Eaton-07-PASP-2}, whereas the black open circles and solid light-blue line represent the same for the observations from \cite{Eaton-82-Ap+SS-3}. 
The solid purple line is the modeled light curve for a completely unspotted K star as adopted from Figure~2 of \cite{Eaton-07-PASP-2}. We also plotted the \swift\ XRT equivalent of the \xmm\ PN detector data (count rate) from the segment Q2 using the open blue diamond in Figure~\ref{fig:phase_folded}. The bottom-right inset shows the close-up view of the phase range 0.56--0.64. We find that the flare peak and the decay of the flare are completely outside the eclipse. The peak of the flare was observed in the 0.14--0.2 phase range. We also identified that the XRT observations got over at phase 0.45, which possibly coincided with the end of the flare (i.e., T0+110~ks). Therefore, the entire flaring episode occurred outside the primary and secondary eclipses. We also identify that the Q1 (observed with \swift) and Q2 (observed with \xmm) segments are at phase ranges of 0.56--0.59 and 0.59--0.64, respectively. Therefore, both of these segments are also out of the secondary eclipse phase. 



\tabstart
\caption{The best-fitted parameters obtained from the spectral fitting of quiescent corona (Q1 and Q2 segments) of \szp. The best-fitted parameters obtained from fitting \apec ~1-T, \apec ~2-T, and \apec ~3-T models to the Q1 (\swift\ XRT spectra) and Q2 (\xmm\ EPIC PN+MOS spectra) segments are given below.}
\label{tab:par_qs_1}
\tabcolsep=0.4cm
\begin{tabular}{lccccccc}
\hline\hline\\[-2.5mm]
Parameters                          &		\multicolumn{3}{c}{Q1 (XRT)$^{**}$}                                               &		\multicolumn{3}{c}{Q2 (PN+MOS)$^{\dagger}$}                                 \\[0.5mm]                                                               
                                    &  APEC-1T                  &  APEC-2T                      &  APEC-3T                    &  APEC-1T      &  APEC-2T                      &  APEC-3T                        \\[1mm]  
                                                                                                                              \hline\\[-2mm]                                                                             
\nh\   ($10^{20}$ atoms cm$^{-2}$)	&  2.38$_{-0.01}^{+0.02}$   &	4.3$\pm$4.2                 &	6$_{-5}^{+7}$             &  2.41         &	2.4$_{-0.3}^{+0.4}$             &	3.3$\pm$0.5           \\ \\[-1.3mm]  
 kT$_1$	(keV)				        &  1.5$_{-0.3}^{+0.2}$      &	0.98$_{-0.12}^{+0.08}$ 	    &	$>$0.18        	          &  1.47	      &	1.024$\pm$0.008             	&	0.36$\pm$0.03	      \\[0.5mm]      
 EM$_1$	($10^{54}$ cm$^{-3}$)		&  1.3$\pm$0.3              &	1.0$_{-0.6}^{+0.5}$ 	    &	$<$1.17                   &  1.62	      &	1.1$\pm$0.2 		&	0.29$\pm$0.08 		  \\[0.5mm]                  
 log$_{10}$(L$_{\rm X1}$ in \ergs)	&  30.99$\pm$0.02           &   30.80$\pm$0.03              &   30.1$_{-0.2}^{+0.1}$      &  31.08	      & 30.873$\pm$0.007      &   30.07$\pm$0.01   \\ \\[-1.3mm]                 
 kT$_2$	(keV)				        &  --	                    &	$>$3.21               		&	1.1$\pm$0.1	          &  --	          &	4.2$_{-0.3}^{+0.5}$ &	1.13$\pm$0.02 		  \\[0.5mm]                      
 EM$_2$	($10^{54}$ cm$^{-3}$)		&  --	                    &	0.3$_{-0.1}^{+0.3}$         &	0.9$_{-0.3}^{+0.4}$       &  --	          &	0.44$\pm$0.09         &	0.9$\pm$0.2         \\[0.5mm]                    
 log$_{10}$(L$_{\rm X2}$ in \ergs)	&  --                       &   30.72$\pm$0.05              &   30.74$\pm$0.03            &  --           & 30.761$\pm$0.008      &   30.818$\pm$0.007   \\ \\[-1.3mm]               
 kT$_3$	(keV)				        &  --                       &  --                           &   $>$3.27                   &  --           &  --           	    &   4.1$_{-0.3}^{+0.5}$         \\[0.5mm]            
 EM$_3$	($10^{54}$ cm$^{-3}$)		&  --                       &  --                           &   0.3$_{-0.1}^{+0.2}$       &  --           &  --           		&   0.44$\pm$0.09         \\[0.5mm]                  
 log$_{10}$(L$_{\rm X3}$ in \ergs)	&  --                       &  --                           &   30.73$_{-0.05}^{+0.04}$            &  --           &  --                 &   30.767$\pm$0.008     \\ \\[-1.3mm]      
 Z	(\zsun)				            &  0.04$_{-0.03}^{+0.04}$	&	0.06$_{-0.03}^{+0.08}$      &	0.07$_{-0.03}^{+0.09}$    &  0.08	      &	0.076$\pm$0.007       &	0.11$\pm$0.01     \\[0.5mm]                      
 log$_{10}$(L$_{\rm X}$ in \ergs)	&  30.99$\pm$0.02           &   31.06$\pm$0.02              &   31.08$\pm$0.02            &  31.08        & 31.122$\pm$0.007      &  31.133$\pm$0.007    \\[0.5mm]                   
 \chisq	(DOF)				        &  1.702 (35)	            &	1.215 (33)    		        &	1.208 (31)   		      &  2.606 (992)  &	1.103 (987)    		&	0.993 (985)    	  \\ \\[-1.3mm]                  
 \hline\hline     
\end{tabular}
\tabnote
\item \textbf{Notes.}
The distance of \szp\ is considered as 89.9$_{-0.6}^{+0.7}$~pc, estimated by \cite{Bailer-Jones-18-AJ-6} using Gaia DR2 observations. 
\item All the errors shown in this table are in a 68\% confidence level.
\item The luminosities L$_{\rm X1}$, L$_{\rm X2}$, L$_{\rm X3}$ and L$_{\rm X}$ are derived in the 0.35--10 keV energy range.
\item  $^{\dagger}$--\xmm~observation was carried out at 2016-05-26 T21:06:32 during the phase range 0.59--0.64. 
\item  $^{**}$--\swift~observations were carried out at 2015-01-15 T09:08:42 during the phase range 0.56--0.59. The best fit was derived in the energy range 0.35--10~keV. 
\item  \nh, kT, and EM are the galactic H~{\sc i} column density, plasma temperature, and emission measures, respectively. $Z$ is the global metallic abundances relative to the solar photospheric abundances \citep[][]{Anders-89-GeCoA-2}.
  \tabend

\section{X-ray Spectral Analysis} \label{sec:xray-spectra}
In the following section, we provide a detailed description of the X-ray spectral analysis carried out in the present work. The spectra for two quiescent segments, Q1 and Q2, have been analyzed independently, whereas time-resolved spectroscopy is performed during the flare. All the uncertainties in the spectral fitting are estimated with a 68\% confidence interval ($\Delta$\chisq = 1), equivalent to $\pm 1\sigma$. In our analysis, the solar photospheric abundances (\zsun) were adopted from \cite{Anders-89-GeCoA-2}, whereas to model \nh, we used the photoionization cross-sections obtained by \cite{WilmsJ-00-ApJ-2}. 

\subsection{Quiescent spectra} \label{subsec:qs}
\subsubsection{\swift\ XRT Spectrum of Q1} \label{subsubsec:swift_q1}
The \swift\ XRT spectrum corresponding to the Q1 segment is shown in Figure~\ref{fig:spectra} with gray asterisks. The 0.35--10~keV spectrum is fitted with the \textit{Astrophysical Plasma Emission Code} \citep[\apec;][]{Smith-01-ApJ-96} available in \xspec. In order to estimate the hydrogen column density, we used the \textit{Tuebingen-Boulder ISM absorption model} \citep[\tbabs;][]{WilmsJ-00-ApJ-2}. Initially, the spectrum of the Q1 segment was fitted with a single (1-T), double (2-T), and triple (3-T) temperature plasma model with solar abundances. None of the plasma models (1-T, 2-T, or 3-T) were found to be acceptable with solar photospheric abundances due to large values of \chisq. In the next stage, the global abundance ($Z$) was left as a free parameter. We found that both the 2-T and 3-T plasma models with the sub-solar abundances fitted well, yielding \chisq\ within the acceptable range. However, spectral fitting with the 3-T model was relatively better than the 2-T model with an F-test value of 1.14 and an F-test probability of 66.7\%. 
Therefore, from the spectral analysis of the \swift\ XRT data, the quiescent corona of \szp\ seems to be represented by the three temperatures plasma. The estimated temperature components are $>$0.18, 1.1$\pm$0.1, and $>$3.27 keV, whereas corresponding Emission Measures (EMs) are estimated to be $<$1.17~\E{54}, 0.9$_{-0.3}^{+0.4}$~\E{54}, and  0.3$_{-0.1}^{+0.2}$~\E{54}~cm$^{-3}$. The best-fit value of global abundances was estimated to be 0.07$_{-0.03}^{+0.09}$~\zsun. In our analysis, \nh ~was a free parameter, and its value was estimated to be 6$_{-5}^{+7}$~\E{20}~atoms~cm$^{-2}$. 
Within a 0.1 degree cone in the direction of \szp, the survey of \cite{HI4PI-Collaboration.-16-A+A-1} suggests that the expected average Galactic H~{\sc i} column density to be 4.8$\pm$0.1~\E{20}~atoms~cm$^{-2}$, which is consistent with our estimated value. The best-fitted 3-T plasma model, along with the residuals, is shown with a solid black line in Figure~\ref{fig:spectra}. The best-fitted values of the derived parameters for all three models are given in the second, third, and fourth columns of Table~\ref{tab:par_qs_1}. In order to estimate the unabsorbed luminosities of individual temperature components, as well as the total unabsorbed luminosity of the quiescent corona of \szp, we used the \cflux\ model\footnote{For \texttt{cflux} model please see \href{https://heasarc.gsfc.nasa.gov/xanadu/xspec/manual}{https://heasarc.gsfc.nasa.gov/xanadu/xspec/manual}}. The unabsorbed luminosity for the Q1 segment is estimated to be 10$^{31.08\pm0.02}$ \ergs. The luminosities corresponding to the first, second, and third temperature components are estimated to be 0.1, 0.5, and 0.4 times the total luminosity. 


\begin{figure}
\begin{center}
  \includegraphics[height=0.48\textwidth, angle=-90, trim={0 1.4cm 0 0},clip]{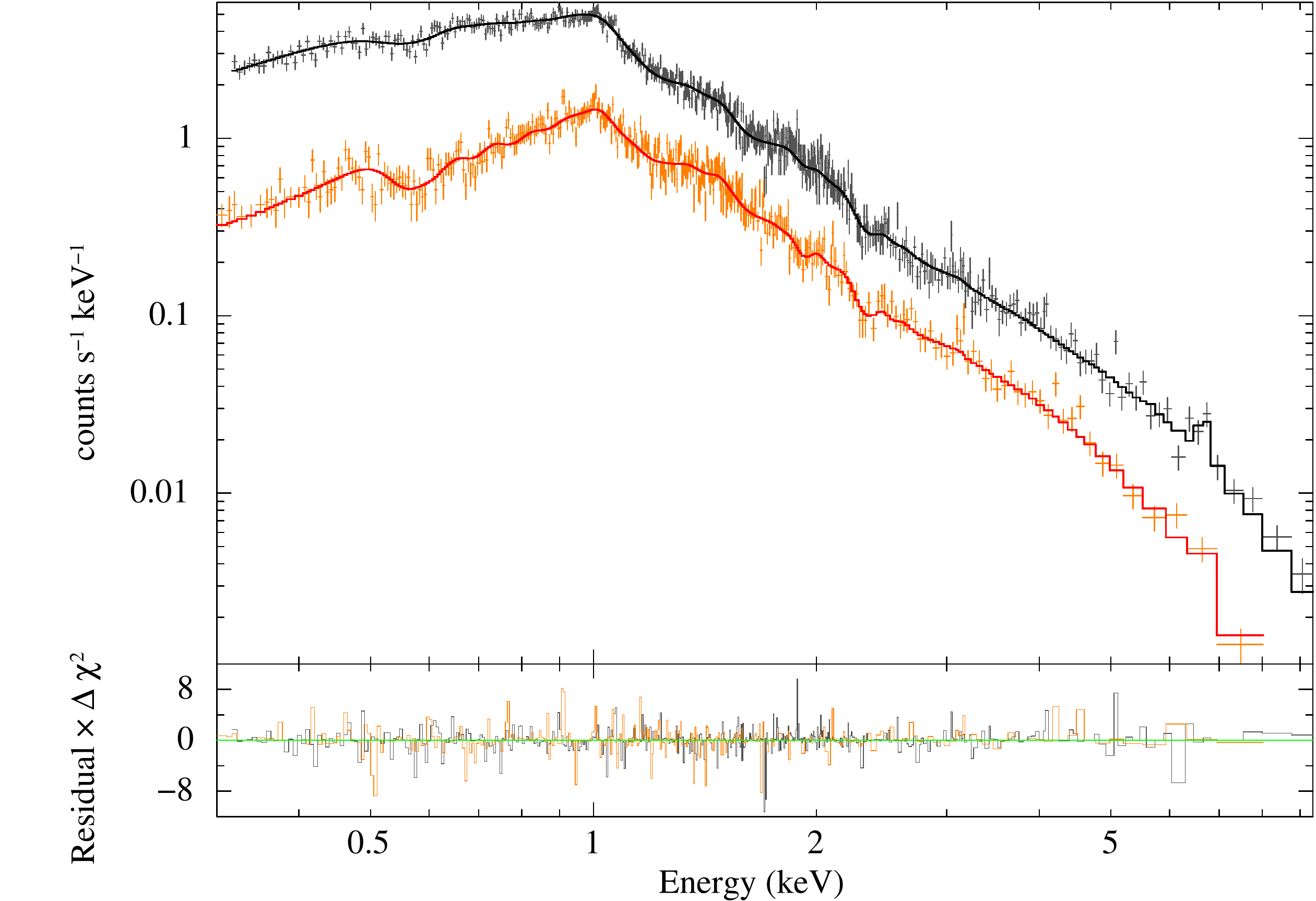}
  \caption{\label{fig:pn-mos_spec} The \xmm\ EPIC PN+MOS simultaneous spectra of \szp\ for the segment Q2 is shown along with the best-fitted 3-T \apec\ model. The gray and orange points in the figure correspond to the data from the PN and MOS instruments, respectively. The black and red solid lines in the upper panel show the corresponding best-fitted 3-T \apec\ models. The bottom panel shows the residuals in units of $\Delta \chi^2$. The best-fitted parameters are given in Table~\ref{tab:par_qs_1}.
}
\end{center}
\end{figure}
\begin{figure*}
\begin{center}
\includegraphics[height=1\textwidth, angle=-90, trim={0 1.cm 0 0},clip]{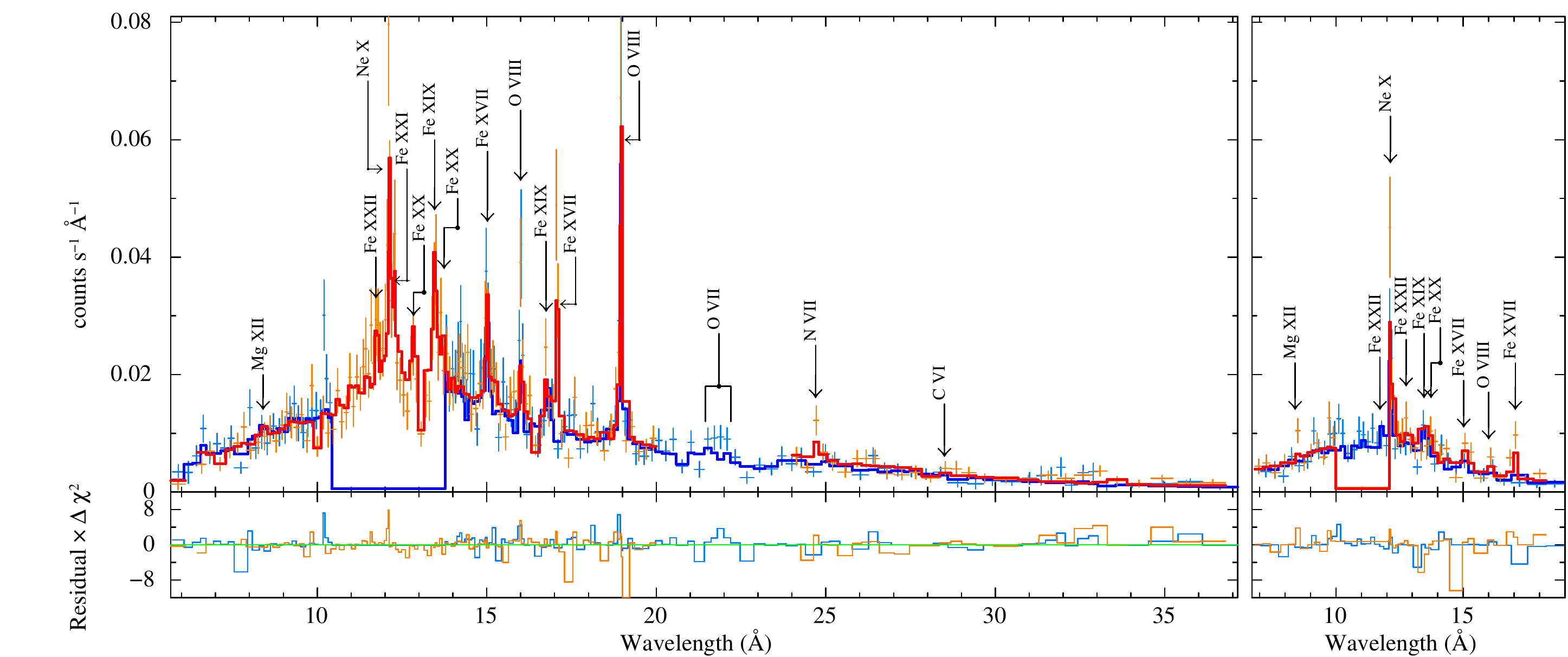}
\caption{\label{fig:rgs_spec} Time-integrated high-resolution \xmm\ RGS spectra of the quiescent segment (Q2) have been shown. The top-left panel shows the first-order spectra of RGS~1 (light-blue plus symbol) and RGS~2 (orange plus symbol). The top-right panel shows the second-order spectra with the same symbols. A simultaneous spectral fitting of RGS~1~+~RGS~2, first order + second order, has been performed. The best-fitted 3-T \vapec\ models are shown with blue and red solid lines for RGS~1 and RGS~2, respectively. The black arrows indicate the emission lines of different elements in different ionization states. The bottom panels of each figure show the corresponding residuals in the unit of $\Delta$\chisq. 
}
\end{center}
\end{figure*}

\subsubsection{\xmm\ EPIC Spectra of Q2} \label{subsubsec:qs_PN-MOS}
\szp\ was observed simultaneously with three EPIC detectors (i.e., PN, MOS~1, and MOS~2). As MOS~1 and MOS~2  are identical detectors and do not show any difference in the spectra of \szp, we extracted the combined spectra of MOS~1~+~MOS~2~(=~MOS) and used them for further spectral analysis. Since both of the EPIC instruments have a similar spectral resolution, the simultaneous spectral fitting has been performed for the EPIC PN and MOS spectra. Figure~\ref{fig:pn-mos_spec} shows PN and MOS spectra of the Q2 segment using gray and orange plus symbols. Initially, the spectra were fitted with 1-T, 2-T, and 3-T plasma models with solar abundances, which did not give an acceptable fit. When we let the global abundances vary freely, it converges to a sub-solar value and gives a better fit. However, the 1-T plasma model with sub-solar abundances does not give an acceptable fit. Although 2-T and 3-T plasma models both give a fit within the acceptable range, we found the 3-T model is relatively better than the 2-T model, with an F-test value of 54.5 and F-test probability $>$99.9\%. The fifth, sixth, and seventh columns of Table~\ref{tab:par_qs_1} summarise the best-fitted values of the derived parameters for all three models. The best-fit temperatures have been estimated as 0.36$\pm$0.03, 1.13$\pm$0.02, and 4.1$_{-0.3}^{+0.5}$~keV. Corresponding emission-measures were estimated to be 0.29$\pm$0.08~\E{54}, 0.9$\pm$0.2~\E{54}, and 0.44$\pm$0.09~\E{54}~cm$^{-3}$. The spectral fitting also gives an estimation of the global metallic abundance to be 0.11$\pm$0.01~\zsun. Since these values are similar to those estimated from the spectral fitting of the Q1 segment, it is more likely that the corona of \szp\ was in a similar activity level during Q1 and Q2 segments. This finding is important, as Q1 and Q2 segments were observed $\sim$1.4~yrs apart. This also suggests that both Q1 and Q2 correspond to the quiescent corona of \szp. It is noteworthy that the PN+MOS joint spectral fitting was performed with the \xmm\ data, which had better statistics than the \swift\ XRT data. The derived parameters for Q2 segments were also more precise than those estimated for the Q1 segment of \swift\ XRT. Therefore, for further analysis, we have considered the parameters estimated from the Q2 segment as the quiescent parameters. The best-fitted 3-T models for the PN and MOS detectors are shown with the black and red solid lines in the upper panel of Figure~\ref{fig:pn-mos_spec}, respectively. The bottom panel of Figure~\ref{fig:pn-mos_spec} shows the residuals from the best-fitted models. 

\tabstartNOSTAR
\caption{Spectral parameters of SZ~Psc for the quiescent phase Q2, derived from the \xmm\ RGS spectra}
\label{tab:par_qs_2}
 \tabcolsep=0.04cm 
\begin{tabular}{lclc}
\hline\hline\\[-2.5mm]
Parameters                      &   Values                  & 	Parameters                          &   Values                      \\[1.0mm]
\hline\\[-2mm]
\nh\   ($10^{20}$ cm$^{-2}$) 	&	3.34                    &   log$_{10}$(L$_{\rm X}$ in \ergs)	&   30.971$\pm$0.009  \\ \\[-1.3mm] 
 kT$_1$	(keV)				    &	0.36 	                &   EM$_1$	($10^{54}$ cm$^{-3}$)		&	0.29                          \\[0.5mm]
 kT$_2$	(keV)				    &	1.13 		            &   EM$_2$	($10^{54}$ cm$^{-3}$)		&	0.90                          \\[0.5mm]
 kT$_3$	(keV)				    &   4.06                    &   EM$_3$	($10^{54}$ cm$^{-3}$)		&   0.44                          \\ \\[-1.3mm]
He                              &   0.46$_{-0.07}^{+0.05}$  &   Si                                  &     $<$0.19      \\[0.5mm]
C                               &   $<$0.30                 &   S                                   &     $<$0.31      \\[0.5mm]
N                               &   0.3$\pm$0.1           &   Ar                                  &     $<$0.37      \\[0.5mm]
O                               &   0.10$\pm$0.01           &   Ca                                  &     $<$0.08     \\[0.5mm]
Ne                              &   0.28$\pm$0.04           &   Fe                                  &     0.076$\pm$0.006  \\[0.5mm]
Mg                              &   0.20$\pm$0.06           &   Ni                                  &     0.2$\pm$0.1    \\[0.5mm]
\chisq	(DOF)				    &	  1.114 (819)    		    \\ \\[-1.3mm]
%
 \hline\hline     
\end{tabular}
\tabnote
\item \textbf{Notes.}
  We have adopted the distance of \szp\ as 89.9$_{-0.6}^{+0.7}$~pc, estimated by \cite{Bailer-Jones-18-AJ-6} using Gaia DR2 observations. All the errors shown in this table are for a 68\% confidence interval.   Abundances relative to solar photospheric \citep[][]{Anders-89-GeCoA-2}.  The luminosities L$_{\rm X}$ are derived in the 0.35--2.5 keV energy range.\\
  \tabendNOSTAR

\begin{figure*}
  \center
  \includegraphics[height=17.5cm,angle=-90,trim={0 35 0 0}, clip]{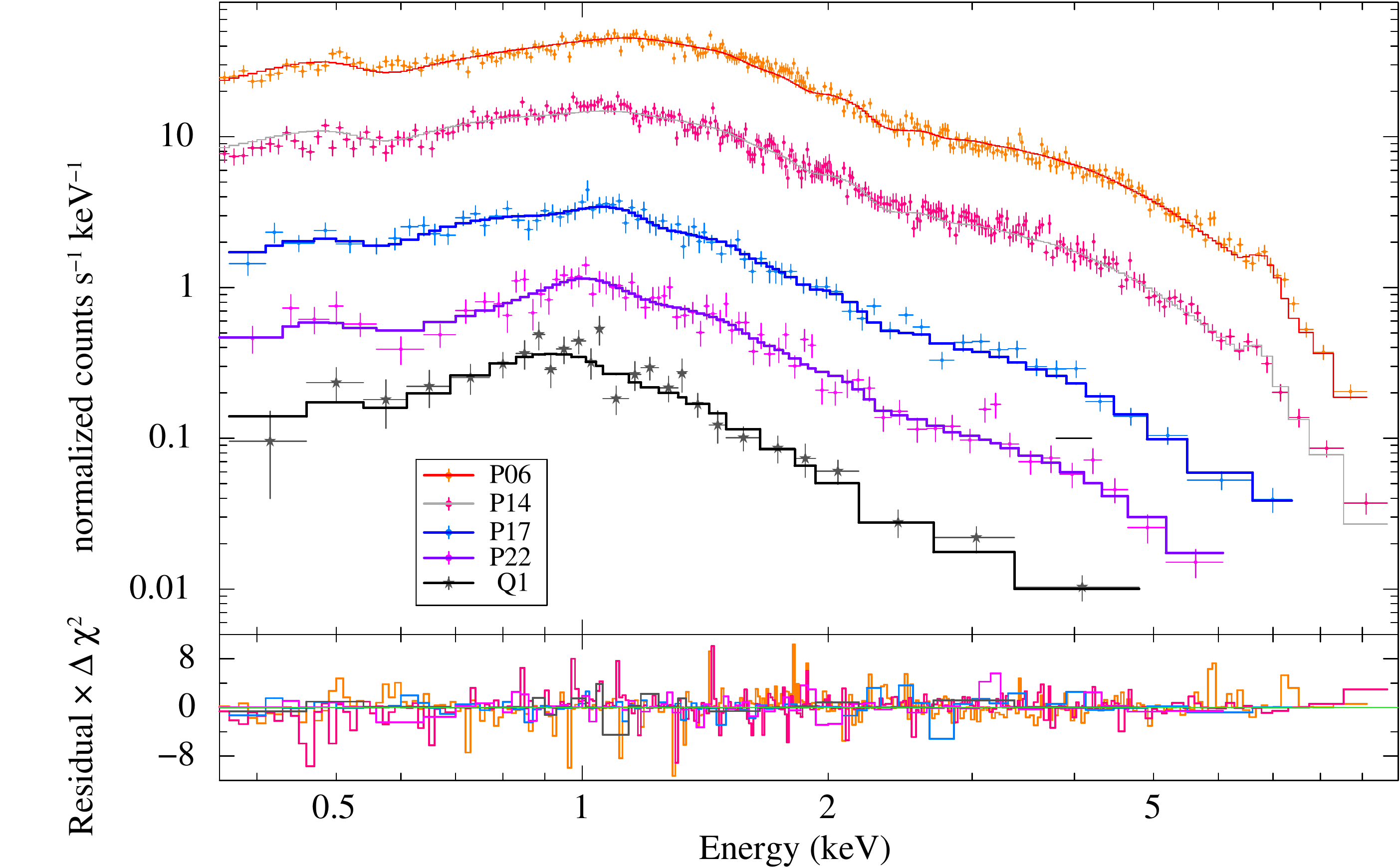}
  \caption{Time-resolved XRT spectra of \szp\ in 0.35--10 keV range are shown. The gray asterisks in the top panel show the spectra for the quiescent phase Q1 best fitted with a 3-T \apec\ model, where the model parameters have been shown in Table~\ref{tab:par_qs_1}. All other spectra are the representative flare spectra of different time segments, as mentioned in the inset of the top panel. The spectra corresponding to the flare segments are best-fitted with a 4-T \apec\ model as described in Section~\ref{subsubsec:trsxrt} and given in Table~\ref{tab:trs_all}. (Please also see the \textbf{\textit{online-only material}} of this paper, where the \swift\ XRT time-resolved spectra of all the segments have been shown along with the best-fitted models). 
}
\label{fig:spectra}
\end{figure*}
\subsubsection{\xmm\ RGS Spectra of Q2} \label{subsubsec:qs_RGS}
Simultaneously observed high-resolution RGS spectra of the Q2 segment in the 0.35--2.5 keV energy band are useful for estimating the elemental abundances and plasma density. In the left panel of Figure~\ref{fig:rgs_spec}, the first order spectra of RGS~1 and RGS~2 are shown with light-blue and orange plus symbols. At the same time, the right panel of Figure~\ref{fig:rgs_spec} shows the second-order spectra with the same colors and symbols. The emission lines are identified and marked with black arrows. Strong emission lines of iron, neon, oxygen, nitrogen, and magnesium are clearly detectable. We carried out simultaneous spectral analysis of both the orders of RGS~1 and RGS~2 data. In order to perform the spectral fitting, we used a 3-T \textit{Variable Astrophysical Plasma Emission Code} \citep[\vapec;][]{Smith-01-ApJ-96}. We fixed the temperatures, emission measures, and the \nh\ as derived from the EPIC PN+MOS simultaneous spectral fitting. The metallic abundances of helium, carbon, nitrogen, oxygen, neon, magnesium, silicon, sulfur, argon, calcium, iron, and nickel were allowed to vary freely and independently. Table~\ref{tab:par_qs_2} shows the derived parameters from the spectral fitting. The best-fitted models are shown in the top panel of Figure~\ref{fig:rgs_spec}, using blue and red solid lines for RGS~1 and RGS~2, respectively. The bottom panel shows the residual in the unit of $\Delta$\chisq. We have discussed it further in Section~\ref{sec:disc_abund}. 

Using the \xmm\ RGS spectra, we also investigated the electron densities of coronal plasma from the density-sensitive line ratios of forbidden to inter-combination lines of the helium-like triplets of \ion{O}{vii} \citep[][]{Gabriel-69-Natur-5}. If the electron collision rate is sufficiently high, the ions in the upper level of the forbidden transition do not return to the ground level radiatively. Instead, the ions are collisionally moved to the upper level of the inter-combination transitions, from where they eventually decay radiatively to the ground state. Therefore, the resulting ratio of the forbidden to the inter-combination line ($f/i$) is sensitive to density. As the He-like triplet of \ion{Ne}{ix} and \ion{O}{vii} were not strong enough in the RGS spectra of the quiescent corona of \szp, we could not estimate the precise coronal density by this method.

\subsection{Flare spectra: Time-Resolved Spectroscopy} \label{subsubsec:trsxrt}
In order to study the temporal evolution of the stellar parameters during the flare, we performed a time-resolved spectral analysis using \swift~ data. The entire flare duration is divided into twenty-two time segments and are shown by vertical shaded regions in Figure~\ref{fig:lc_swift}. These divisions are chosen in such a way that each segment contains a sufficient and similar number of total counts. The length of the time bins varies from 0.33 to 5.95~ks. Among those segments, one segment (P01) has BAT observation but no XRT observation, two segments (P02 and P03) have both XRT and BAT observations, whereas nineteen segments (P04 -- P22) have only XRT observations. The P01 segment that corresponds to only BAT observation is shown with the blue shaded vertical region in the top panel of Figure~\ref{fig:lc_swift}. All other segments are shown with alternate green and orange shaded vertical regions in the top two panels of Figure~\ref{fig:lc_swift}. The time intervals for which the X-ray spectra were accumulated are given in the second column of Table~\ref{tab:trs_all}. In this section, we discuss XRT, BAT, and XRT+BAT spectral analysis separately.

\subsubsection{XRT Spectral Analysis}
The \swift\ XRT spectra for a few representative time intervals\footnote{A complete set of best-fitted \swift\ XRT spectra corresponding to all the flare segments (P02--P22) and Q1 segment are provided in the \textit{online-only material}.} during the flare and the quiescent Q1 are shown in Figure~\ref{fig:spectra}. The spectral evolution is clearly visible throughout the flare. The X-ray emission during the flaring event has a contribution from both the flare and quiescent emission. Therefore, in order to estimate the `effective' contribution only from the flare, we performed spectral analysis by taking into account the quiescent emission as the frozen background emission. We fitted a four-temperature (4-T) astrophysical plasma model (\apec) to each of the flare segments keeping the first three temperature components fixed to the quiescent values as estimated in Section~\ref{subsec:qs}. A 4-T plasma model gives the best fit with the reduced \chisq\ in the acceptable range. Initially, in the spectral fitting, the \nh\ was a free parameter. The value of \nh\ was found to be constant during the flare and comparable with the quiescent state value (within a 1$\sigma$ uncertainty level). Therefore, in the next stage of spectral fitting, \nh\ was fixed to the quiescent state value along with the parameters of the first three temperature components. The time evolution of derived spectral parameters is shown in Figure~\ref{fig:trs_all}, and the values are given in Table \ref{tab:trs_all}. The temperature, corresponding emission measure, and abundances were found to vary during the flare. The peak value of global abundances ($Z$) was derived to be 0.66$\pm$0.09 \zsun\, which is $\sim$11 times that of the minimum value observed towards the end of the flare. The flare temperature (kT$_4$) peaked at a value of 17.1$^{+1.1}_{-0.9}$ keV, which is $\sim$3.4 times that derived at the end of the flare. Corresponding emission measure (EM$_4$) was found to follow the flare light curve, and the highest value was derived to be 2.13$\pm$0.05~\E{56}~cm$^{-3}$, which is almost ten times the minimum value observed at the end of the flare. The peak X-ray luminosity of the flare in the 0.35--10 keV energy range was derived using the {\tt cflux} model and found to be 4.16~\E{33}~erg~s$^{-1}$, which is $\sim$346 times as luminous as that of the Q1 segment. From Figure~\ref{fig:trs_all}, it is evident that the temperature peaks before the luminosity, emission measure, and abundances peak. This phenomenon is consistent with the idea of the hydrodynamic model \citep[see][]{Reale-07-A+A-2}. This indicates that the coherent plasma evolution and the heating cause the evaporation of the chromospheric gas and increase the metal abundances in the flaring loop. 
\begin{figure}
\centering
\includegraphics[height=8.2cm, angle=-90, trim={0 1.4cm 0 0}, clip]{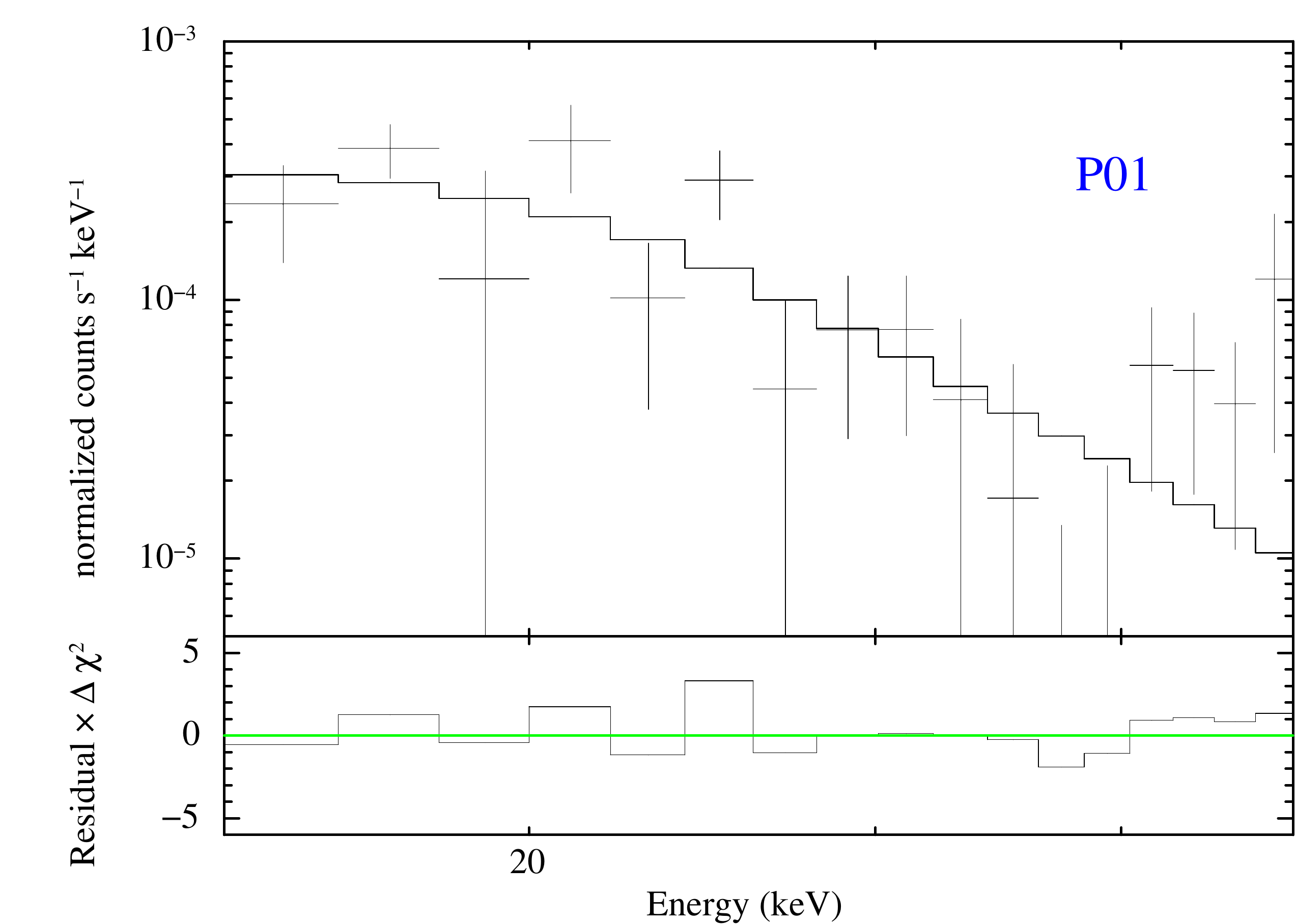}
\vspace{2mm}
\includegraphics[height=8.2cm, angle=-90, trim={0 1.4cm 0 0}, clip]{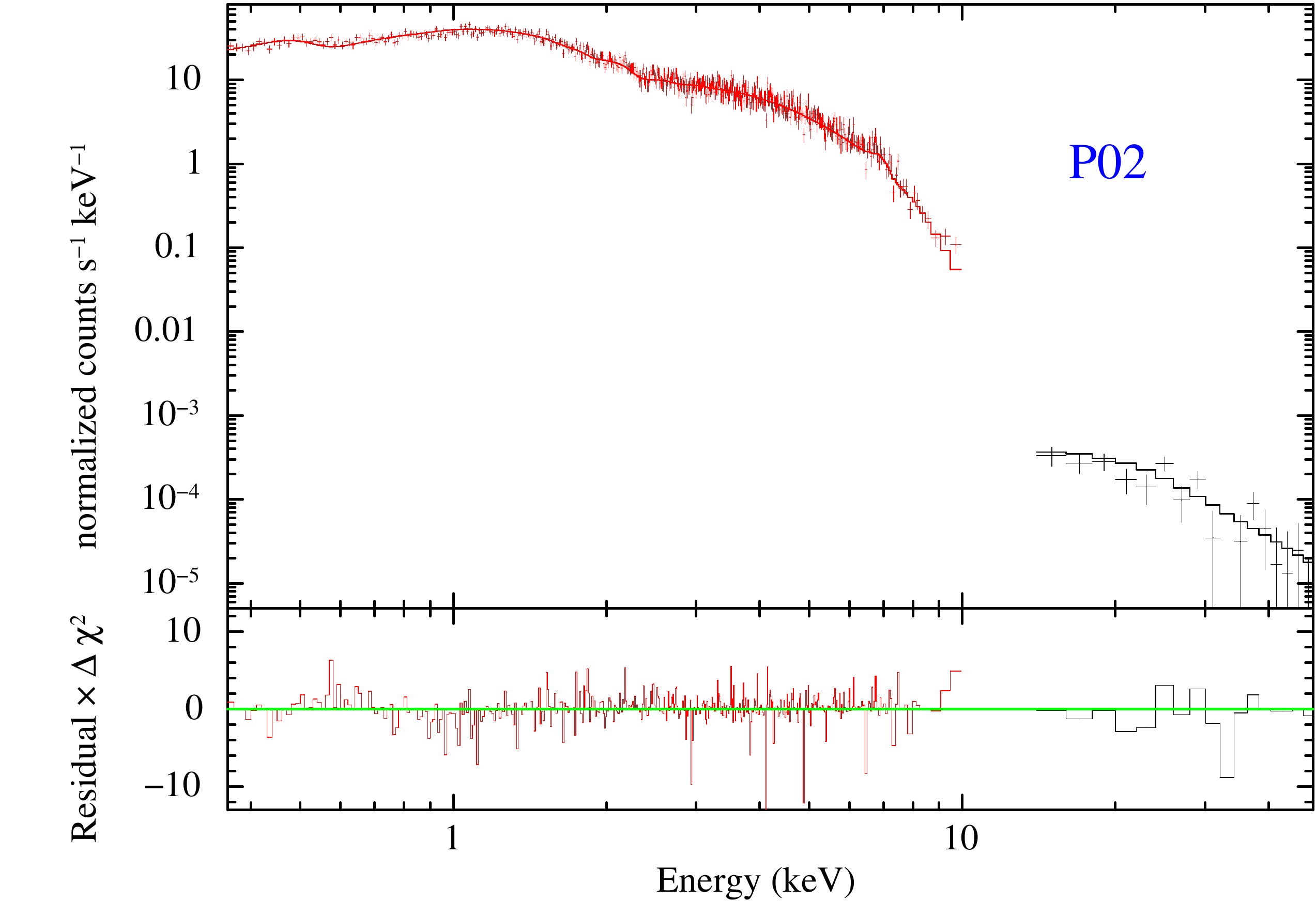}
\vspace{2mm}
\includegraphics[height=8.2cm, angle=-90, trim={0 1.4cm 0 0}, clip]{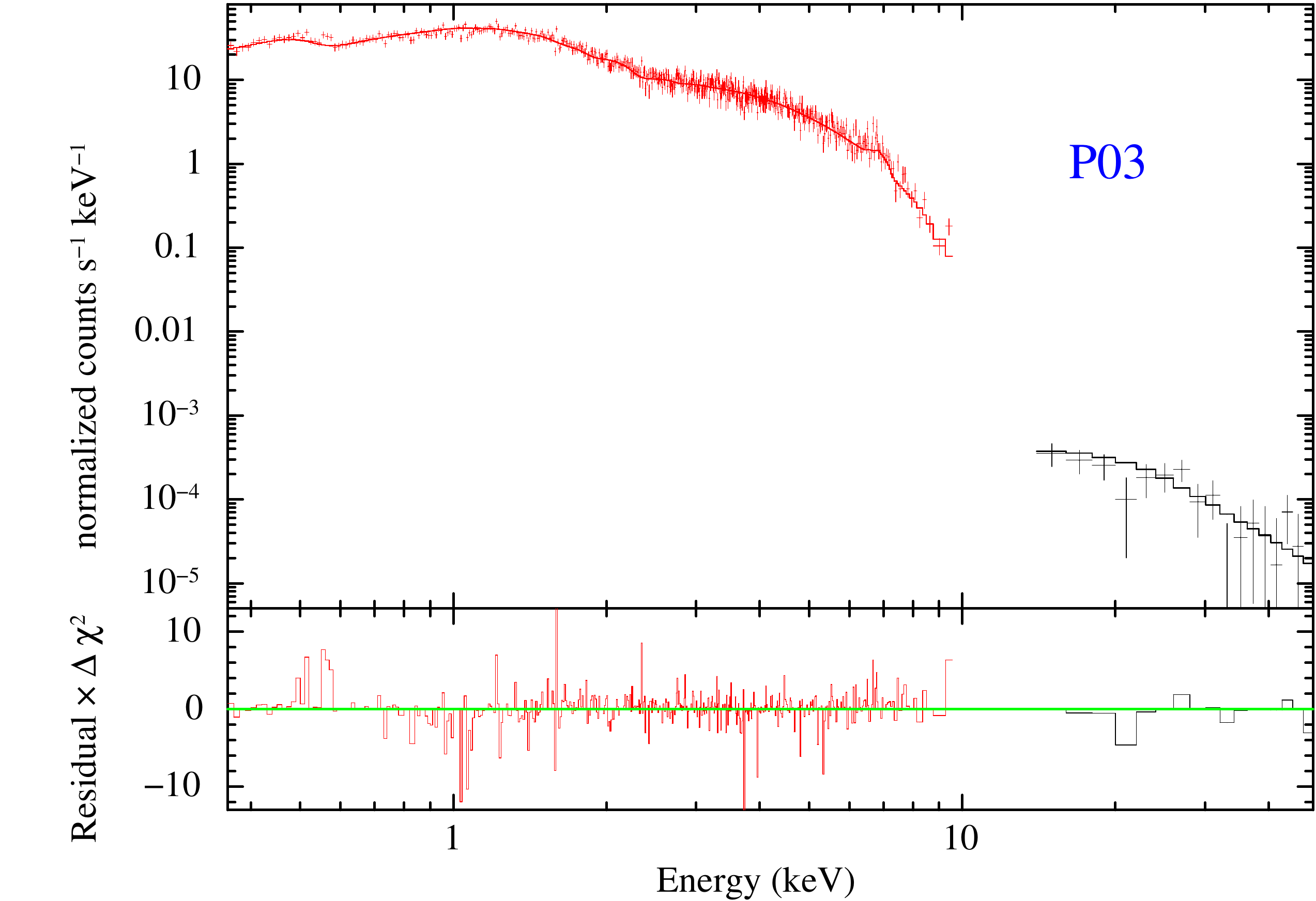}
\caption{Spectral fitting of the {\it Swift} BAT and XRT+BAT data. The figure at the top shows the  BAT spectra for segment P01. The middle and bottom figures show the combined XRT and BAT spectra for the segments P02 and P03, respectively. In the top panel of each figure, the spectra and best-fit 4-T model are shown. The bottom panel of each figure shows corresponding residuals.}
\label{fig:spec_BAT_XRT+BAT}
\end{figure}

\tabstart
\tabcolsep=0.4cm
\center
\caption{Time-resolved spectral parameters during the flare event of SZ~Psc}
\label{tab:trs_all}
\begin{tabular}{ccccccccccccccccccccc}
\toprule
\toprule
\multirow{2}{*}{$\mathbf{Parts}$}&
$\mathbf{Time~Interval}$&
$\mathbf{kT_4}$&
$\mathbf{EM_4}$&
$\mathbf{Z}$&
$\mathbf{log_{10}(L_{x})}$&
\multirow{2}{*}{$\mathbf{\chi^2~(DOF)}$}\\

$\mathbf{}$&
$\mathbf{(ks)}$&
$\mathbf{(keV)}$&
$\mathbf{(10^{55}~cm^{-3})}$&
$\mathbf{(\rm{Z}_{\odot})}$&
$\mathbf{(in~cgs)}$&
$\mathbf{}$\\

\midrule  
\midrule  
P01$^*$	        &   T0-0.127  :	T0+0.381  	 &		$14_{-4}^{+6}$	        &		$18_{-7}^{+11}$	&		0.41$^{**}$	                        &		$33.06_{-0.07}^{+0.06}$	&		1.136 (15)		\\[1mm]
\midrule  
P02$^{\dagger}$	&   T0+0.381  :	T0+0.771  	 &		15.5$_{-0.5}^{+0.6}$	&		19.5$\pm$0.5	&		0.44$_{-0.12}^{+0.13}$&	33.768$\pm$0.007	&		1.053 (532) 		\\[1mm]
P03$^{\dagger}$	&   T0+0.771  :	T0+0.957  	 &		15.0$\pm$0.6	        &		20.0$\pm$0.5	&		0.57$\pm$0.13	&		33.784$\pm$0.007	&		1.125 (528)		\\[1mm]
\midrule                                                                                                                                   
P02	&  T0 + 0.381	    :	T0 + 0.771       &		17.1$_{-0.9}^{+1.1}$	&		19.8$_{-0.5}^{+0.6}$&	0.41$\pm$0.14	&		33.574$\pm$0.007	&		1.028 (515)			\\[1mm]
P03	&  T0 + 0.771	    :	T0 + 1.151       &		15.7$\pm$0.8	        &		20.1$\pm$0.5	&		0.57$_{-0.13}^{+0.14}$&	33.592$\pm$0.007	&		1.133 (511)			\\[1mm]
P04	&  T0 + 1.151		:	T0 + 1.521       &		14.5$_{-0.6}^{+0.7}$	&		20.7$\pm$0.5	&		0.52$\pm$0.12	&		33.602$\pm$0.007	&		1.035 (513)			\\[1mm]
P05	&  T0 + 1.521		:	T0 + 1.881       &		14.1$_{-0.9}^{+0.7}$	&		21.3$\pm$0.5	&		0.43$\pm$0.11	&		33.606$\pm$0.007	&		1.079 (505)			\\[1mm]
P06	&  T0 + 1.881		:	T0 + 2.211       &		15.0$\pm$0.8	        &		21.2$\pm$0.5 	&		0.62$\pm$0.14	&		33.619$\pm$0.007	&		1.107 (494)			\\[1mm]
P07	&  T0 + 5.451		:	T0 + 5.941       &		12.9$_{-0.2}^{+1.3}$	&		19.9$_{-0.4}^{+0.7}$&	0.57$\pm$0.09	&		33.595$\pm$0.007	&		1.285 (553)			\\[1mm]
P08	&  T0 + 5.941		:	T0 + 6.441       &		11.6$\pm$0.4	        &		19.4$\pm$0.4	&		0.52$\pm$0.08	&		33.578$\pm$0.007	&		1.187 (550)			\\[1mm]
P09	&  T0 + 6.441		:	T0 + 6.951       &		12.0$_{-0.4}^{+0.5}$	&		19.0$\pm$0.4 	&		0.56$\pm$0.09	&		33.574$\pm$0.007	&		0.964 (553)			\\[1mm]
P10	&  T0 + 6.951		:	T0 + 7.471       &		11.8$\pm$0.4	        &		18.8$\pm$0.4 	&		0.49$\pm$0.08	&		33.563$\pm$0.007	&		1.051 (545)			\\[1mm]
P11	&  T0 + 7.471		:	T0 + 7.971       &		11.3$\pm$0.4        	&		17.7$\pm$0.4 	&		0.66$\pm$0.09	&		33.550$\pm$0.007	&		1.077 (539)			\\[1mm]
P12	&  T0 + 11.231		:	T0 + 11.691      &		12.4$\pm$0.5	        &		16.5$\pm$0.4 	&		0.53$\pm$0.10	&		33.510$\pm$0.007	&		1.179 (509)			\\[1mm]
P13	&  T0 + 11.691		:	T0 + 12.141      &		10.5$\pm$0.4	        &		16.1$\pm$0.3 	&		0.57$\pm$0.09	&		33.497$\pm$0.007	&		1.110 (503)			\\[1mm]
P14	&  T0 + 28.790		:	T0 + 29.750      &		7.8$\pm$0.3		        &		5.9$\pm$0.1	    &		0.42$\pm$0.06	&		33.040$\pm$0.007	&		1.133 (441)			\\[1mm]
P15	&  T0 + 29.750		:	T0 + 30.770      &		7.5$\pm$0.2		        &		5.7$\pm$0.1     &		0.44$\pm$0.06	&		33.015$\pm$0.007	&		0.987 (440)			\\[1mm]
P16	&  T0 + 57.686		:	T0 + 58.666      &		5.1$\pm$0.3		        &		1.10$\pm$0.04	&		0.35$_{-0.07}^{+0.09}$&	32.294$\pm$0.009	&		1.020 (198)			\\[1mm]
P17	&  T0 + 62.976		:	T0 + 63.966      &		5.0$_{-0.3}^{+0.4}$		&		0.99$\pm$0.04	&		0.42$\pm$0.06	&		32.221$\pm$0.009	&		0.785 (173)			\\[1mm]
P18	&  T0 + 68.743		:	T0 + 71.233      &		5.3$_{-0.6}^{+0.7}$		&		0.41$\pm$0.02	&		0.20$\pm$0.05	&		31.91$\pm$0.01	&		1.427 (106)			\\[1mm]
P19	&  T0 + 74.523		:	T0 + 75.383      &		4.3$\pm$0.4		        &		0.61$\pm$0.03	&		0.06$\pm$0.05	&		31.98$\pm$0.01	&		0.926 (99 )  		\\[1mm]
P20	&  T0 + 81.038		:	T0 + 86.988      &		4.7$\pm$0.4		        &		0.51$\pm$0.03	&		0.25$\pm$0.05	&		31.98$\pm$0.01	&		1.016 (139)			\\[1mm]
P21	&  T0 + 87.968		:	T0 + 91.998      &		3.8$_{-0.5}^{+0.6}$		&		0.51$\pm$0.05	&		0.17$\pm$0.08	&		31.93$\pm$0.01	&		0.875 (77 )  		\\[1mm]
P22	&  T0 + 104.574	    :	T0 + 110.024     &		5.0$_{-0.5}^{+0.7}$		&		0.22$\pm$0.01	&		0.15$\pm$0.03	&		31.68$\pm$0.01	&		0.969 (113)			\\[1mm]
                                                                                                                                          
\bottomrule                                                                                                                               
\bottomrule                                                                                                                               
\end{tabular}

\tabnote
\item \textbf{Notes.}
 All the errors shown in this table are for a 68\% confidence interval. The spectral fitting is performed in the 0.35--10~keV range for XRT observations (4th row to end of the table). 
\item $^* -$ In this time segment, only BAT spectra are available and fitted with an \apec\ 4-T plasma model. The L$_{x}$ for this segment indicates the derived luminosity corresponding to the 14--50~keV range.
\item $^{**} -$ Abundances in this time segment are fixed at the nearest abundance value of segment P02 as derived from fitting the XRT spectra.
\item $^{\dagger} -$ In these time segments, XRT+BAT spectra are fitted with an \apec\ 4-T plasma model. The L$_{x}$ in this segment indicates the derived luminosity corresponding to the 0.35--50~keV range.

\normalsize
\tabend

\subsubsection{Spectral Analysis of BAT Data}
\begin{figure*}
  \centering
\includegraphics[height=15cm,angle=0]{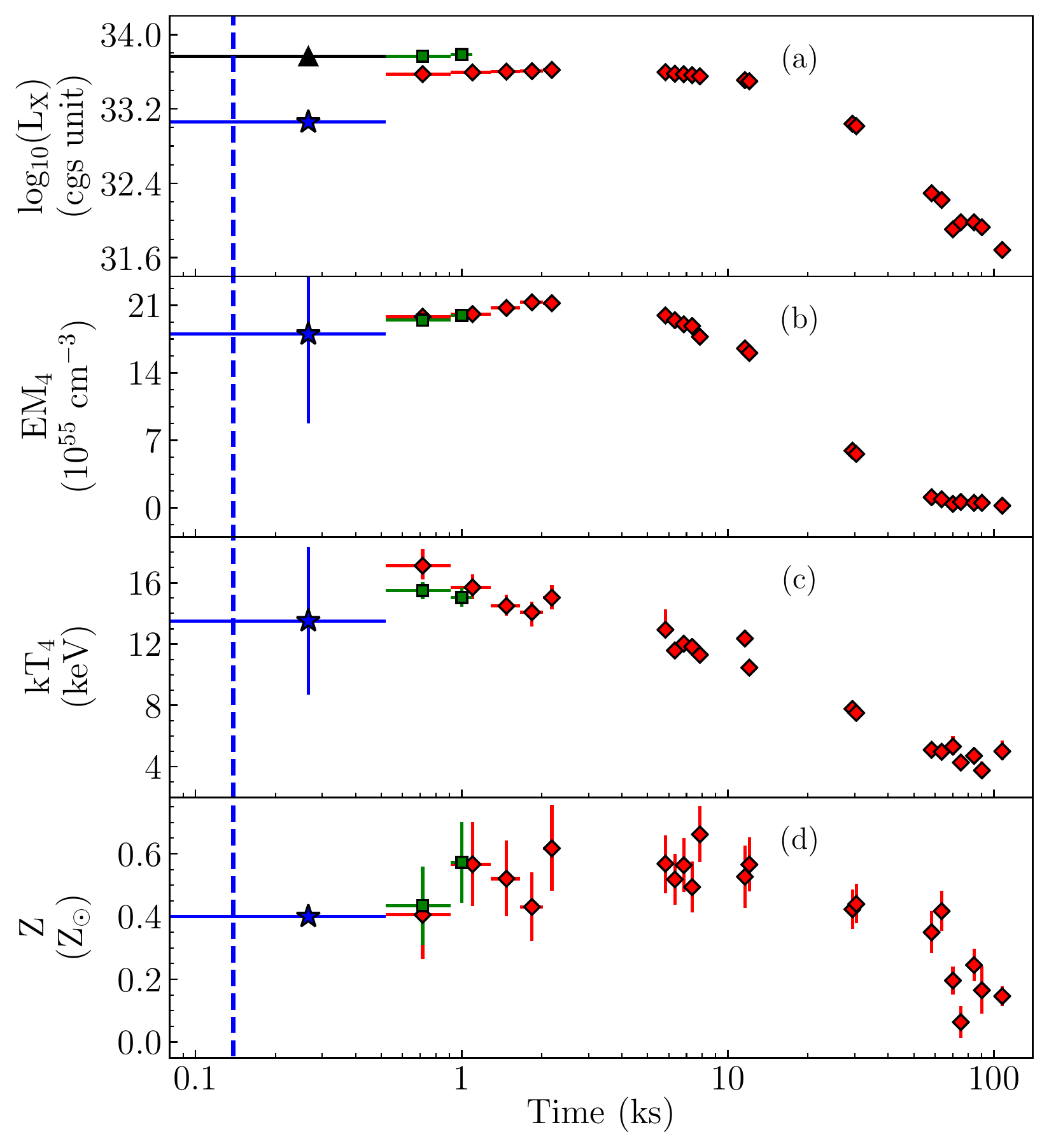}
\caption{
 The evolution of spectral parameters during the flare on \szp\ is shown. In all the panels, the parameters derived from the BAT, XRT, and XRT+BAT spectral fitting are represented by the solid blue asterisk, solid red diamonds, and solid green squares, respectively. In the top panel (panel-a), the X-ray luminosities are derived in the 0.35--10~keV (red solid diamond), 14--50~keV (blue solid star) and 0.35--50~keV (green solid squares) ranges. For the first segment, the 14--50~keV BAT luminosity is extrapolated to a 0.35--10 keV energy range (solid black triangle). Panels (b)–(d) display the variations of emission measure, plasma temperature, and abundances, respectively. The blue dashed vertical line indicates the trigger time of the flare. Horizontal bars give the time range over which spectra were extracted; vertical bars show a 68\% confidence interval of the parameters.}
\label{fig:trs_all}
\end{figure*}

The spectral analysis of `only BAT' spectra is performed for the segment P01 using the 4-T \apec\ plasma model. As in the case of XRT spectral fitting, the first three temperature components in this fitting were frozen at the quiescent values. Due to poor statistics, it was very difficult to estimate the spectral parameters. Therefore, for this analysis, we assumed that the global abundance is not very different from the nearest time segment (P02) as derived from XRT analysis. The fourth temperature and corresponding emission measure are kept as free parameters. The derived values of the temperature and emission measure are 14$^{+6}_{-4}$ keV and 1.8$^{+1.1}_{-0.7}$~\E{56}~cm$^{-3}$, respectively. We also used the {\tt cflux} model to estimate the X-ray luminosity. The X-ray luminosity in the 14--50~keV range is derived to be $\sim$1.14~\E{33}~\ergs. The BAT spectra and the best-fit model are shown in the top panel of Figure~\ref{fig:spec_BAT_XRT+BAT}. The estimated values from this analysis are given in the first row of Table~\ref{tab:trs_all}. In Figure~\ref{fig:trs_all}, the BAT observations are indicated with solid blue asterisks. In order to compute the equivalent XRT luminosity in the 0.35--10~keV range, we converted the BAT luminosity using \texttt{webpimms}\footnote{\href{https://heasarc.gsfc.nasa.gov/cgi-bin/Tools/w3pimms/w3pimms.pl}{https://heasarc.gsfc.nasa.gov/cgi-bin/Tools/w3pimms/w3pimms.pl}}. For this conversion, the best possible multi-temperature model is considered, where the three \apec\ temperatures are taken from the three hotter components from our analysis. The equivalent X-ray luminosity in the 0.35--10 keV band is estimated to be 5.8~\E{33}~\ergs\ and is shown with the black triangle in Figure~\ref{fig:trs_all}. This value is marginally higher than the luminosity during the segment P02 derived from XRT analysis.

\subsubsection{XRT+BAT Spectral Analysis}
At the beginning of the \swift\ XRT observations, \szp\ was observed with both XRT and BAT instruments until T0+0.957~ks, which comprises two segments: P02 and P03. We carried out an XRT+BAT joint spectral fitting for these two segments. As in the case of XRT spectral analysis, the XRT+BAT observations were best fitted with a 4-T \apec\ plasma model, keeping the first three temperature components fixed to the quiescent values. The fourth temperature, emission measure, and global abundances are kept as free parameters. Since galactic H~{\sc i} column density was not found to be variable in the `only XRT' spectral analysis, we fixed \nh\ to the quiescent value. The best-fit XRT+BAT spectra are shown in the two bottom panels of Figure~\ref{fig:spec_BAT_XRT+BAT}. In Figure~\ref{fig:trs_all}, the derived parameters are shown with green squares. These parameters are also given in the second and third rows of Table~\ref{tab:trs_all}. The temperatures, emission measures, and abundances estimated from XRT+BAT analysis were found to have similar values as those derived from XRT spectral fitting within the 1$\sigma$ uncertainty level. The luminosity in the 0.35--50~keV energy range was estimated to be 5.86~\E{33} and 6.08~\E{33}~\ergs\ for the P02 and P03 segments, respectively. These values are $\sim$1.6 times the corresponding estimated luminosities in the 0.35--10~keV energy band using only XRT observations.

%
\begin{figure}
  \includegraphics[width=8.9cm, trim={0 0 1.5cm 0}, clip, angle=0]{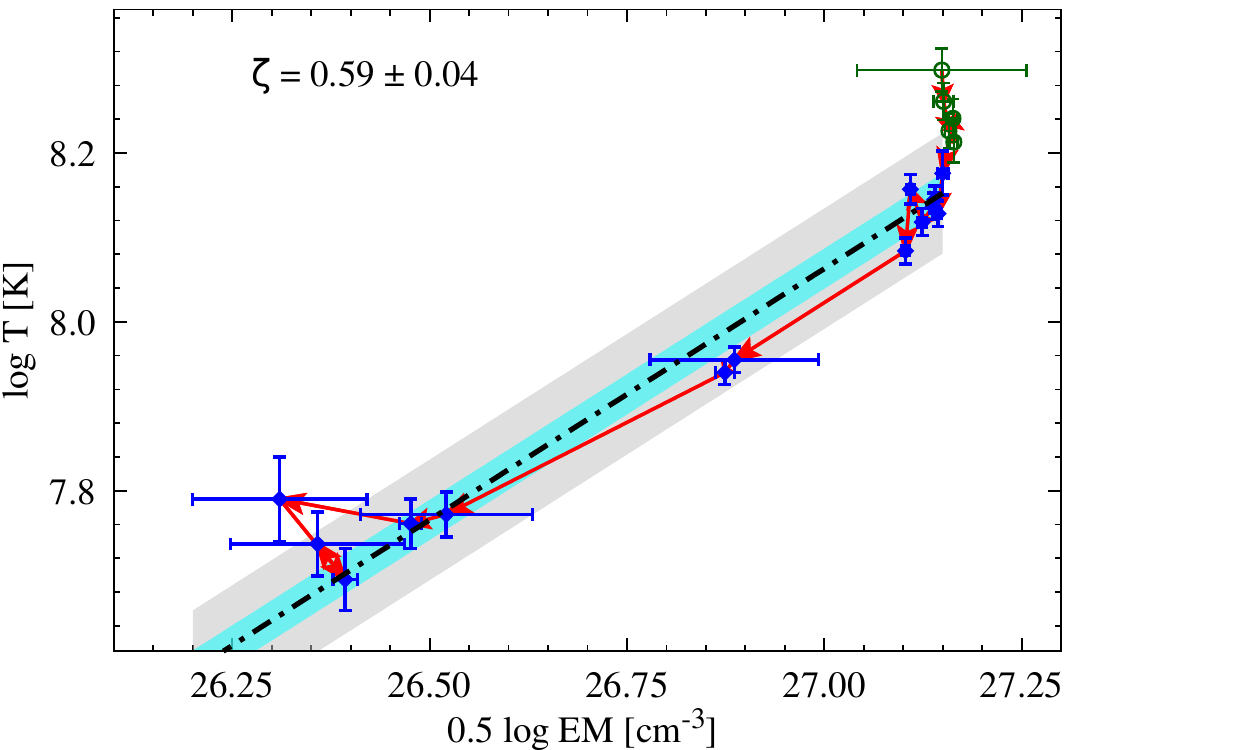}
  \caption{Evolution of the flare in [log $\sqrt{EM}$ -- log T] plane. The green open circles and solid blue diamonds indicate the variation during the rising and decay phase of the flare. The red arrows indicate the chronological order in which the flare evolved in the [log $\sqrt{EM}$ -- log T] plane. The black dot-dashed line shows the slope ($\zeta$), best-fitted in the decay phase of the flare. The cyan and grey shaded regions show 1$\sigma$ and 3$\sigma$ uncertainty levels.}
\label{fig:EM-kT}
\end{figure}

\section{Hydrodynamic Loop Modeling} \label{sec:loop-modeling}
The magnetic loop structures of the flaring events are well studied on the Sun as they are spatially resolved. On the other hand, the stellar flares are neither spatially resolved nor direct measurements of coronal loop parameters are yet possible. Due to having several similarities of the stellar flares with the solar flares, it is possible to estimate the physical size of the loop structures involved in the stellar flares. Several approaches have been developed throughout the years to model the decay of the flares. 

In this study, we used the state-of-the-art time-dependent one-dimensional hydrodynamic loop model of the stellar flares \citep{Reale-97-A+A-3, Reale-07-A+A-2}. This model assumes that flare occurs in a single dominant loop with a constant cross-section and semi-circular loop geometry. Followed by a magnetic reconnection process, a heat pulse is supposed to be released near the loop apex at the beginning of the flare. The thermodynamic cooling timescales are proportional to the length of the flaring loop. Moreover, studies on the solar flares show that the slope ($\zeta$) in the density-temperature plane of the flare decay path provides a diagnostic of sustained heating \citep{Sylwester-93-A+A-3}. The hydrodynamic model also considers the prolonged decay timescales due to the sustained heating during the decay phase of the flare, which, in turn, provides a more realistic estimation of the flaring loop length within a single-loop scenario.

However, we know from large solar flares that the flaring events in most cases do not occur in a uniform, single, pre-defined semi-circular, and steady loop as assumed in the hydrodynamic loop model of \cite{Reale-97-A+A-3} and \cite{Reale-07-A+A-2}. Instead, large flares are often associated with loop arcades of multiple loops. Therefore, for a superflare similar to that of \szp, it is essential to carefully consider the applicability of the hydrodynamic loop model.

Initially, the hydrodynamic loop model was validated on 20 single-loop solar M- and C-class flares imaged by the Yohkoh SXT telescope \citep[][]{Reale-97-A+A-3}. Later, this model was further applied to larger flares, including five solar X-class limb flares \citep{GetmanK-11-ApJ-5}. In EUV TRACE images, those flares were found to be associated with arcades of multiple loops. In contrast to the general expectations that the single-loop model would over-estimate flaring loop heights, \cite{GetmanK-11-ApJ-5} found that the loop heights were comparable to or less than the individual loops measured from the TRACE images.

 Further, the single-loop Hydrodynamic modeling was argued to be applicable for the multiloop arcades assuming the presence of a dominant loop \citep[see][]{Reale-04-A+A-2}. However, the single-loop method was also found to be applicable in cases of multiloop structures where individual loop events have similar temporal temperature and emission measure profiles and firing nearly simultaneously \citep{GetmanK-11-ApJ-5}. If the multiple individual loops do not firing near-simultaneously, it is expected to produce multi-modal light curves with shorter subflare components \citep[][]{Aschwanden-01-SoPh-9}.

 In this work, we understand that the frequent data gap (as shown in Figure~\ref{fig:lc_swift}) has limited the information about the time variation during the superflare on \szp. However, the light curve seems to show a long duration and unimodal structure. Therefore, in the case of the superflare on \szp, we consider that the application of the hydrodynamic loop modeling is valid, although with substantial inaccuracies.

Therefore, utilizing the empirical formula as estimated using hydrodynamic simulations \citep{Reale-07-A+A-2} the length of the flaring loop ($\mathrm{L_{HD}}$) can be derived as, 
\begin{equation}
\label{eq:looplength}
\mathrm{L_{HD}}~[cm] = \frac{\tau_{d}\sqrt{T_\mathrm{max}}}{1.85 \times 10^{-4} F(\zeta)}~~~~~~~~~~F(\zeta) \ge 1
\end{equation}
\noindent
where, $T_\mathrm{max}$ is the maximum temperature (K) of loop apex,  $\tau_{d}$ is the \mbox{e-folding} decay time (s), and  $F(\zeta)$ is a \mbox{non-dimensional} factor that depends on the band-pass and spectral response of the instruments. For the \swift\ XRT observations, the maximum loop apex temperature $T_\mathrm{max}$ was calibrated by \cite{Osten-10-ApJ-5} as $T_{\rm max} = 0.0261 \times T_{\rm obs}^{1.244}$, where both temperatures are in the unit of Kelvin. Using the peak flare temperature as estimated in Section~\ref{subsubsec:trsxrt}, the maximum loop-apex temperature during the flare of \szp\ was estimated to be 549$\pm$39 MK. The factor $F(\zeta)$ was also calibrated by \cite{Osten-10-ApJ-5} as $F(\zeta) = {{1.81} / {(\zeta - 0.1)}} + 0.67$, which is valid for $0.4 \lesssim \zeta \lesssim 1.9$. These limits of applicability correspond to the heating timescale being very long on one end and no heating on the other end.

For the \swift\ XRT observations of the flare of \szp, since no density determination was available, we used the quantity $\sqrt{E\!M}$ as a proxy for the density, where the geometry of the flaring loop is assumed to remain constant during the decay. Figure~\ref{fig:EM-kT} shows the path in the \mbox{[log $\sqrt{E\!M}$ -- log $T$]} diagram for the flare on SZ~Psc. The green open circles in the diagram show the rising phase of the flare, whereas the decay phase is shown with the blue diamonds. A linear fit to the decay phase of the flare is shown with the black dot-dashed line. The uncertainties are estimated considering the uncertainties in both axes, and the 1$\sigma$ and 3$\sigma$ uncertainties are shown by cyan and grey-shaded regions, respectively. The slope $\zeta$ is estimated to be $0.59 \pm 0.04$, indicating the presence of sustained heating during the decay phase of the flare. Using above quantities and the values of \td\ mentioned in \S~\ref{sec:lc}, the loop length of the flare was derived to be 6.3$\pm$0.5~\E{11}~cm by using Eqn.(\ref{eq:looplength}). 

\begin{table}
\centering
\begin{threeparttable}

\tabcolsep=0.17cm
\caption{Derived loop parameters for the flare on \szp}
\label{tab:loop_params}
\begin{tabular}{lcccccccccccccccc}
\toprule
\textbf{Ref.} &\textbf{Parameters}            &   \textbf{Units}            &   \textbf{Flare}  \\
\hline
1  & $\tau_\mathrm{r}$                                      &   (ks)                                    &   {14.4 $\pm$ 0.5}\\%
2  & $\tau_\mathrm{d}$                                      &   (ks)                                    &   {21.4 $\pm$ 0.3}\\%
3  & ${\rm log}_{10}({\rm L}_\mathrm{X, max})$              & (cgs unit)                                &   {33.619 $\pm$ 0.007}\\%
4  & kT$_\mathrm{4, max}$                 & (keV)                                     &   {17 $\pm$ 1}\\%
5  & EM$_\mathrm{4, max}$                                   & (10$^{56}$ cm$^{-3}$)                     &   {2.13 $\pm$ 0.05}\\%
6  & T$_\mathrm{obs}$                                  & (MK)                                      &   {199 $\pm$ 11}\\%
7  & T$_\mathrm{max}$                                 & (MK)                                      &   {549 $\pm$ 39}\\%
8  & $\zeta$                                                & --                                        &   {0.59 $\pm$ 0.04}\\%
9   & F($\zeta$)                                             & --                                        &   {4.3 $\pm$ 0.3}\\%
10  & L$_{\rm HD}$                                           &  (10$^{11}~\mathrm{cm}$)                  &   {6.3 $\pm$ 0.5}\\
11 & L$_{\rm QS}$                                           &  (10$^{11}~\mathrm{cm}$)                  &   {7.6 $\pm$ 0.5}\\%
12 & $\beta_{peak}$                                         &   --                                      &   {0.21 $\pm$ 0.03}\\
13 & $V$                                                    &  ($10^{34}~\mathrm{cm^{3}}$)              &   {4 $\pm$ 1}      \\
14 & n$_{e}$                                                &  (10$^{10}~\mathrm{cm^{-3}}$)             &   {8 $\pm$ 2}\\%
15 & $p$                                                    &  ($10^{3}~\mathrm{dyn~cm^{-2}}$)          &   {3.5 $\pm$ 0.7}\\
16 & $B$                                                    &  ($\mathrm{G}$)                          &   {298 $\pm$ 29}\\
17 & HR$_\mathrm{V}$                                        & ($\mathrm{erg~s^{-1}~cm^{-3}}$)           &   {0.20 $\pm$0.06}    \\
18 & HR                                                     & ($10^{33}~\mathrm{erg~s^{-1}}$)           &   {7$\pm$1}  \\
19 & $\mathrm{E_{tot}}$                                    & ($10^{38}~\mathrm{erg}$)                  &   {2.1$\pm$0.3}\\
20 &   $B_\mathrm{tot}$                                     & ($\mathrm{G}$)                           &   {490 $\pm$ 60}\\
\hline
\end{tabular} 
\begin{tablenotes}
\item  \textbf{Note.}
  \item (1, 2) -- The e-folding rise and decay times, respectively, derived from the light curve.
\item (3) -- Luminosity at the flare peak (\mbox{0.35--10~keV} band).
\item (4, 5) -- Peak flare temperature and emission measure, respectively, as estimated from spectral fitting.
\item (6, 7) -- Observed maximum temperature and loop apex temperature, respectively, in MK.
\item (8) -- Slope of the decay path in the density-temperature diagram.
\item (9) -- The function F($\zeta$) indicates the presence of sustained heating.
\item (10) -- Length of the flaring loops estimated using hydrodynamic loop model \citep[][]{Reale-97-A+A-3}.
\item (11) -- Length of the flaring loops estimated using quasi-static loop model \citep[][]{van-den-OordG-89-A+A-1}.
\item (12) -- Loop aspect ratio at flare peak.
\item (13, 14, 15) -- Loop volume, density, and pressure at flare peak.
\item (16) -- Estimated magnetic field.
\item (17) -- Heating rate per unit volume at the flare peak. 
\item (18) -- Heating rate at the flare peak.
\item (19) -- Total radiated energy (\mbox{0.35--10~keV} band).
\item (20) -- Total magnetic field required to produce the flare.
\end{tablenotes}
\end{threeparttable}
\end{table}

\section{Coronal loop properties} \label{sec:loop-prop}
As described in Section~\ref{sec:loop-modeling}, the hydrodynamic model assumes single loop with a constant cross-section. The ratio of the loop's cross-sectional radius and the loop length ($\beta = r/\mathrm{L_{HD}}$) is a measure of the thickness of the loop. For many solar and stellar flares the $\beta$ is often fixed at $\beta$ = 0.1 (see \citealt{Reale-97-A+A-3} for solar flares; \citealt{Favata-05-ApJS-6} and \citealt{Getman-08-ApJ-2} for Orion superflares; \citealt{Crespo-Chacon-07-A+A} for small stellar flare and \citealt{Karmakar-17-ApJ-5} for stellar X-ray superflares). However, recent studies have shown that $\beta$ can be estimated from the data within the hydrodynamic model framework \citep[see][]{GetmanK-11-ApJ-5, GetmanK-21-ApJ-1}.

The dominant cooling mechanisms of a coronal plasma are thermal conduction and radiation. The following equations show the corresponding timescales expressed in terms of the $\beta$, loop length $\mathrm{L_{HD}}$, and temporal profiles of temperature $T(t)$, X-ray luminosity $L_X(t)$, and emission measure $EM(t)$ \citep[as adopted from][]{GetmanK-11-ApJ-5}.
\begin{equation} \label{eq:cooling_timescales}
\tau_{con}(t) = \frac{3 k_b}{\kappa_0 T(t)^{5/2} \beta} \sqrt{\frac{EM(t) \mathrm{L_{HD}}}{2 \pi}} ,
\end{equation}
\begin{equation} \label{eq:radiation_timescales}
\tau_{rad}(t) = \frac{3 k_b T(t) \beta}{L_{X,0.01-50}(t)} \sqrt{EM(t) 2 \pi \mathrm{L_{HD}}^3} ,
\end{equation}
where $\kappa_0$ = 9.2\E{-7}~erg~s$^{-1}$~cm$^{-1}$~K$^{-7/2}$ is the coefficient of thermal conductivity \citep[][]{SpitzerL-65-pfig-3}, and $k_b$ is the Boltzmann constant. A combined cooling time ($\tau_{th}$) can be defined as an exponential folding time that combines both conduction and radiation processes, as given in the following equation. 
\begin{equation}  \label{eq:combined_timescales}
  \frac{1}{\tau_{th}}  = \frac{1}{\tau_{con}} + \frac{1}{\tau_{rad}} 
\end{equation}
From the equations \ref{eq:cooling_timescales}, \ref{eq:radiation_timescales}, and \ref{eq:combined_timescales}, the value of $\beta$  can be estimated such that the combined cooling time $\tau_{th}$ would be the closest to the observed flare decay timescale (corrected for possible sustained heating, i.e. $\tau_{d} / F(\zeta)$). In above equations, we have used flare emission measure and temperatures as estimated in Section~\ref{subsubsec:trsxrt} and given in Table~\ref{tab:trs_all}. In order to estimate the X-ray luminosity in a wide energy band (i.e. 0.01--50~keV, $L_{X,0.01-50}$), we have used the flare luminosity as given in Table~\ref{tab:trs_all}. In order to convert the X-ray luminosity from 0.35--10~keV to 0.01--50~keV, we have used the multi-component Mission Count Rate Simulator tool {\tt webpimms}\footnote{\href{https://heasarc.gsfc.nasa.gov/cgi-bin/Tools/w3pimms/w3pimms_pro.pl}{https://heasarc.gsfc.nasa.gov/cgi-bin/Tools/w3pimms/w3pimms\_pro.pl}}. For this conversion, we have used the parameters of the three highest temperature components for each segment. Throughout the decay phase of the flare, the value of the loop thickness is found to be consistent within the limit $\beta$ = 0.23$\pm$0.03. Whereas, at flare peak, the value of the loop thickness is estimated to be $\beta_{peak} = 0.21$. Therefore the loop volume at the flare peak is estimated to be $V = \pi \beta_{peak}^{2} \mathrm{L_{HD}}^3$ = 4$\pm$1~\E{34}~cm$^3$.

  Using the flare peak emission measure from Section~\ref{subsubsec:trsxrt}, and the volume of the loop at flare peak, the desnity of the flaring plasma can be estimated as 
\begin{equation}
  n_e~[{\rm cm^{-3}}] = \sqrt{{\rm EM_4} \over V}
\label{eq:vol}
\end{equation}
 We found the density of the loop at flare peak was 8$\pm$2~\E{10}~cm$^{-3}$. Assuming totally ionized hydrogen plasma, from the relation $p = 2 n_e kT_{max}$, the pressure in the loop is estimated to be 3.5$\pm$0.7~\E{3}~dyne~cm$^{-2}$ at the flare peak.

For a static non-flaring coronal loop, \cite{Rosner-78-ApJ-1} derived a scaling law (hereafter RTV law), assuming equilibrium between a spatially uniform heating rate and the conductive and radiative loss rates. While the validity of the RTV scaling law applies to a coronal loop in hydrostatic equilibrium, \cite{Aschwanden-08-ApJ-2} suggested that it might also be applicable to a flaring loop near the peak time because both the energy and momentum equations are nearly balanced near the flare peak. Therefore, using RTV scaling law, the heating rate per unit volume (HR$_\mathrm{V}$) at the flare peak can be estimated by the following relationship.
\begin{equation}
  \mathrm{HR_V}~[{\rm erg~s^{-1} cm^{-3}}] = \frac{d H}{d V d t} \simeq 10^5 ~ p^{7/6} ~ \mathrm{L_{HD}}^{-5/6}
\label{eq:E_H}
\end{equation}
Utilizing the derived loop length and pressure at the flare peak, we estimated an approximate value of HR$_\mathrm{V}$ for the flare on \szp\ to be 0.20$\pm$0.06~$\rm erg ~ s^{-1} ~ cm^{-3}$. Using Equations~\ref{eq:vol} and \ref{eq:E_H}, the total heating rate (HR = $\frac{d H}{d V d t} \times V$) is estimated to be 7$\pm$1~\E{33} $\mathrm{erg~s^{-1}}$, which is $\sim$1.7 times than the flare maximum X-ray luminosity. The heating rate during the flare was found to be only $\sim$0.4\% of the bolometric luminosity of the primary, whereas it was $\sim$1.0\% of the bolometric luminosity of the secondary component of the \szp\ system. Assuming the constant heating rate throughout the rise and decay phase of the flare, the total energy associated with the heating mechanism is estimated as $\mathrm{E_{tot}}~=~(\tau_{\rm r}+\tau_{\rm d}) \times {\rm HR}$. The total energy during the flare is derived to be 2.1$\pm$0.3~\E{38}~erg, which is also derived to be $\sim$78~s of the bolometric energy output of the primary component of \szp.

If the heating mechanism is responsible, the present flare is essentially due to some form of dissipation of magnetic energy. The estimated magnetic field in the newly generated loop is estimated by using the equation B = $\sqrt{8\pi~p}$, which is estimated to be 298$\pm$29~G. However, if we assume that the energy released during the flare is indeed of magnetic origin, the total non-potential magnetic field $B_{\rm tot}$ involved in the flare energy release within an active region of the star can be estimated using the following equation:
  \begin{equation}
    \mathrm{E_{tot}}~[{\rm erg}] = \frac{(B_{\rm tot}^2-B^2)}{8 \pi} V
\label{eq:Etot_B}
\end{equation}
If we assume that the loop geometry does not change during the flare, using the values of $\mathrm{E_{tot}}$, $B$, and $V$, the total magnetic field required to produce the flare is estimated to be 490$\pm$60~G. The derived parameters of the flaring loop are summarized in Table~\ref{tab:loop_params}.

\section{Discussion and Conclusions} \label{sec:discussion}

\subsection{Flaring loop size and location}
\label{sec:disc_loop}
In this study, we carried out a detailed time-resolved spectral analysis and estimated the loop length and other flare parameters. In order to compare with other loop models, we also estimated the loop lengths using another extensively used loop model \textit{viz.} the quasi-static loop model \citep[][]{van-den-OordG-89-A+A-1}. According to this model, the energy loss is entirely due to radiative cooling, while the conduction redistributes the energy along the loop. In the quasi-static approximation, the cooling process occurs through a sequence of stationary states of constant pressure. With the assumption of the shape of the flaring loop to be semi-circular with a constant cross-section throughout the flare, the quasi-static decay time scale can be estimated using Equation~26 of \cite{van-den-OordG-89-A+A-1}. In our analysis, the quasi-static decay time scales estimated from the EM$_{4}$ and kT$_{4}$ data were found to be similar (within 1$\sigma$ uncertainty level). The quasi-static decay time was estimated to be $\sim$11.3~ks, which results in a loop length of 7.6$\pm$0.5~\E{11}~cm. This value is within the 1.3$\sigma$ uncertainty level from that estimated from the hydrodynamic loop modeling as described in Section~\ref{sec:loop-modeling}. However, a slightly lower value of loop length as estimated from the hydrodynamic loop modeling can be well explained by the consideration of sustained heating as a model parameter during the decay phase of the flare. Therefore, in this study, we consider the loop length estimated from the hydrodynamic loop modeling as more accurate than that from the quasi-static loop model. The estimated loop-length is also found to be of a similar order to other RS~CVn type binaries \citep[see][]{Agrawal-86-MNRAS-6, Gudel-99-ApJ, OstenR-03-ApJ-5, Pandey-12-MNRAS-8, Tsuboi-16-PASJ-4, SasakiR-21-ApJ-2}.

\begin{figure*}
  \centering
\includegraphics[height=8.4cm, angle=-90]{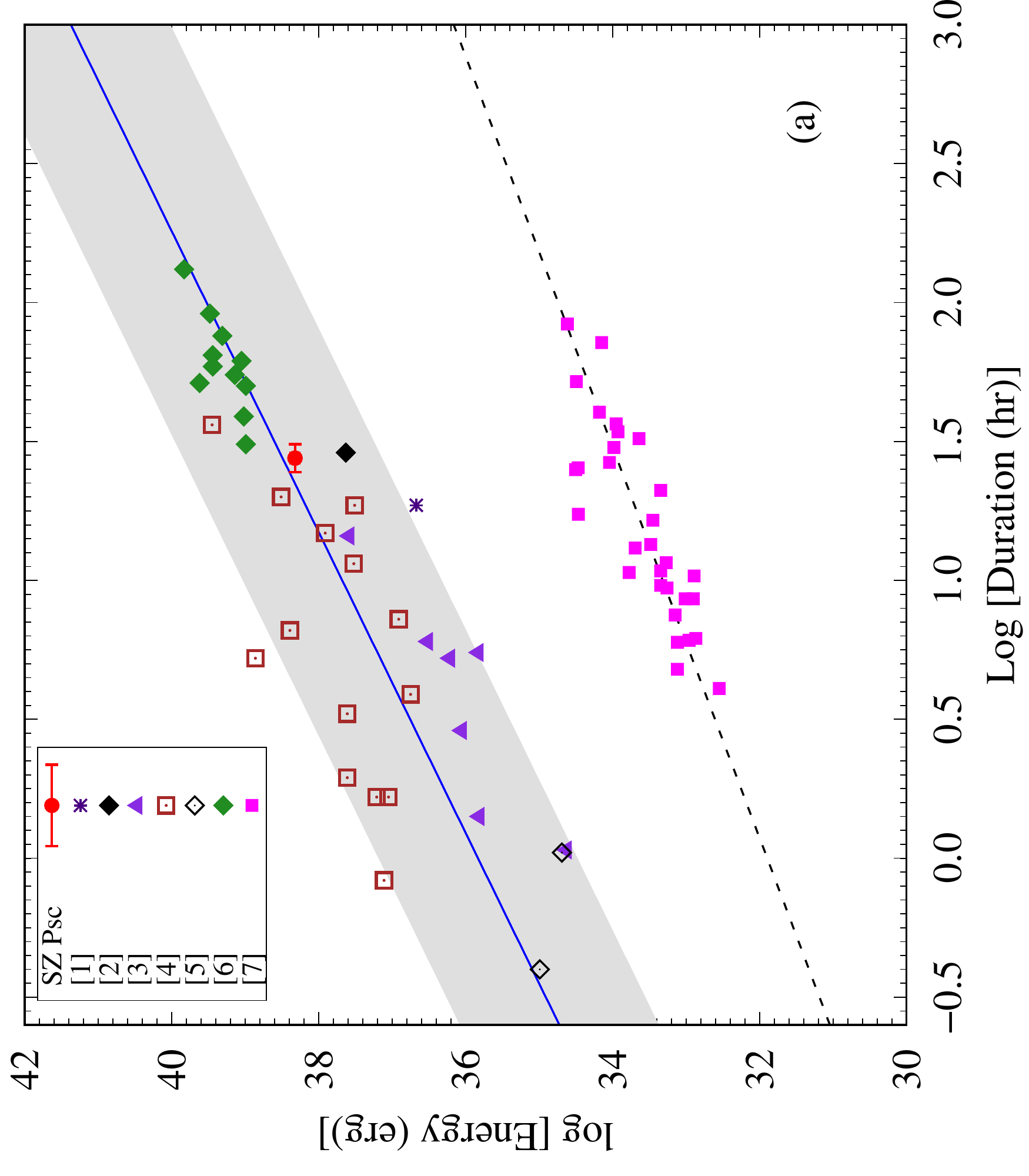}
\hspace{0.2mm}
\includegraphics[height=8.4cm, angle=-90]{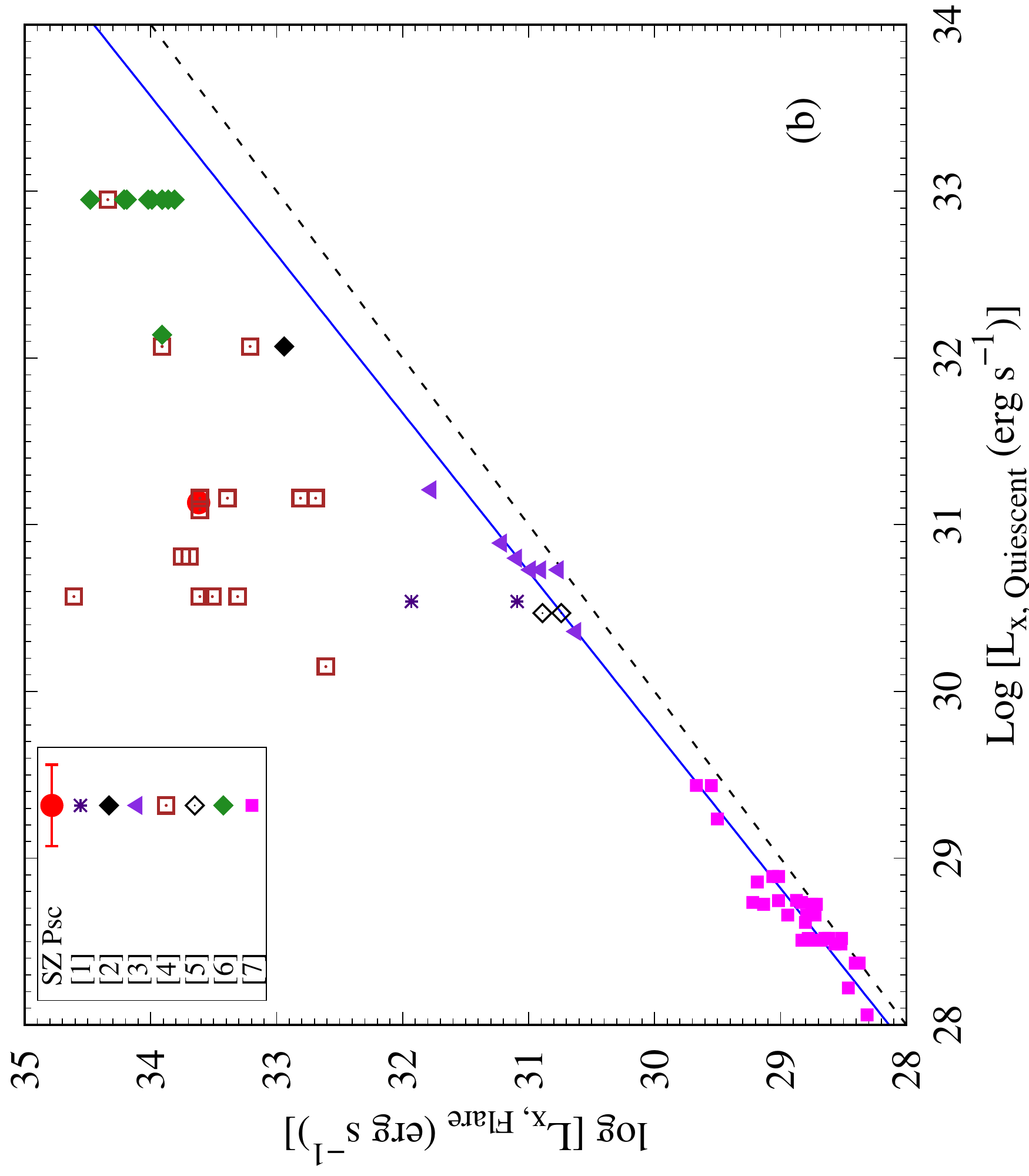}
\caption{
(a) Flare duration vs. Flare energy and (b) Flare luminosity vs. Quiescent luminosity are shown. The flare on \szp, investigated in this study, is shown with the solid red circle. We have compared the flares on the RS~CVn type binary systems from the literature with different symbols.  
Figure legends: [1]: Large X-ray flare on HU~Vir (\citealt{EndlM-97-A+A}), [2]: BeppoSAX observation of a large flare on UX~Ari (\citealt{Franciosini-01-A+A-3}), [3]: Seven flares from five RS~CVn-type binaries (UZ~Lib, $\sigma$~Gem, $\lambda$~And, V711~Tau, and EI~Eri; \citealt{Pandey-12-MNRAS-8}), [4]: Thirteen flares eight RS~CVn systems (VY~Ari, UX~Ari, HR1099, GT~Mus, V841~Cen, AR~Lac, SZ~Psc, and II~Peg; \citealt{Tsuboi-16-PASJ-4}), [5]: Two flares on XY~Uma (\citealt{Gong-16-RAA-33}), [6]: Eleven flares on GT~Mus (\citealt{SasakiR-21-ApJ-2}), [7]: Thirty flares observed with {\it Extreme Ultraviolet Explorer} (EUVE) in nine RS~CVn type binary systems ($\sigma^2$ CrB, V824~Ara, V815~Her, AR~Lac, BH~CVn,  V711~Tau, UX~Ari, II~Peg, $\lambda$~And, VY~Ari, and AR~Psc; \citealt{Osten-99-ApJ}). 
All the X-ray luminosities and energies are converted to 0.35--10~keV energy range using the \texttt{webpimms} with multi-temperature \apec\ model, considering the nearest corresponding temperature components. In panel (a), the solid blue line indicates the derived relation between flare duration vs. flare energy in soft X-ray (0.35--10~keV energy band) as discussed in Section~\ref{sec:discussion}. The black dashed line shows the empirical limit for the flares on RS~CVn binaries as observed in EUV observations (\citealt{Osten-99-ApJ}). In panel (b), the solid blue line indicates the empirical limit for EUV observations adopted from \pcite{Osten-99-ApJ}, whereas the black dashed line is a line of equality between the two luminosities.
}
\label{fig:discuss}
\end{figure*}

We can not spatially resolve the flare location on the \szp\ system. However, with the estimated loop length, we can have some idea about the flare location. The separation of the binary components, as estimated using Gaia DR2 observations, is found to be 3.53~AU \citep[see][]{KervellaP-19-A+A-3}, which is $\sim$84 times that of the estimated loop-length of the flare. It is unlikely that the magnetic field lines responsible for the flare were inter-binary loops as found in a few RS~CVn binaries \citep[see ][]{Uchida-83-ASSL-2}. The flare is likely to be associated with just one of the stellar components. The flaring loop height of the flare on \szp\ was estimated to be 57\% of the radius of the K1~IV primary, whereas it is 192\% of the radius of the F8~V secondary. As the estimated total magnetic field strength that would be required to accumulate the emitted energy and to keep the plasma confined in a stable magnetic loop configuration is $\sim$490~G, it is expected that a strong magnetic dynamo is present on the stellar component that is associated with the flare. Several studies have shown that the F-type stars are less active than the K-type stars due to their thinner convective envelop \citep[e.g. ][]{SchrijverC-83-A+A, Savanov-16-AcA-3, Savanov-18-AstBu-13, SavanovI-19-AstL-8,  Karmakar-16-MNRAS-8, Karmakar-18-BSRSL-10, KarmakarS-19-BSRSL, KarmakarS-19-PhDT-2}. Therefore, it is more likely that the K-type primary component is responsible for such a high level of magnetic activity. Moreover, with the high-rotational velocity of each component due to being a tidally locked system, since the subgiant primary component has a thicker convection zone than the main-sequence secondary component, it is more likely that this large flare was associated with the primary component. 


\subsection{Flare duration vs energy on RS~CVn-type binaries} 
\label{sec:disc_dn-vs-E}

As described in Section~\ref{sec:lc}, at the beginning of the observation, the flare on \szp\ was already in the rising phase with a \swift\ XRT count rate of  $\sim$79~\cts. Therefore, we could not estimate the total duration of the flare precisely. However, with the estimated flare duration of  $>$31 hours, this flare is identified as one of the longest duration X-ray flares on \szp. Using MAXI/GCS instrument, another long-duration flare on \szp\ was detected in 2--20~keV band \citep[see Figure~1 of][where FN14 represents the flare on SZ~Psc]{Tsuboi-16-PASJ-4}. However, due to the large temporal uncertainty ($\sim$8 hrs) in the MAXI/GCS observation, the duration of the flares is difficult to compare quantitatively. 

The observed flare duration in SZ~Psc is large as compared to those of solar-type stars or main-sequence binaries \citep[see][]{Pandey-08-MNRAS-16, Pandey-15-AJ-6, Savanov-18-ARep-8, KarmakarS-22-MNRAS}. However, it is well among the flare duration for RS~CVn type binaries which ranges from a few minutes to several days \citep[e.g.][]{Franciosini-01-A+A-3, Pandey-12-MNRAS-8, Gong-16-RAA-33, SasakiR-21-ApJ-2}. We also estimated the total energy released during the flare is 2.1$\pm$0.3\E{38}~erg. This is very large in comparison to the total energy of the X-ray superflares on the solar-type stars \citep[see][]{Favata-00-A+A-3, Osten-10-ApJ-5, Karmakar-17-ApJ-5}. Recent studies have shown that the RS~CVn stars release energy in the range of \Pten{34} to  \Pten{40}~erg during flares \citep[see][]{EndlM-97-A+A, Franciosini-01-A+A-3, Pandey-12-MNRAS-8, Tsuboi-16-PASJ-4,  SasakiR-21-ApJ-2}.  

In Figure~\ref{fig:discuss}(a), we have plotted the flare duration of the RS~CVn-type binaries with the flare energies as derived in the literature (see legends 1--6 in Figure~\ref{fig:discuss}). We converted the observed X-ray flux into the 0.35--10 keV energy band using \texttt{webpimms}\footnote{\href{https://heasarc.gsfc.nasa.gov/cgi-bin/Tools/w3pimms/w3pimms.pl}{https://heasarc.gsfc.nasa.gov/cgi-bin/Tools/w3pimms/w3pimms.pl}}. For the conversion, we considered the best-known coronal parameters and utilized those in the multi-temperature astrophysical plasma models. We found that a positive correlation exists between the flare duration and flare energy of the RS~CVn binaries in the soft X-ray band. We computed a linear Pearson correlation coefficient of 0.84 between the duration and flare energies, with a null hypothesis probability of 9.13~\E{-9}. We further derived the relationship between the flare energy (in the 0.35--10~keV band) and the flare duration of the RS~CVn-type binary systems as follows.
\begin{equation}
  \mathrm{log_{10}\left(\frac{E_{tot}}{erg}\right)} = (1.8\pm0.2) \times \mathrm{log_{10} \left(\frac{Dn}{hr}\right)} + (35.8\pm0.2)
  \label{eq:disc}
\end{equation}
The blue solid line in Figure~\ref{fig:discuss}(a) shows the best-fitted line with reduced $\chi^{2}$  value of 0.7 (for 34 dof). The grey shaded region represents the 1$\sigma$ variations. It is found that the flare on \szp\ is also following the relationship between energy and flare duration. In Figure~\ref{fig:discuss}(a), the solid pink squares (legend 7) indicate the energy vs. duration relation observed in the EUV energy band adapted from \cite{Osten-99-ApJ}. The black dashed line shows the derived relationship between the flare energy in EUV and the flare duration corresponding to E$_\mathrm{flare}$ $\alpha$ Dn$^{1.42}$. This power-law index is in good agreement (within 1$\sigma$) with the derived relationship given in Equation~\ref{eq:disc}. However, the energy budget for the soft X-ray (0.35--10~keV) in the flares in RS~CVn binaries is found to be a hundred times more than the EUV counterpart.

\subsection{Quiescent vs flaring X-ray emission}  \label{sec:disc_Lx}
Using \xmm\ and \swift\ observations in the 0.35--10~keV energy band, we found that the quiescent corona of \szp\ can be well described by three temperature plasma with estimated temperatures of 4.1, 13.1, and 47.6~MK. This is the first time a detailed analysis of the quiescent corona of \szp\ has been performed in the soft X-ray energy band. Using the imaging proportional counter (IPC) of the \textit{Einstein Observatory} in the 0.2--4~keV energy range, \cite{Majer-86-ApJ-2} assumed that the corona of \szp\ consists of a single-temperature plasma, and the temperature was estimated to be 7.44--7.58~MK. From the \rosat\ all-sky survey observations in the 0.1--2.4~keV energy band, \cite{Dempsey-93-ApJ-4} reported that the corona of \szp\ consists of two temperature plasma, with the derived temperatures of 1.2 and 16.6~MK. A few RS~CVn type binaries are also found to have three temperature quiescent corona  \citep[e.g., UX~Ari, V711~Tau, and EI~Eri;][]{Gudel-99-ApJ, Pandey-12-MNRAS-8}.

The time-resolved spectroscopy (present work) shows that the flare temperature varies with time, and the observed peak flare temperature is derived to be 199~MK. Using MAXI/GCS observations \cite{Tsuboi-16-PASJ-4} estimated the temperature of a flare on \szp\ to be $>$23.2~MK. For RS~CVn binaries, the flare temperatures have been estimated to be  25 -- 300 MK \citep[][]{Gudel-04-A+ARv-4, Osten-07-ApJ-2, Pandey-12-MNRAS-8}, whereas the temperature is also similar to those of the superflares \citep[][]{Maggio-00-A+A, Favata-00-A+A-3, Osten-10-ApJ-5, Karmakar-17-ApJ-5}. We found a delay between the temperature and the emission measure peaks. Similar delays have been observed in the solar and stellar flares \citep[][]{Sylwester-93-A+A-3, van-den-OordG-89-A+A-1, Stelzer-02-A+A-5}.

At the beginning of the flare, when the possible reconnection event leads to the release of the stored magnetic energy as kinetic energy, the temperature of the newly formed loop reaches its maximum near the loop top. In the later part of the flare, the temperature generally decreases as the heat is transmitted via the radiative and conductive cooling process. Thus the temperature peaks at the beginning of the flare. When the heat pulse reaches the chromospheric footpoint and the denser chromospheric material gets evaporated, the density of the coronal loop and hence the emission measure is found to be increased. Therefore, emission measure peak is found to be delayed with respect to the temperature peak. The emission measure near the flare peak is found to be $\sim$97 times that of the minimum value. The increase in density indicates a large amount of evaporation of chromospheric material within the flaring loop during the flare.  

The peak X-ray luminosity of the flare on \szp\ is estimated as 4.2~\E{33}~\ergs, which is $\sim$306 times more than its quiescent luminosity. Previously, two flares have been observed on \szp\ using MAXI/GSC instrument in the 2--20~keV band. The peak luminosity during those flares were estimated as 5~\E{32}~\ergs\ \citep[detected on 28 September 2009;][]{Tsuboi-16-PASJ-4}, and    3~\E{33}~\ergs\  \citep[detected on 5 November 2011;][]{Negoro-11-ATel-7}. We plotted the flare luminosities vs. quiescent luminosities of RS~CVn binaries in Figure~\ref{fig:discuss}(b). The legends [1--6] show the quiescent and flare luminosity of RS~CVn binaries in the 0.35--10~keV band, whereas EUV observations from \cite{Osten-99-ApJ} are also plotted with pink squares (legend 7). We did not find any correlation between the luminosities of the flares and the quiescent states in X-ray bands. A similar thing was also noticed for the solar case, where the flare luminosities vary within a wide range of distribution. The solid blue line in Figure~\ref{fig:discuss}(b)  showing a linear relationship between the EUV flaring and quiescent luminosities, as adopted from \cite{Osten-99-ApJ}, might be due to a selection or detection bias.

\begin{figure}
\begin{center}
\includegraphics[height=8.5cm, angle=-90]{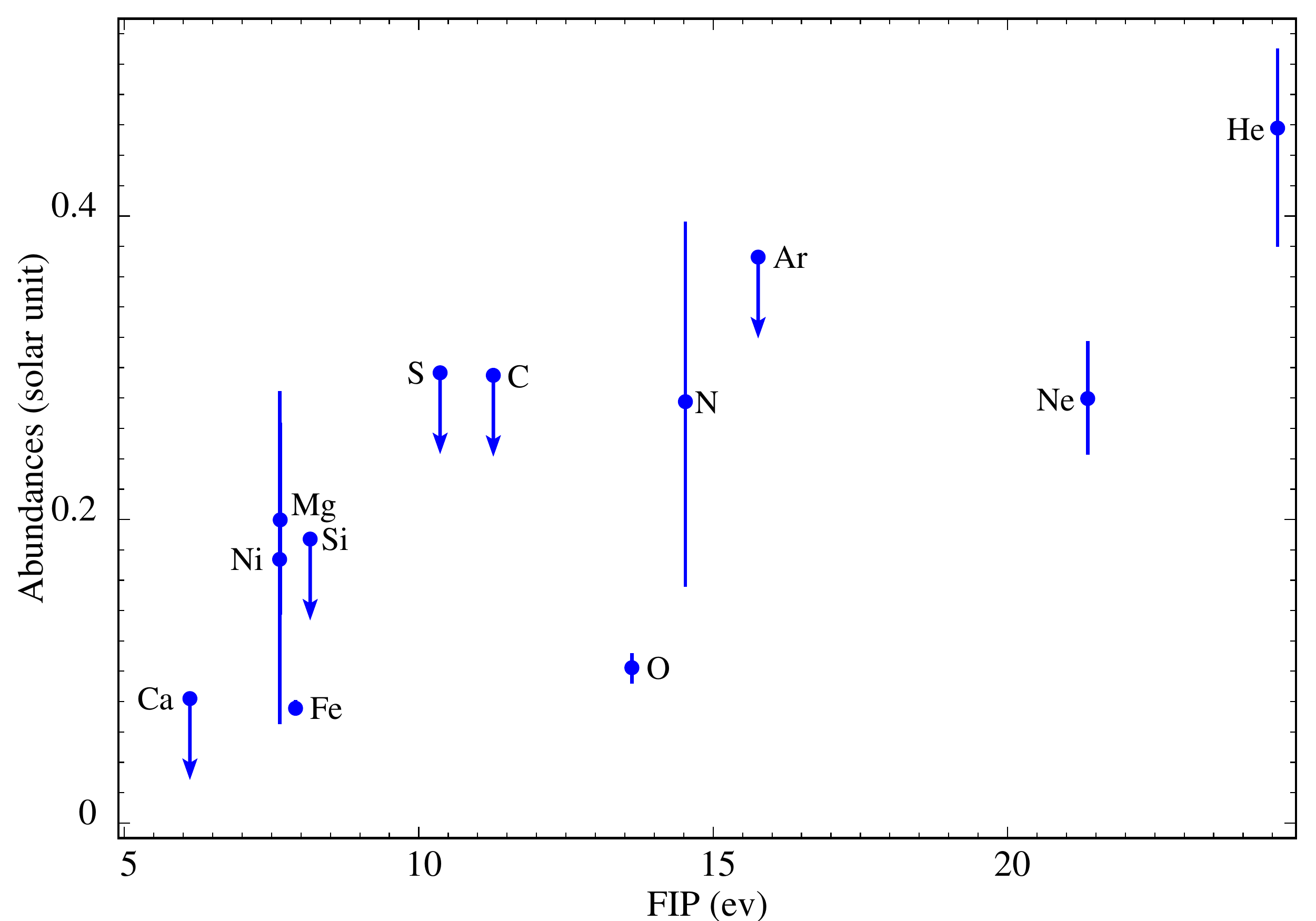}
\caption{\label{fig:fip} Elemental abundances relative to solar photospheric values (\citealt{Anders-89-GeCoA-2}) as a function of the first ionisation potential (FIP) during quiescence (Q2 segment) of \szp. The elemental abundances show an inverse FIP effect.
}

\end{center}
\end{figure}

\subsection{Metallic abundances} \label{sec:disc_abund}

Using \xmm\ RGS spectra, we estimated the elemental abundances of heavy elements. In Figure~\ref{fig:fip}, the elemental abundances with respect to solar photospheric abundances are plotted against the First Ionisation Potential (FIP) of the corresponding element. For the stars like the Sun, it has been observed that the low-FIP elements like Fe, Si, and Mg are 3--4 times more abundant in the corona when compared to high-FIP elements like C, N, O, and Ne. This phenomenon is known as coronal FIP bias \cite[e.g., see][]{FeldmanU-92-PhyS-2, Del-ZannaG-14-A+A-7}. A reversed pattern (i.e., enhanced high-FIP elements when compared to the low-FIP elements) is frequently observed in active stars \citep[e.g., see][]{BrinkmanA-01-A+A-3, AudardM-03-A+A-8}. From Figure~\ref{fig:fip}, we identified that the abundance pattern of the quiescent corona of \szp\ indicates an inverse FIP effect. Several theoretical studies attempted to explain the FIP bias \citep[see][]{HenouxJ-98-SSRv-2}. The latest model is based on the Ponderomotive force model \citep[][]{Laming-15-LRSP-3, LamingJ-21-ApJ}, which is able to reproduce the overall features of both the FIP and inverse FIP bias. The findings of the inverse FIP effect on the quiescent corona of \szp\ can be explained using this model.

Due to the non-availability of high-resolution spectra during the flare, we could not carry out a quantitative study of the temporal evolution of the individual chemical abundances during the flare. However, the global metallic abundance of the flaring corona is found to vary during the flare. In the rising phase of the flare, the abundance seems to increase from 0.4~\zsun\ to 0.6~\zsun. However, considering the uncertainty level, this increase is not significant. In the decay phase of the flare, the abundance is found to decrease significantly up to $\sim$0.1~\zsun. In the case of other RS~CVn binaries such as $\sigma$~Gem, V711~Tau, EI~Eri, II~Peg, UX~Ari, and $\sigma^2$~Crb, the metallic abundances have also been found to vary during the flaring events \citep[see][]{MeweR-97-A+A-1, Gudel-99-ApJ, OstenR-03-ApJ-5, Nordon-08-A+A-1, Pandey-12-MNRAS-8}. The variation of metallic abundances during the flare on \szp\ is found to be consistent with the previously observed behavior. At the beginning of the event, fresh chromospheric material is supposed to be evaporated in the flaring loops, which results in the enhancement of its abundance. During the decay phase of the flare, the fractionation mechanism is considered to be responsible for lower abundances observed in the coronal plasma until it reaches the quiescent state value \citep[][]{Favata-03-SSRv-7}.

\section*{Acknowledgments}
We thank the anonymous reviewer for a careful review that greatly improved the manuscript.
This research has made use of data and software provided by the \textit{High Energy Astrophysics Science Archive Research Center} (HEASARC), which is a service of the Astrophysics Science Division at NASA/GSFC. This work made use of data supplied by the \textit{Swift Science Data Centre} at the University of Leicester, UK. This research has also made use of the \textit{XRT Data Analysis Software} (XRTDAS) developed under the responsibility of the \textit{ASI Science Data Center} (ASDC), Italy. This work is also based on observations obtained with \textit{XMM-Newton}, an ESA science mission with instruments and contributions directly funded by the ESA Member States and the USA (NASA). SK acknowledges Drs Wm. Bruce Weaver and Craig Chester from Monterey Institute for Research in Astronomy for useful discussion. ISS acknowledges the support of the Ministry of Science and Higher Education of the Russian Federation under grant 075-15-2020-780 (N13.1902.21.0039).

\section*{Data availability}
The data underlying this article are available in the HEASARC archive (\href{https://heasarc.gsfc.nasa.gov/db-perl/W3Browse/w3browse.pl}{https://heasarc.gsfc.nasa.gov/db-perl/W3Browse/w3browse.pl}), \xmm\ archive (\href{http://nxsa.esac.esa.int/nxsa-web/#search}{http://nxsa.esac.esa.int/nxsa-web/\#search}), and \swift\ archive (\href{https://heasarc.gsfc.nasa.gov/cgi-bin/W3Browse/swift.pl}{https://heasarc.gsfc.nasa.gov/cgi-bin/W3Browse/swift.pl}).

\bibliography{SK_collections}{}
\label{draft_end}

\clearpage
\setcounter{page}{1}
\begin{figure*}
  \vspace{2cm}
  \center
  \LARGE\textbf{\sc online-only material}\par\medskip
  \includegraphics[height=8.8cm,angle=-90,trim={0 35 0 0}, clip]{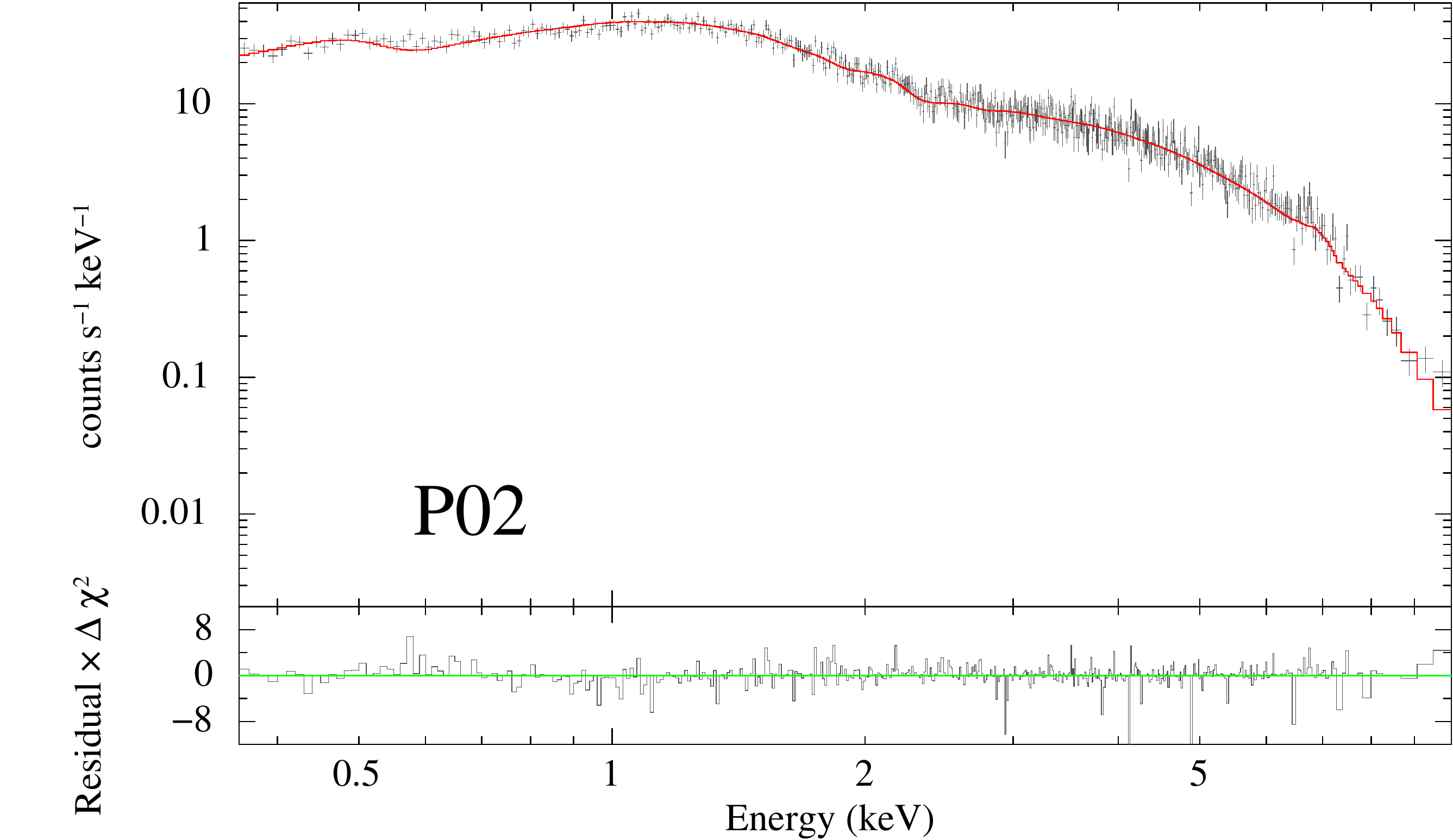}
  \includegraphics[height=8.8cm,angle=-90,trim={0 35 0 0}, clip]{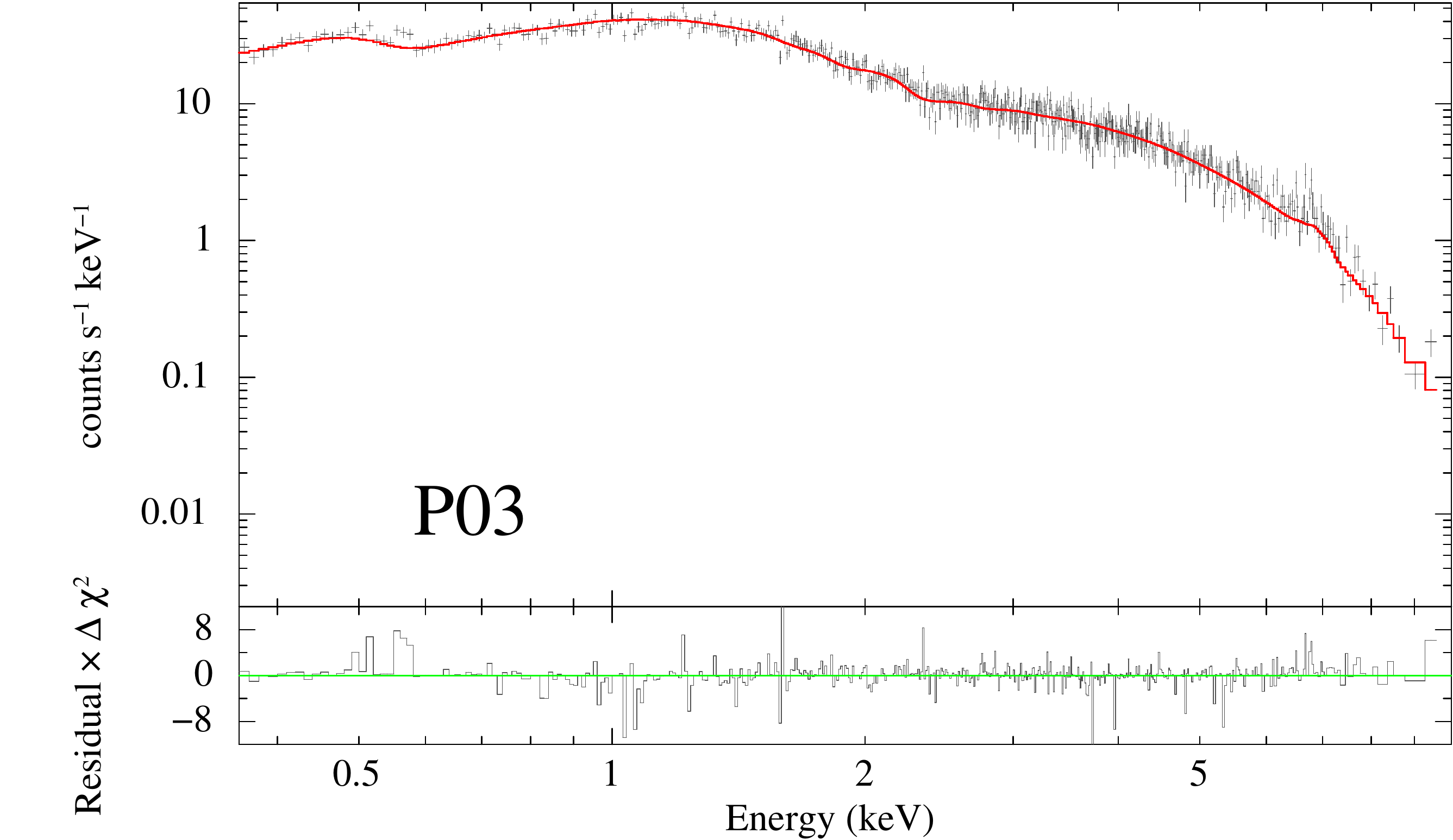}
  \includegraphics[height=8.8cm,angle=-90,trim={0 35 0 0}, clip]{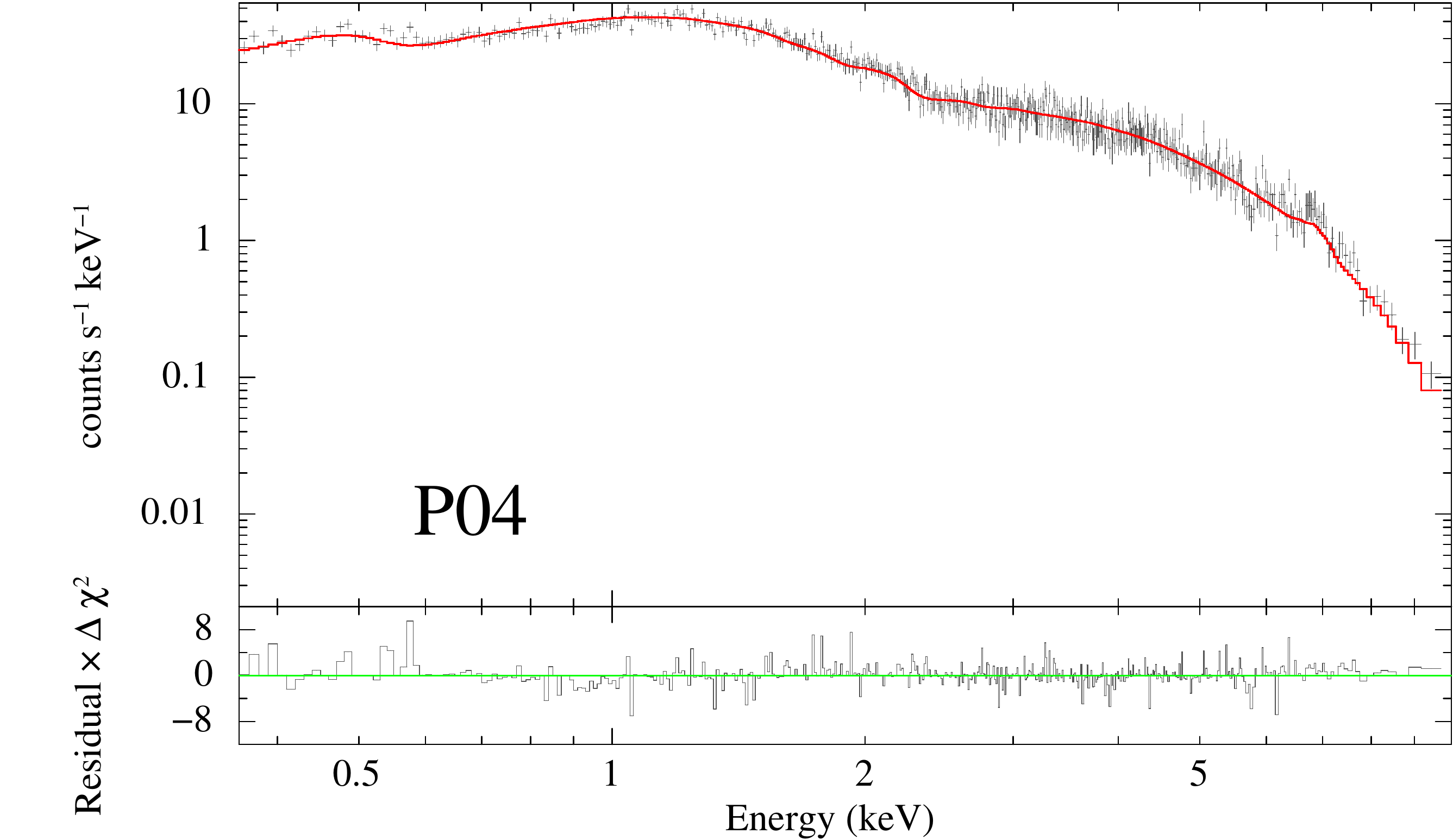}
  \includegraphics[height=8.8cm,angle=-90,trim={0 35 0 0}, clip]{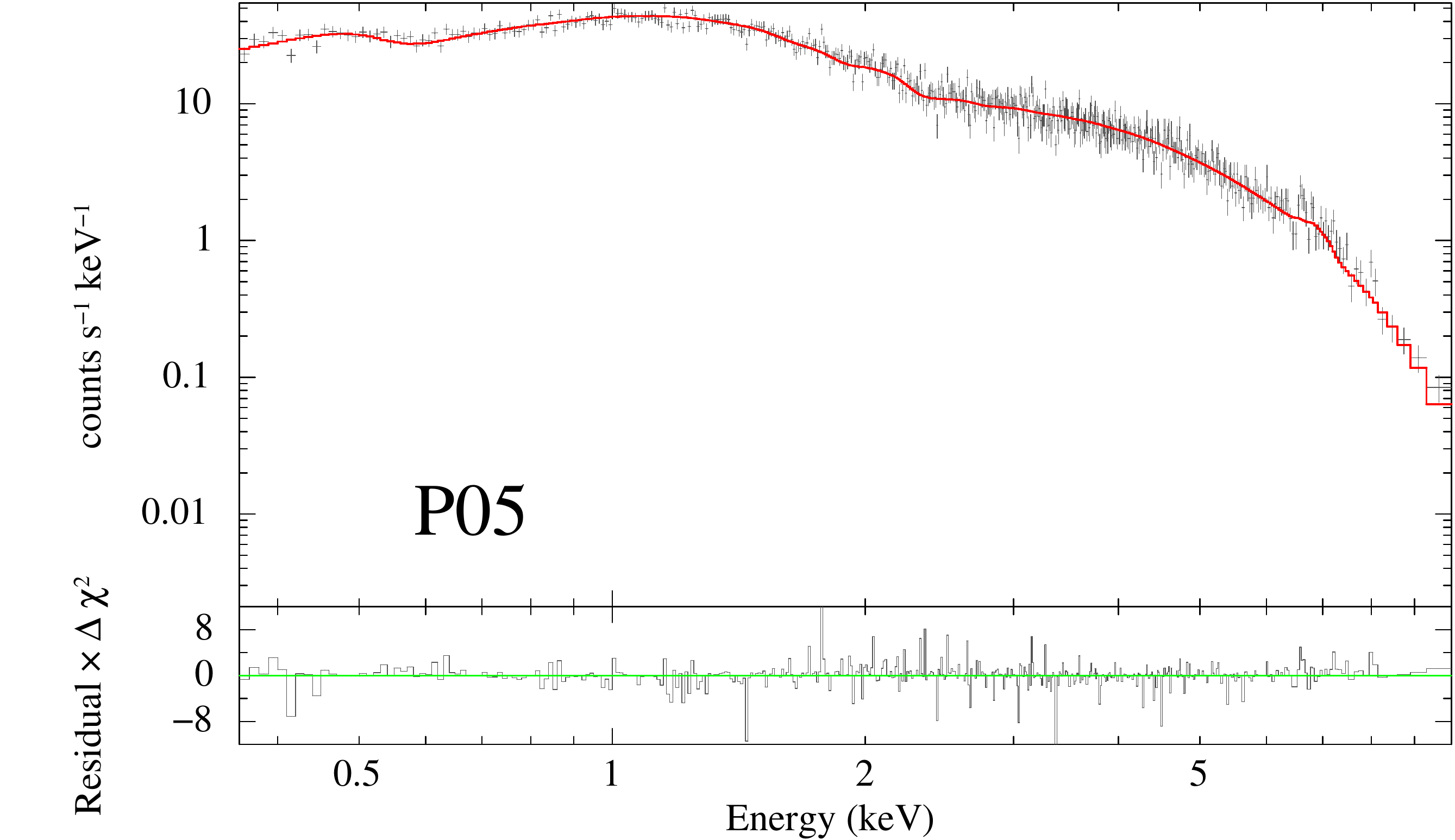}
  \includegraphics[height=8.8cm,angle=-90,trim={0 35 0 0}, clip]{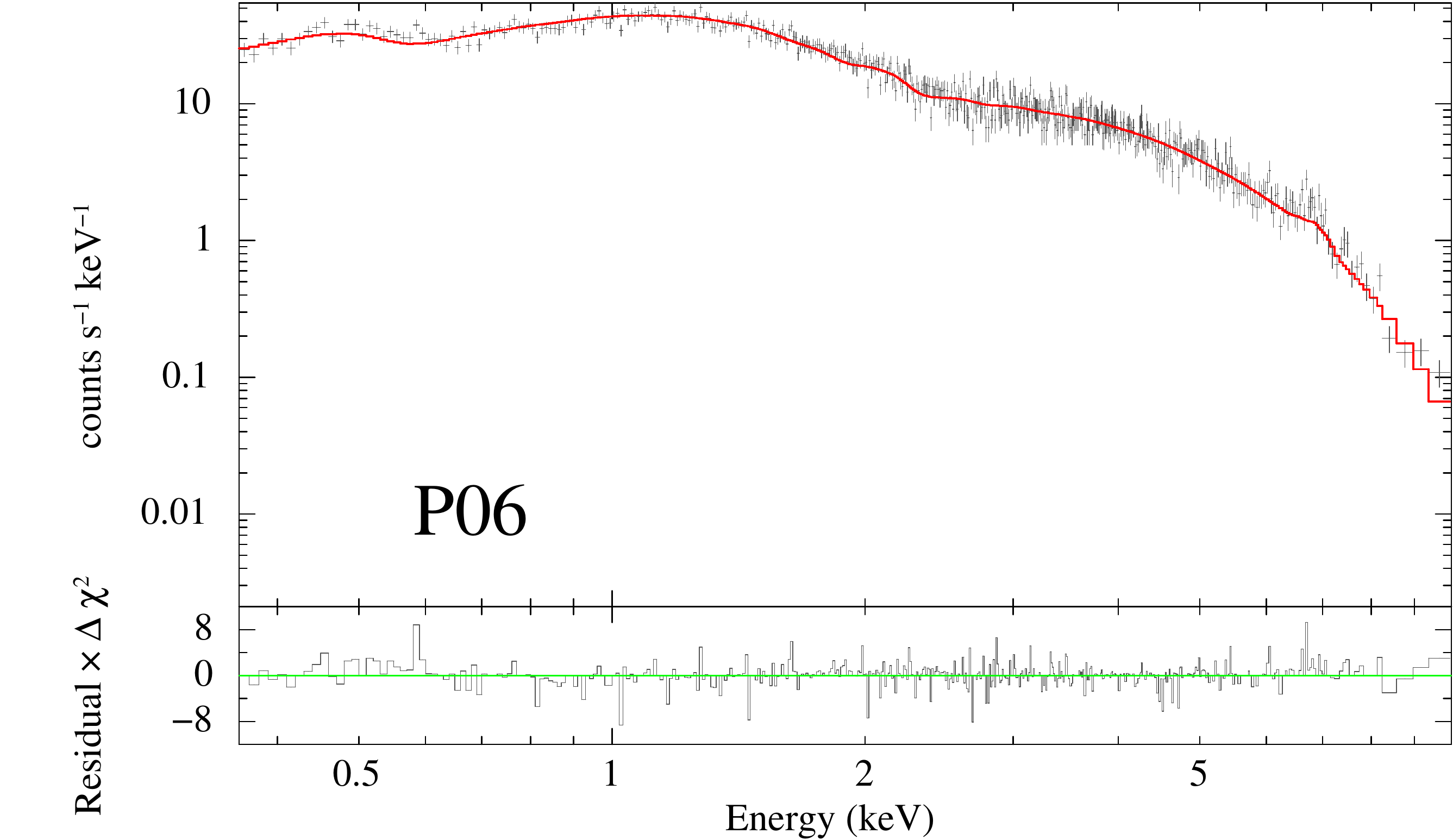}
  \includegraphics[height=8.8cm,angle=-90,trim={0 35 0 0}, clip]{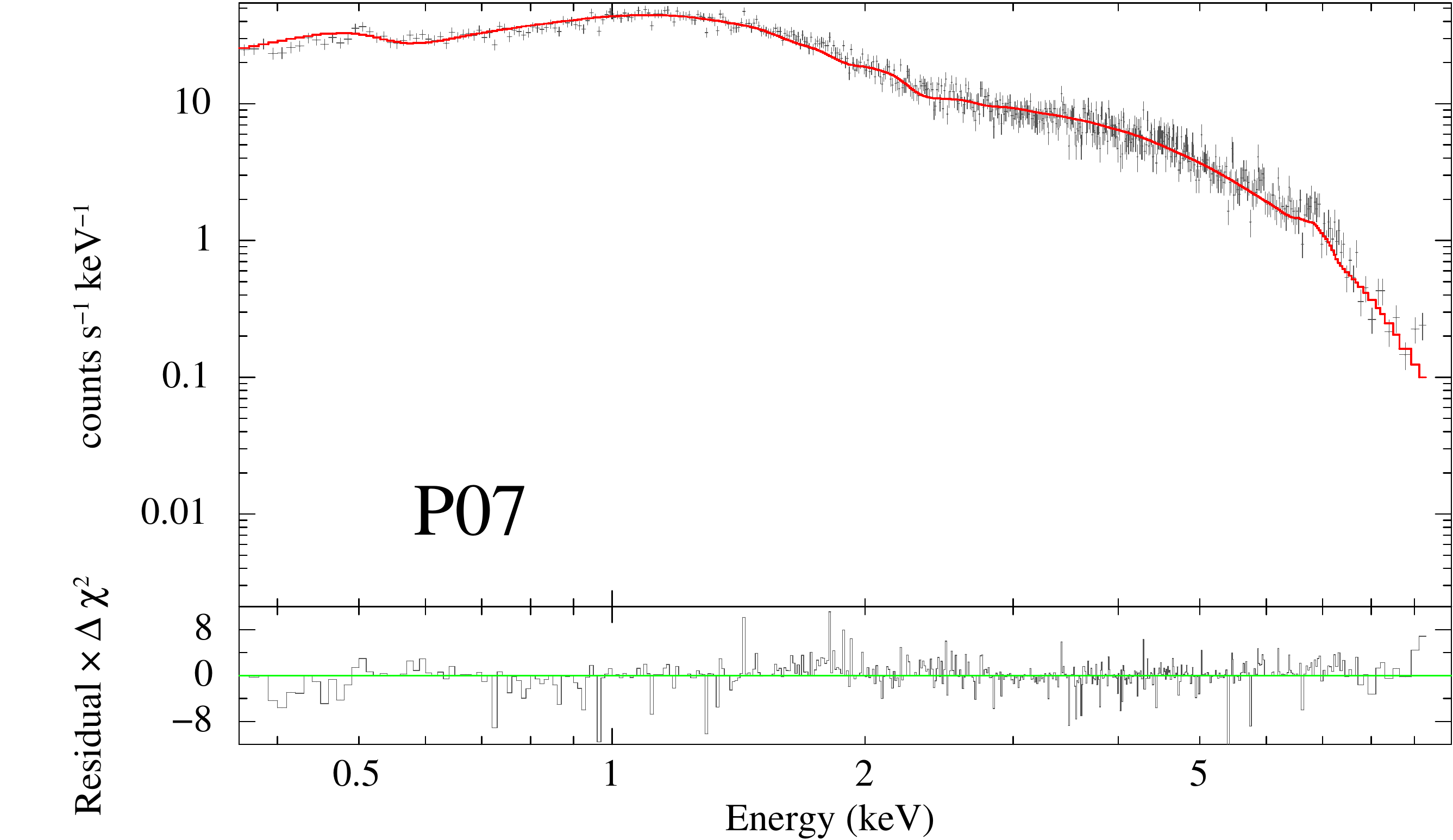}
  \caption{\textbf{The online-only material related to Figure~\ref{fig:spectra}}: Time-resolved \swift\ XRT spectra of \szp\ in 0.35--10 keV energy band are shown for all the individual segments. In the top panel of each figure, the spectra have been shown with gray plus symbols. The spectra have been extracted for a minimum of 20 counts per bin. The labels (black) in the left-bottom corner of each panel indicate the segment numbers. The start-time and end-time corresponding to each time segment is given in the second column of Table~\ref{tab:trs_all}. The time evolution of the spectra in each segment is clearly visible. The quiescent segment `Q1' has been best fitted with a 3-T astrophysical plasma model (\apec). Whereas all the time-resolved segments during the flare have been fitted with a 4-T \apec\ model. The best-fitted spectra have been shown with the solid red line. The bottom panel shows the residual in the unit of $\Delta$\chisq. Please see the text for a detailed explanation.  
}
\label{fig:spectra_online_only_1}
\end{figure*}

\begin{figure*}
  \center
  \LARGE\textbf{\sc online-only material}\par\medskip
  \includegraphics[height=8.8cm,angle=-90,trim={0 35 0 0}, clip]{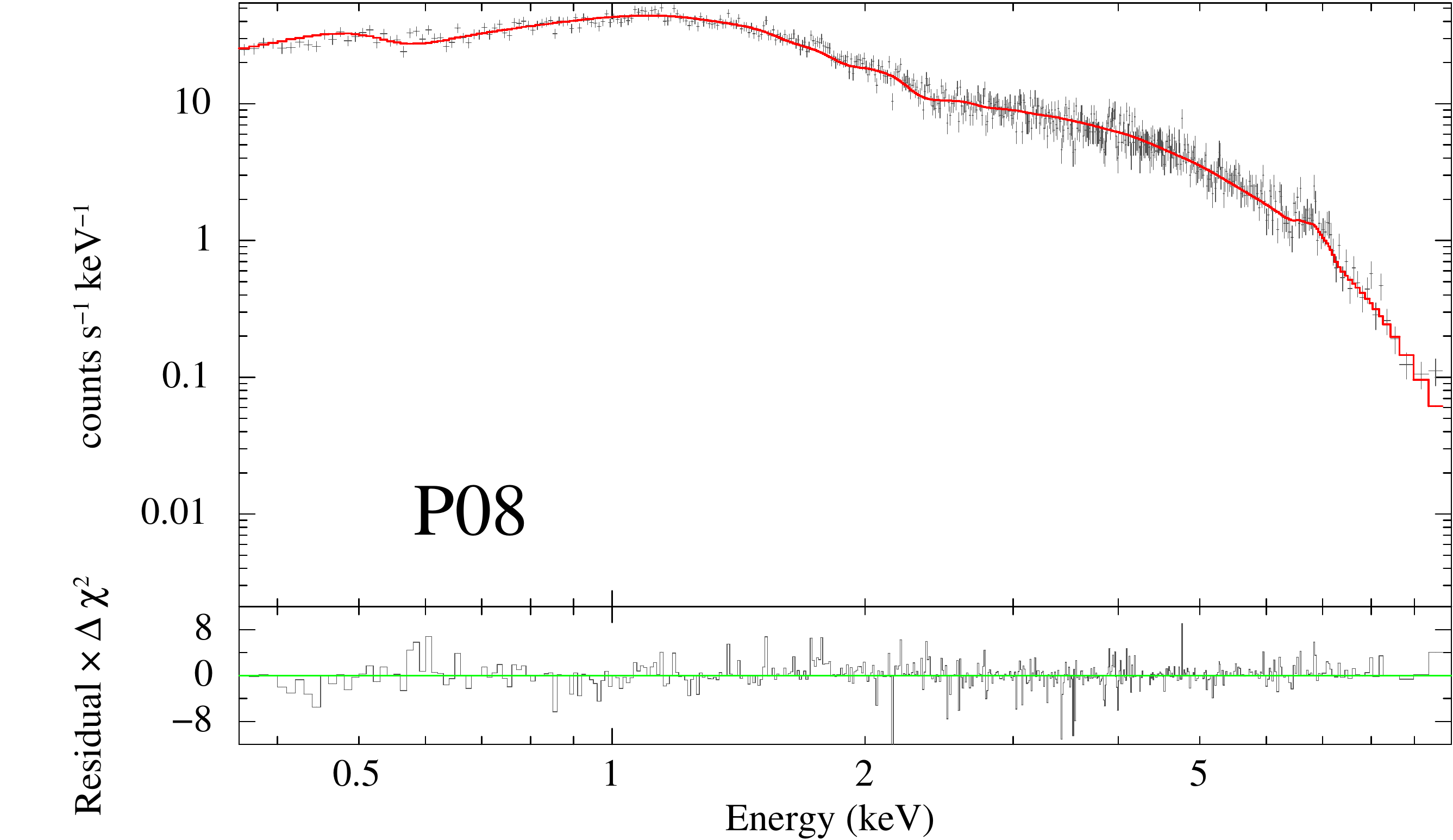}
  \includegraphics[height=8.8cm,angle=-90,trim={0 35 0 0}, clip]{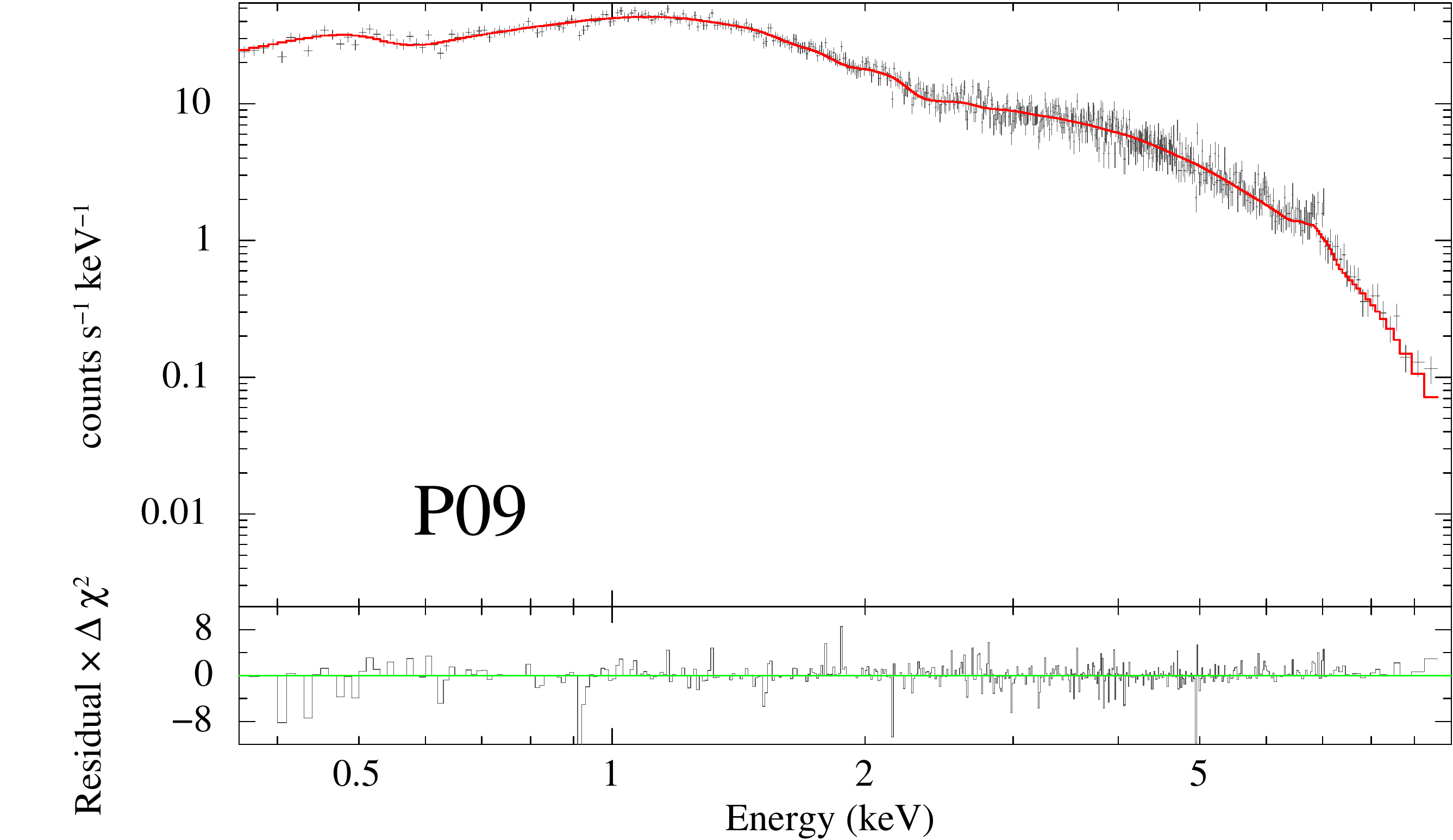}
  \includegraphics[height=8.8cm,angle=-90,trim={0 35 0 0}, clip]{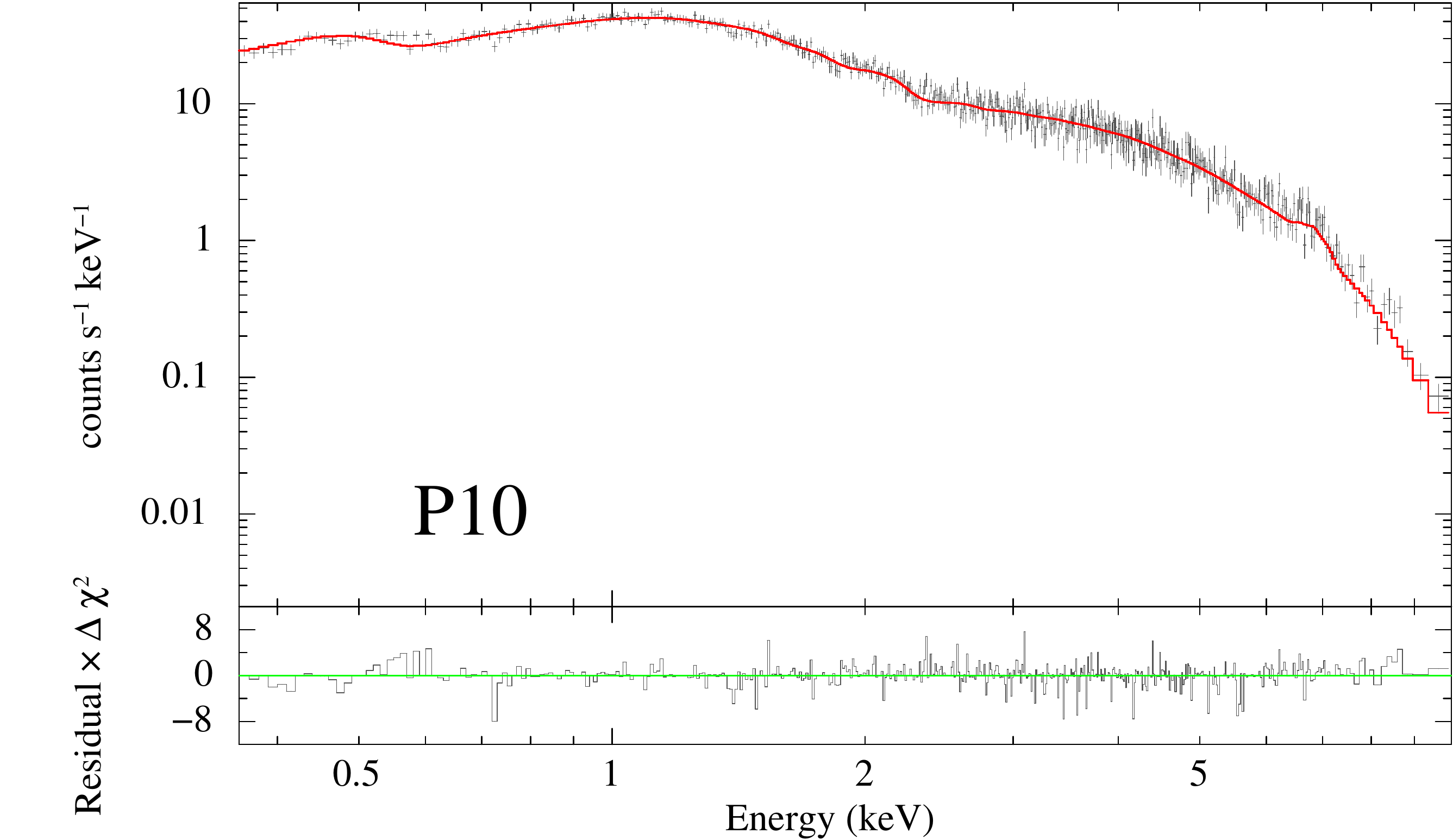}
  \includegraphics[height=8.8cm,angle=-90,trim={0 35 0 0}, clip]{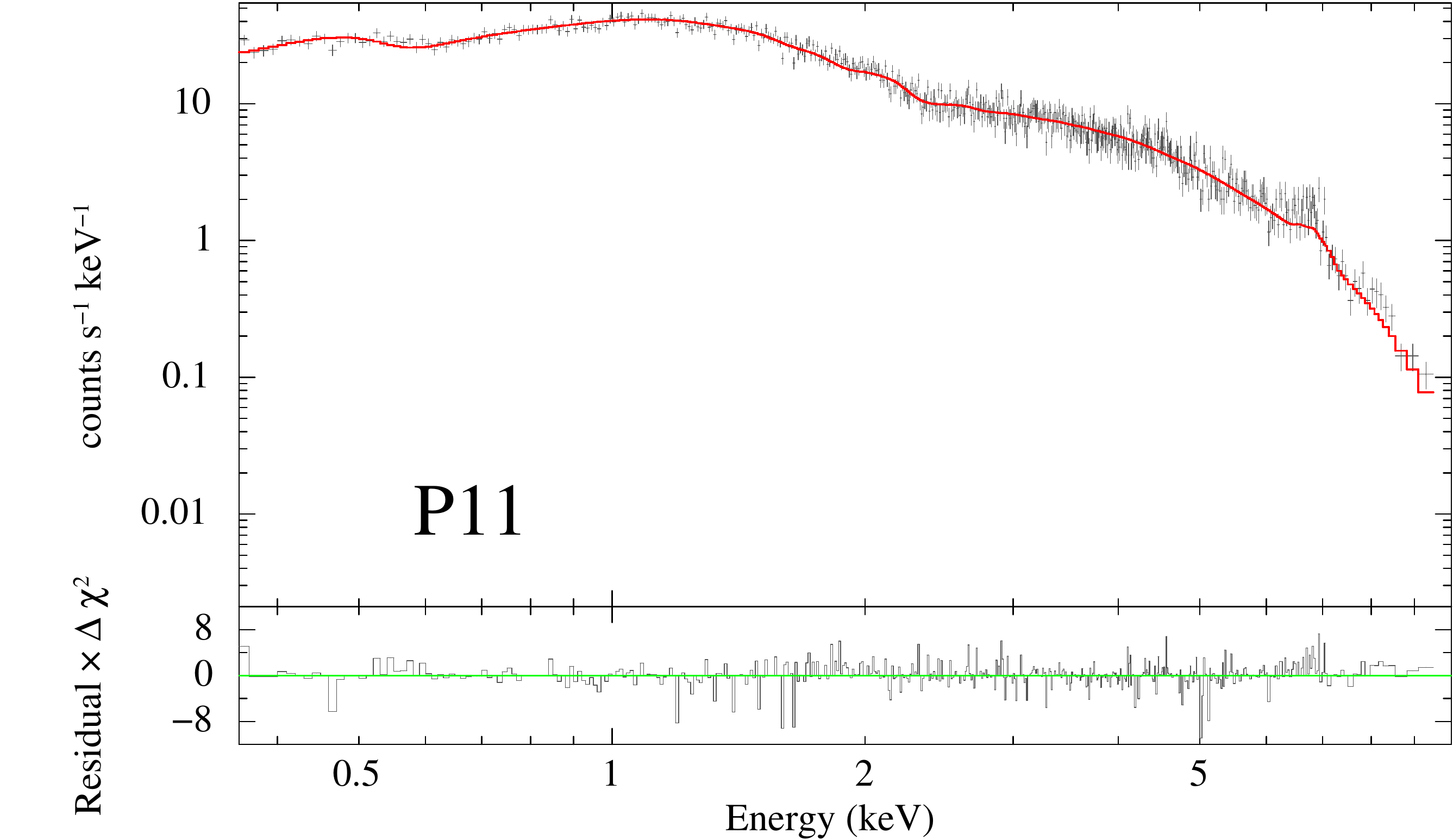}
  \includegraphics[height=8.8cm,angle=-90,trim={0 35 0 0}, clip]{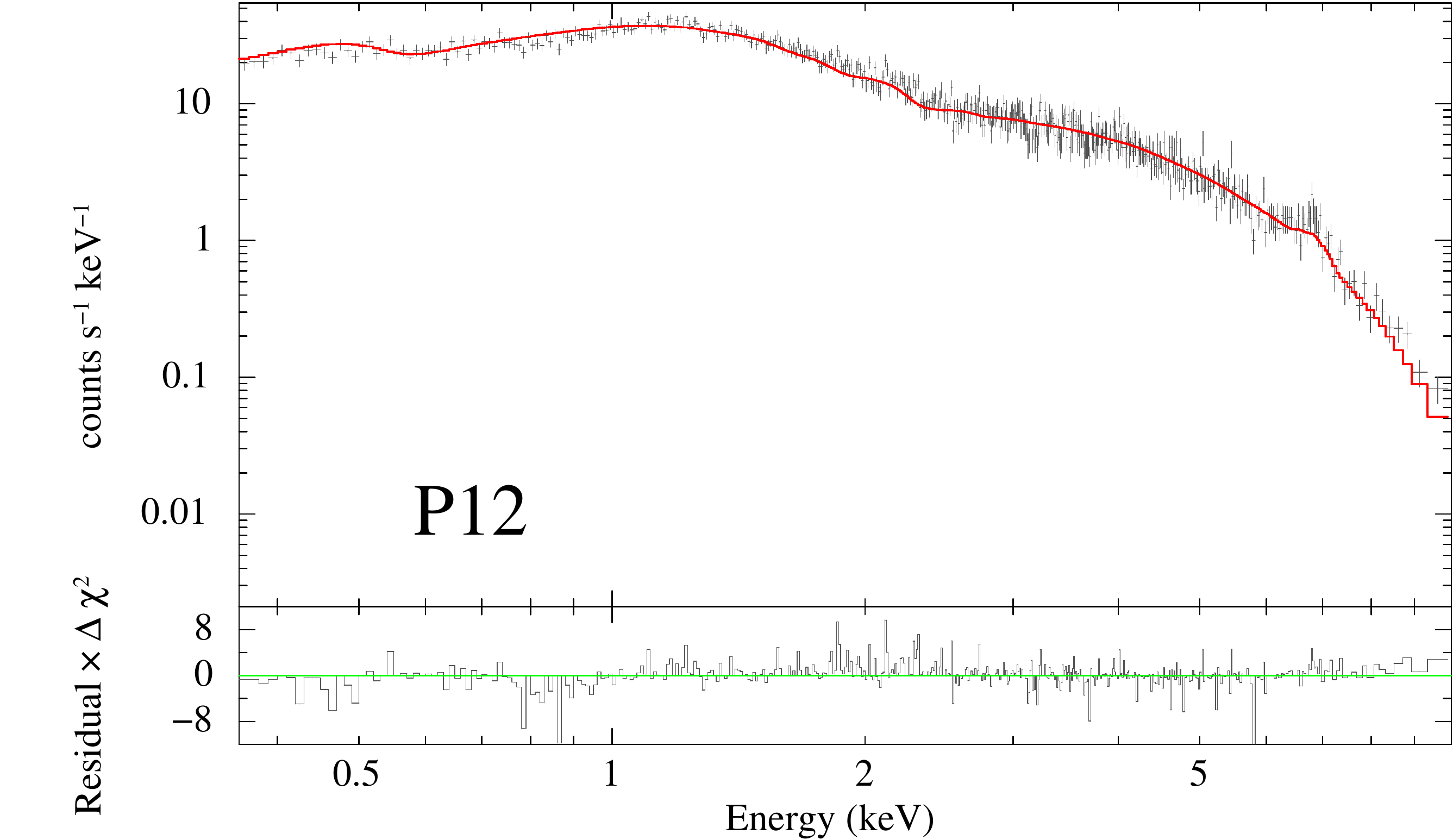}
  \includegraphics[height=8.8cm,angle=-90,trim={0 35 0 0}, clip]{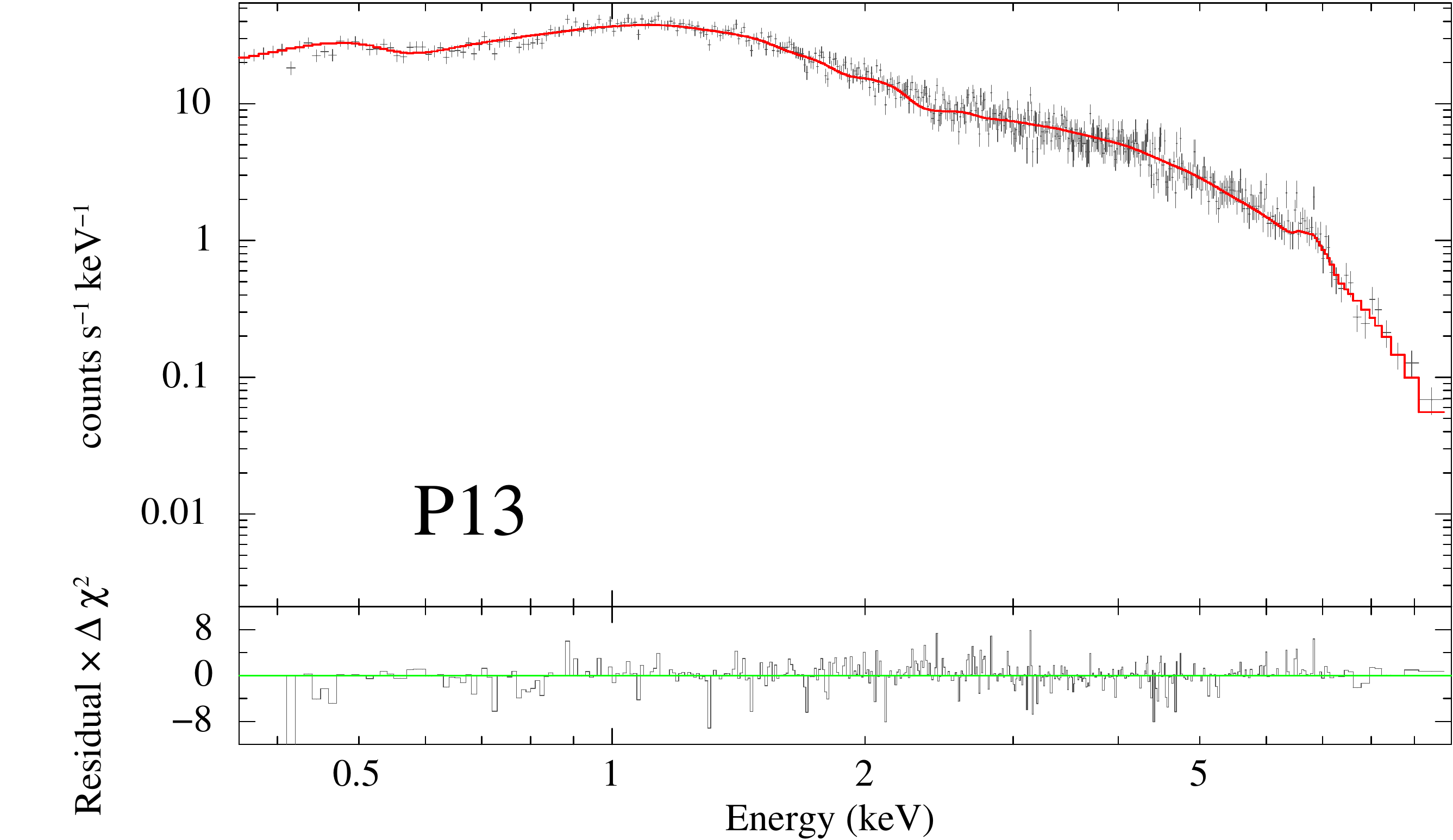}
  \includegraphics[height=8.8cm,angle=-90,trim={0 35 0 0}, clip]{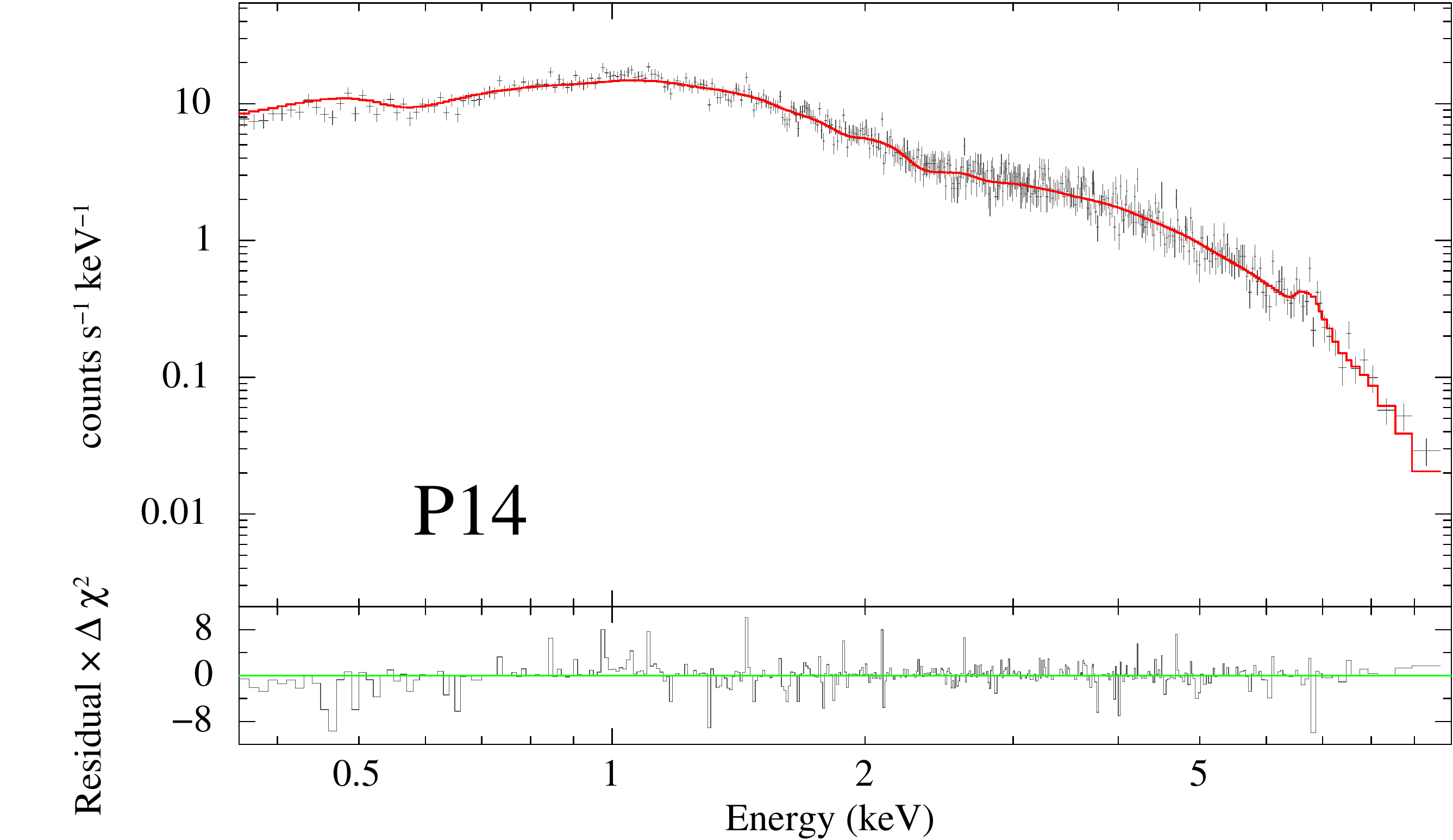}
  \includegraphics[height=8.8cm,angle=-90,trim={0 35 0 0}, clip]{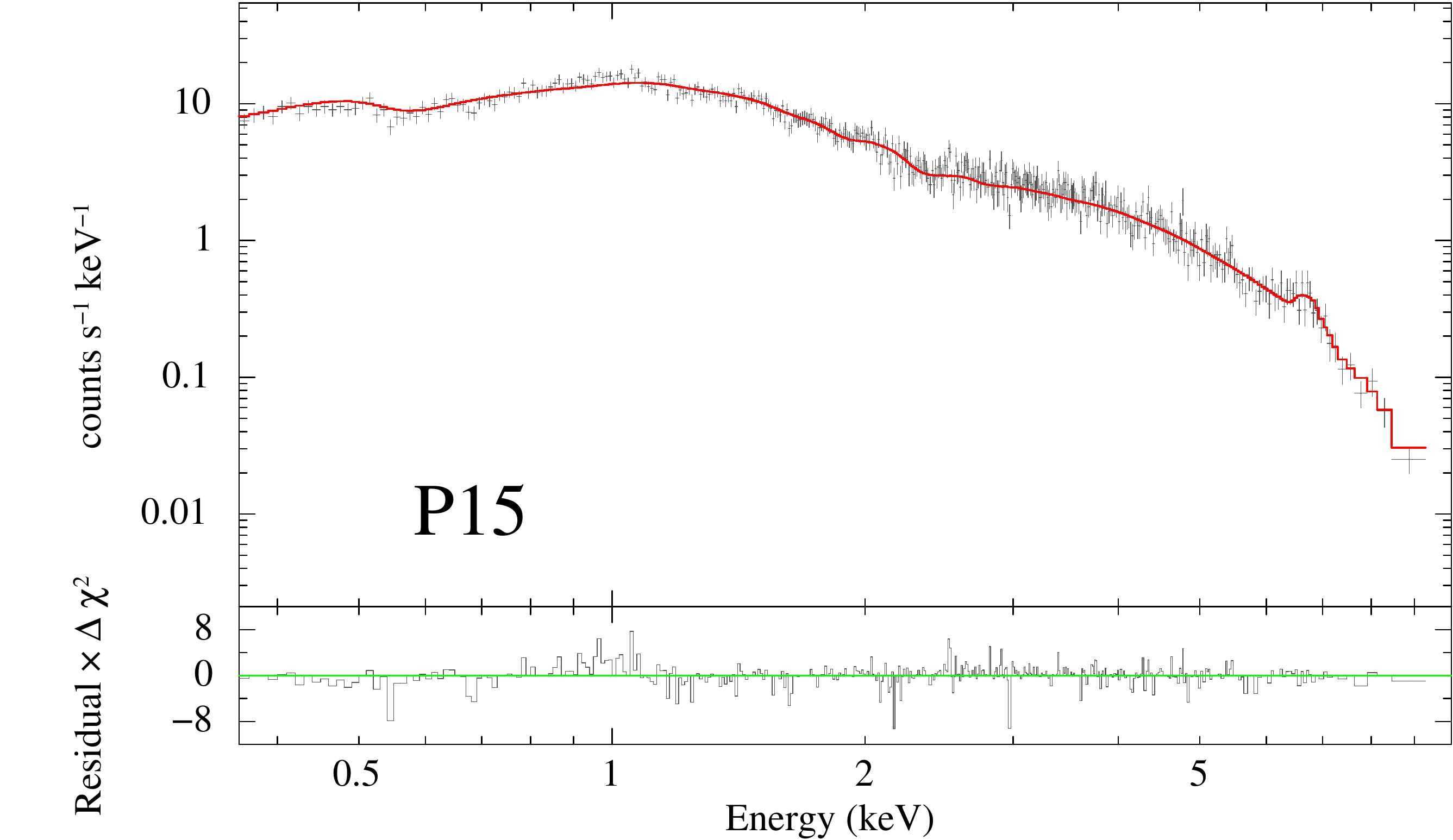}
\contcaption{The online-only material related to Figure~\ref{fig:spectra}}
\label{fig:spectra_online_only_2}
\end{figure*}

\begin{figure*}
  \center
  \LARGE\textbf{\sc online-only material}\par\medskip
  \center
  \includegraphics[height=8.8cm,angle=-90,trim={0 35 0 0}, clip]{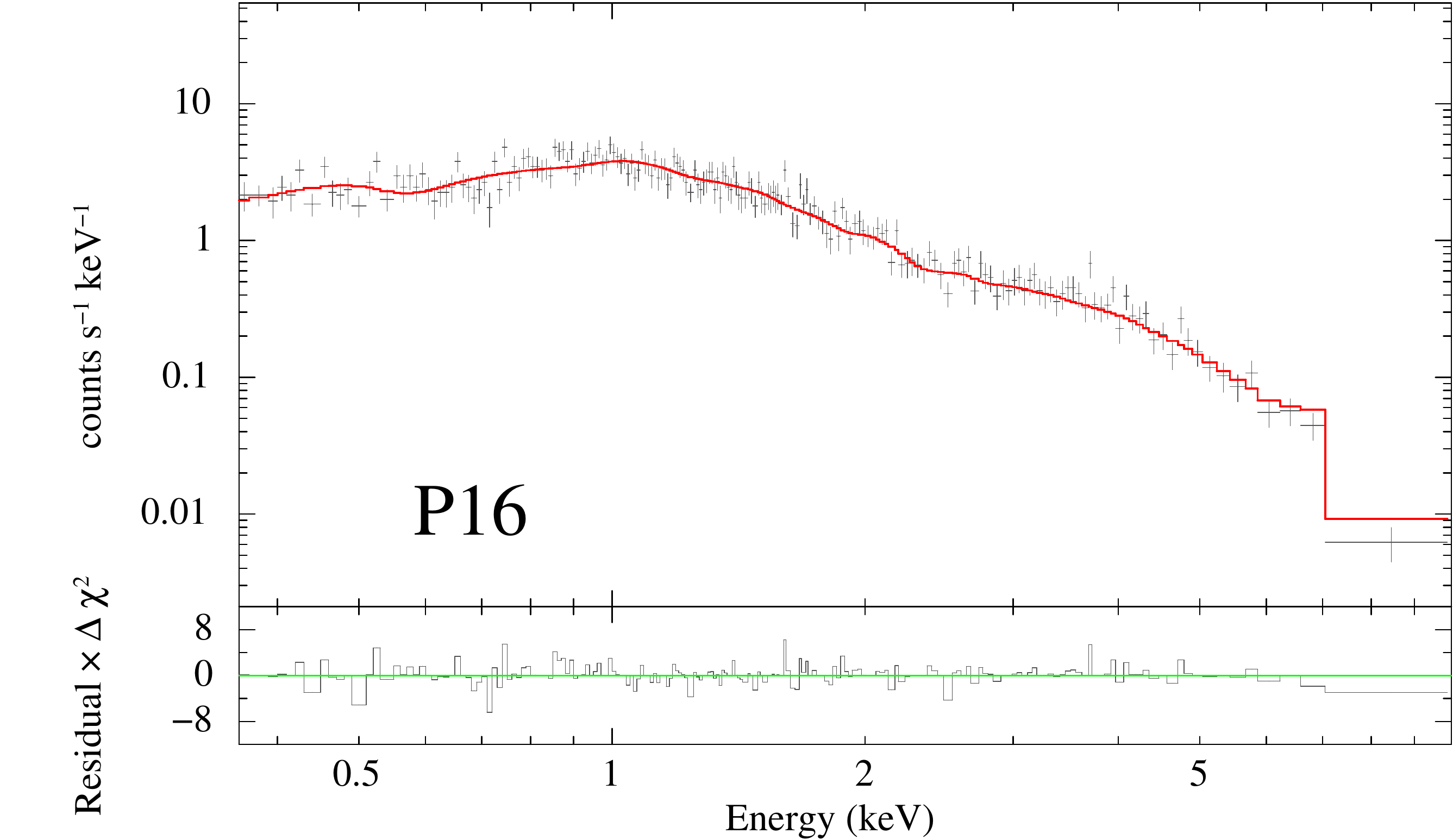}
  \includegraphics[height=8.8cm,angle=-90,trim={0 35 0 0}, clip]{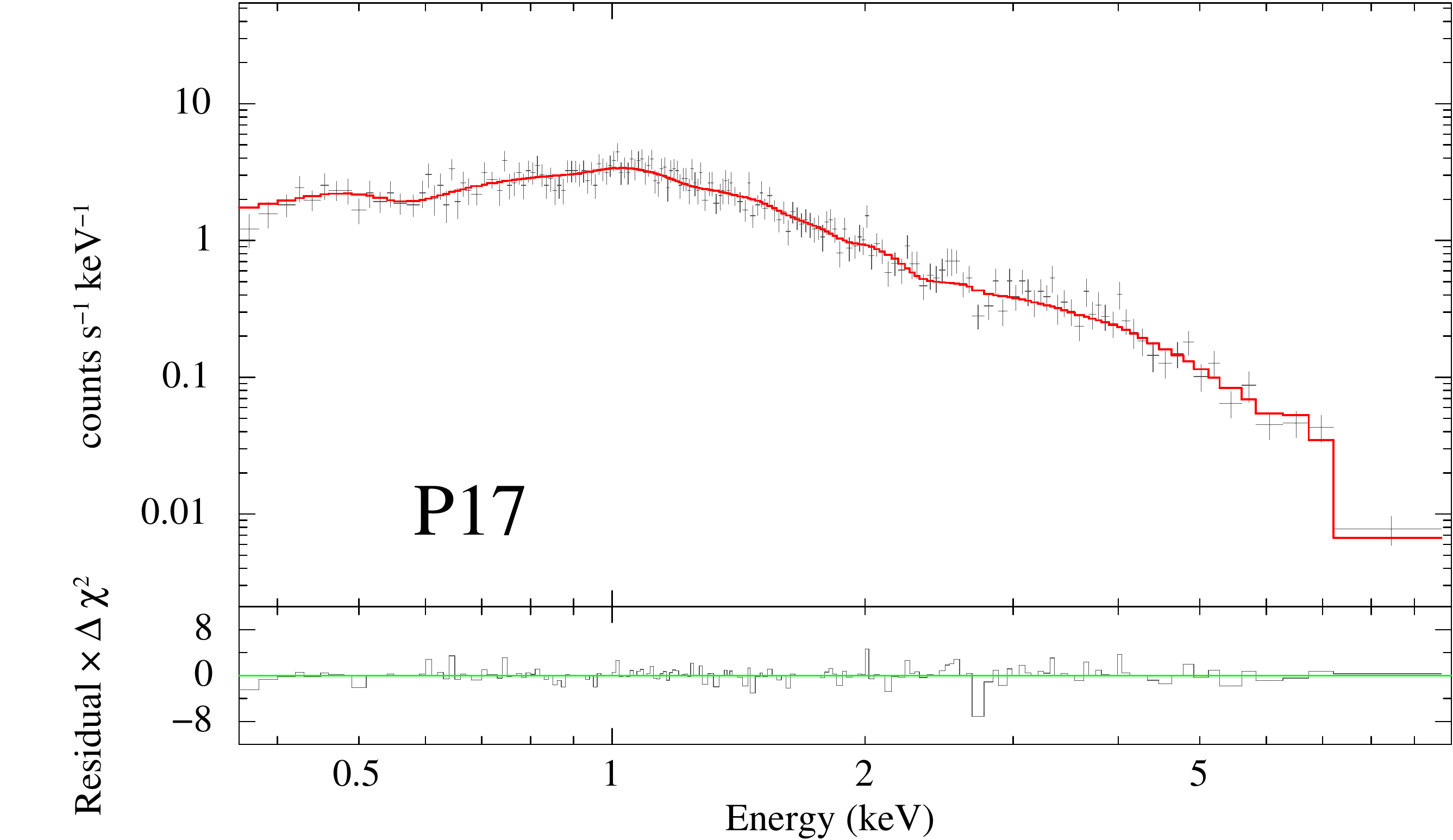}
  \includegraphics[height=8.8cm,angle=-90,trim={0 35 0 0}, clip]{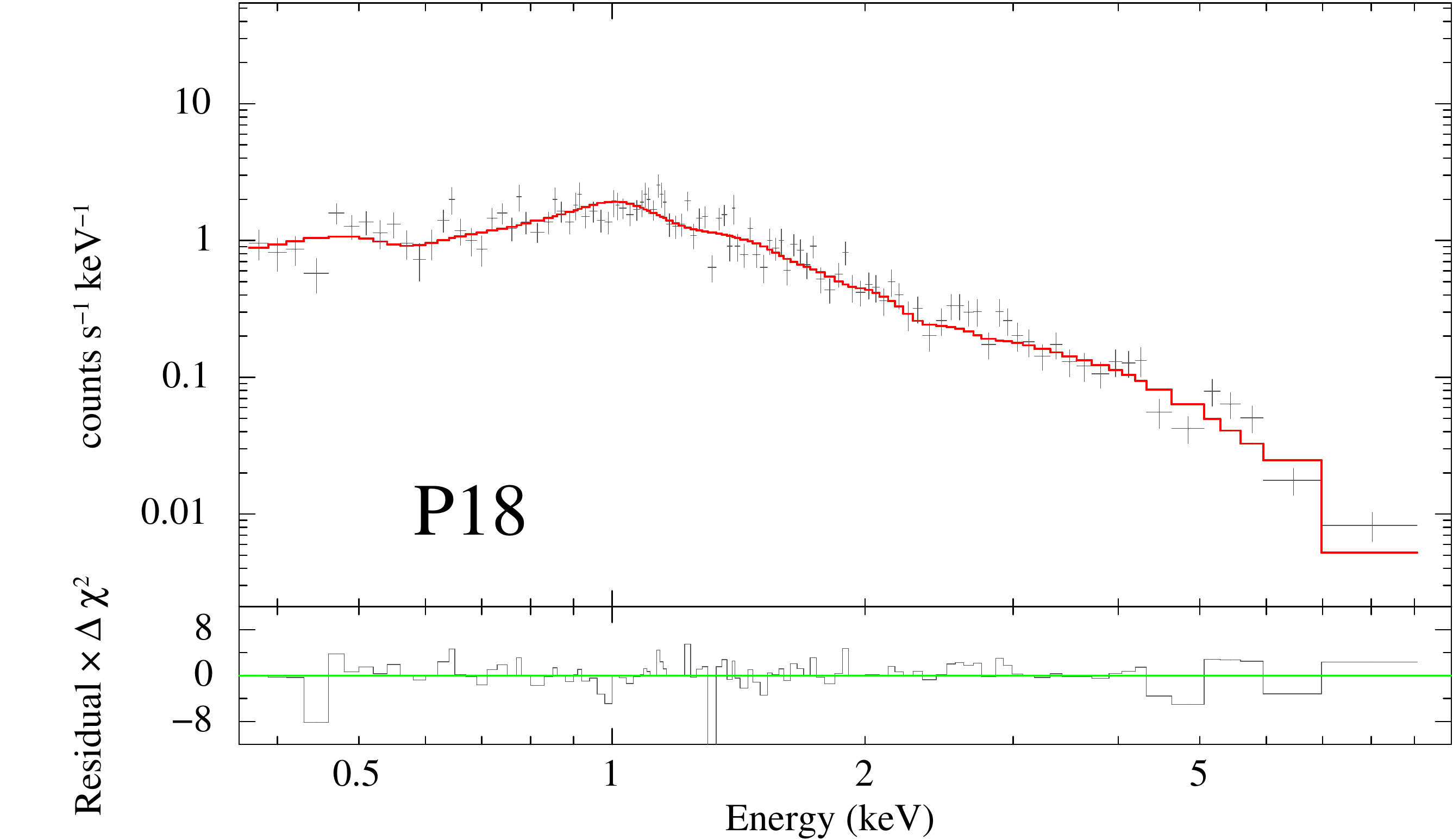}
  \includegraphics[height=8.8cm,angle=-90,trim={0 35 0 0}, clip]{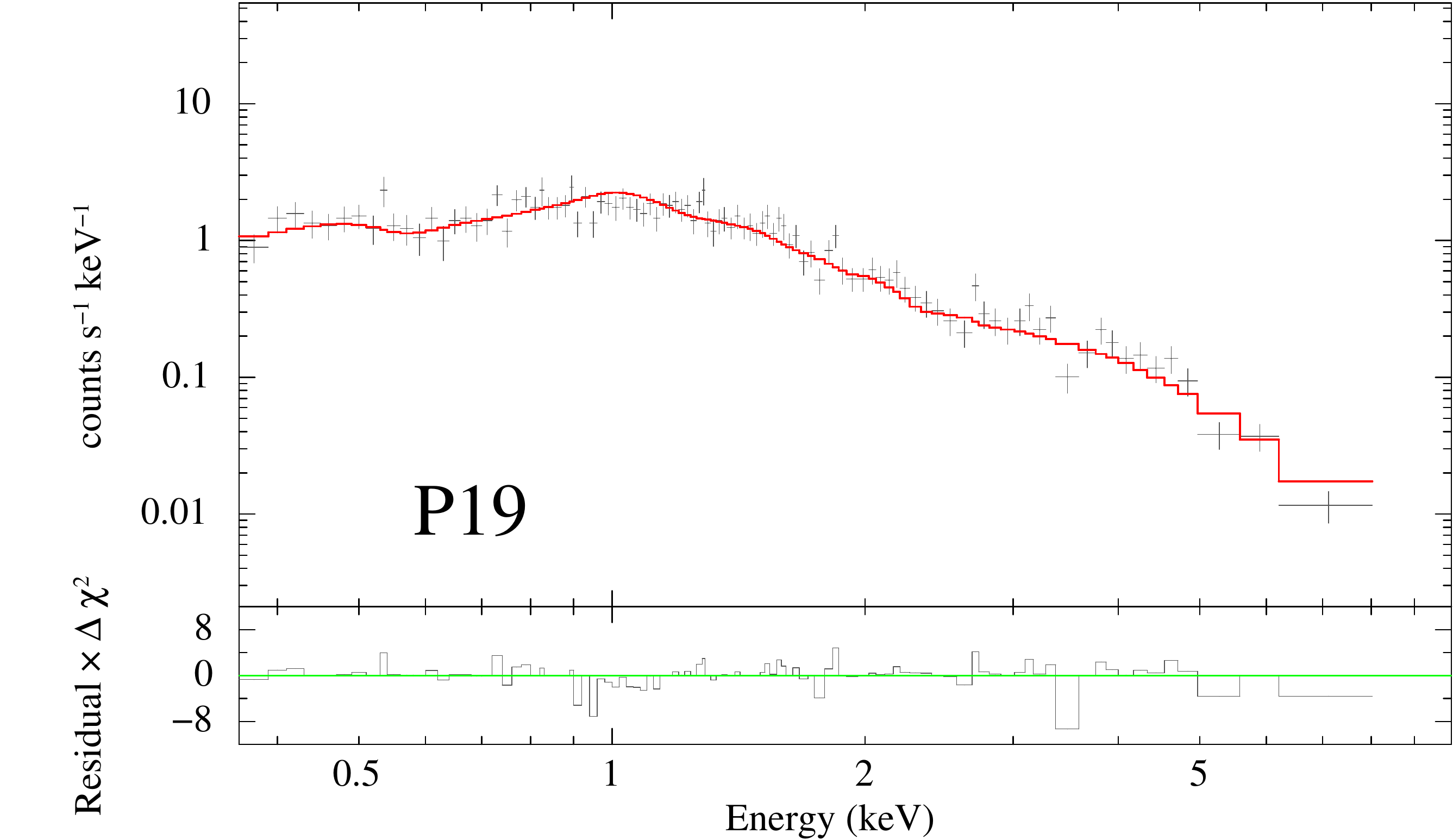}
  \includegraphics[height=8.8cm,angle=-90,trim={0 35 0 0}, clip]{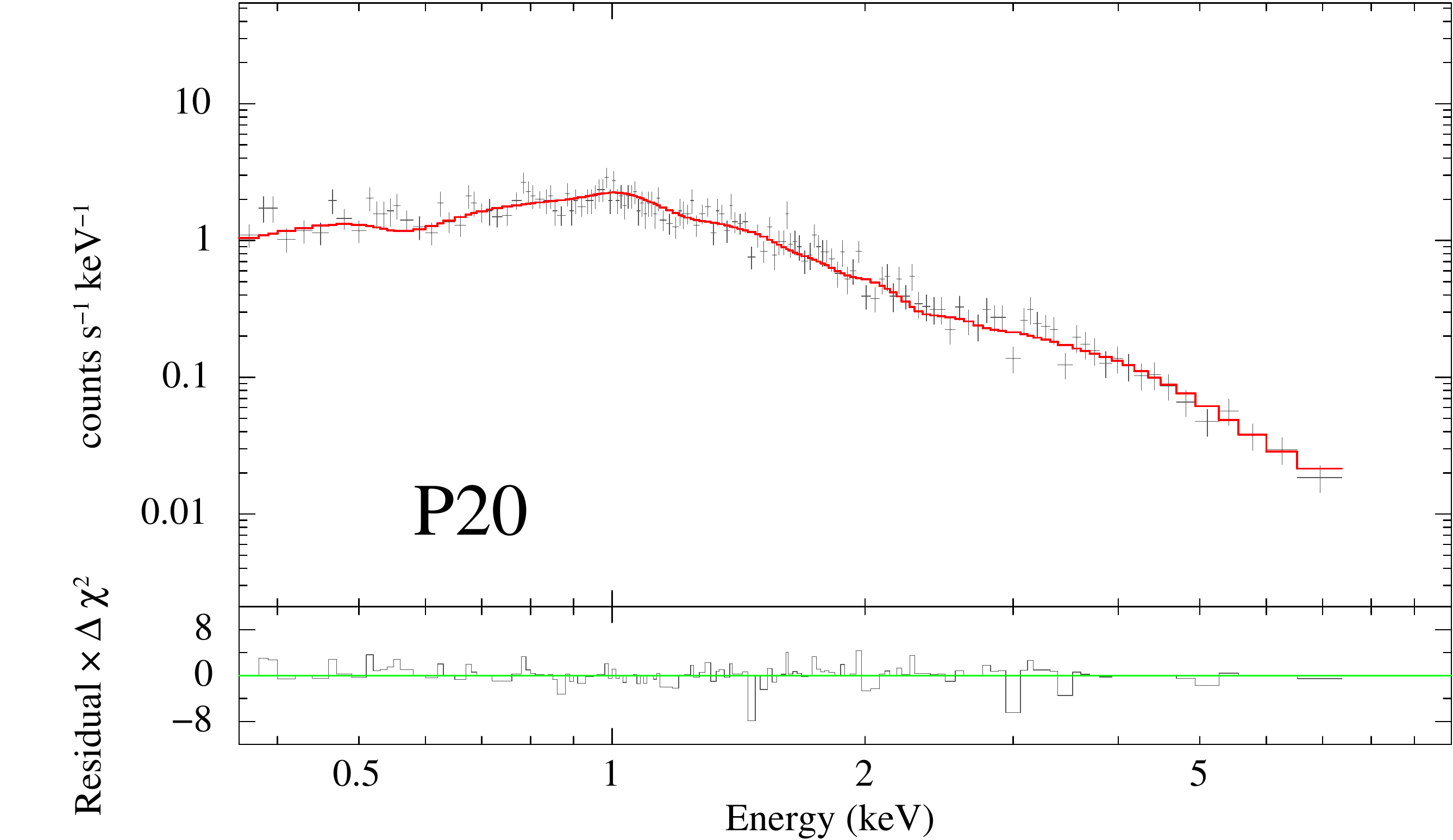}
  \includegraphics[height=8.8cm,angle=-90,trim={0 35 0 0}, clip]{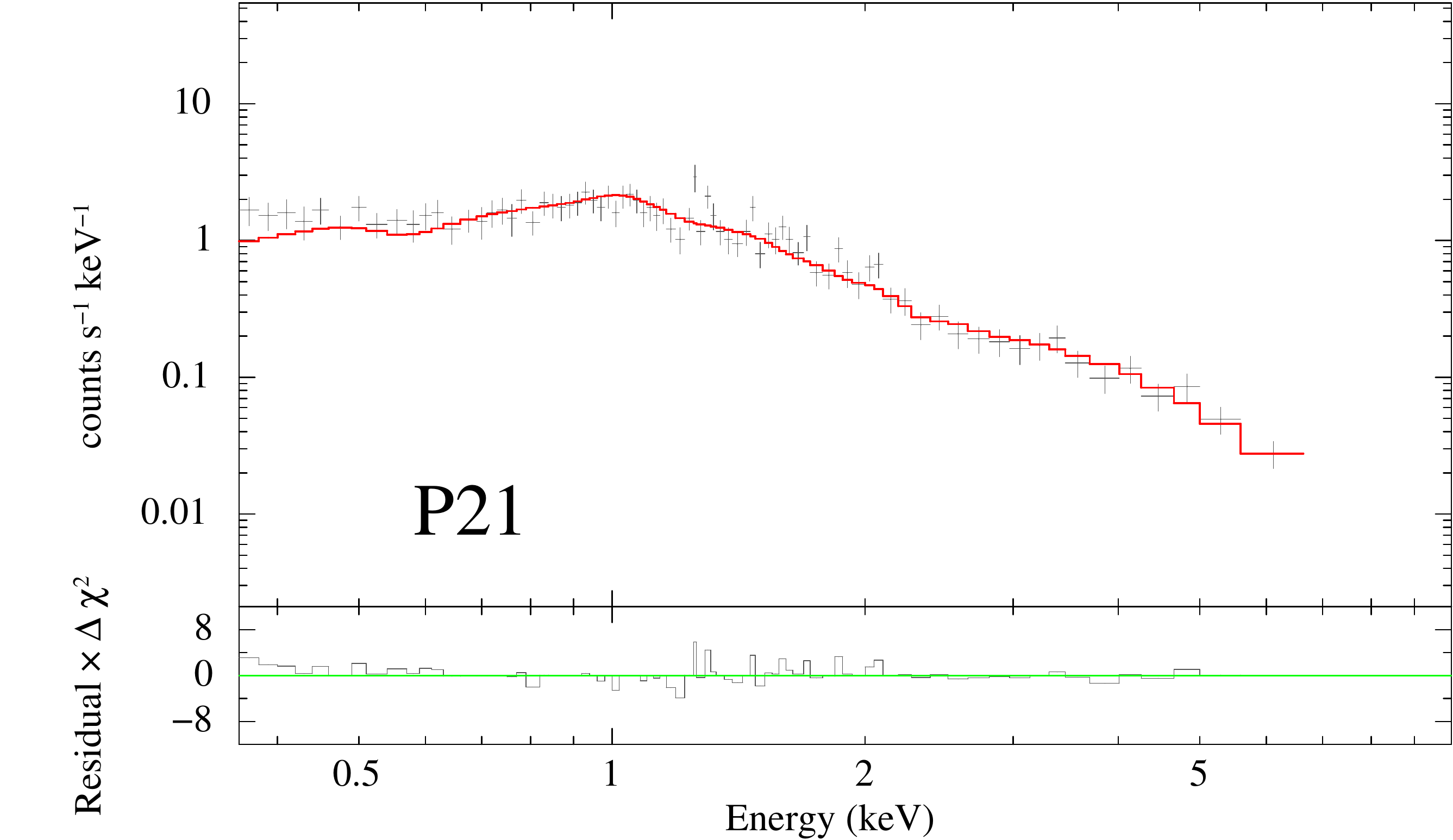}
  \includegraphics[height=8.8cm,angle=-90,trim={0 35 0 0}, clip]{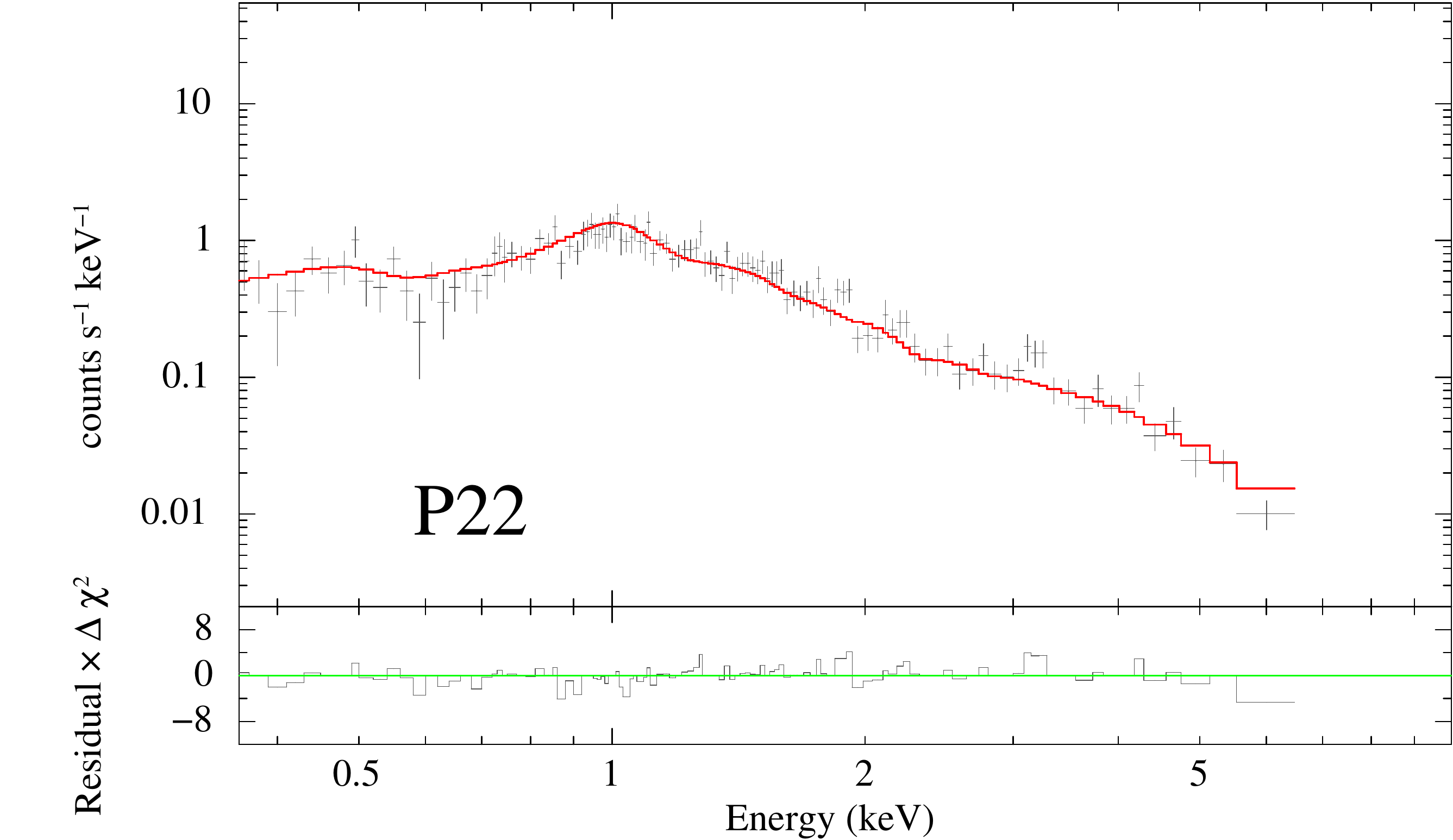}
  \includegraphics[height=8.8cm,angle=-90,trim={0 35 0 0}, clip]{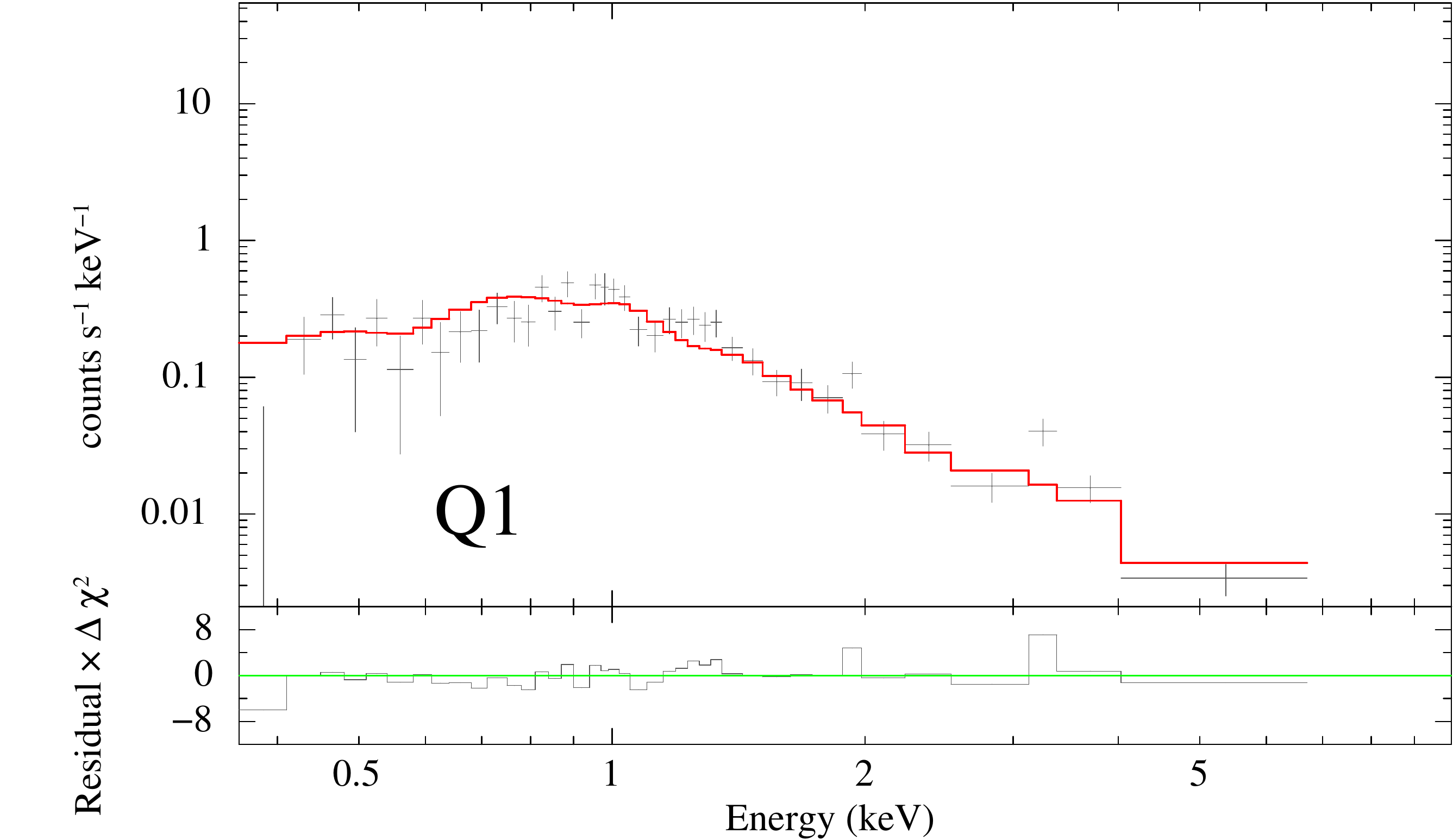}
\contcaption{The online-only material related to Figure~\ref{fig:spectra}}
\label{fig:spectra_online_only_3}
\end{figure*}
\clearpage
\end{document}